# Quantifying disparities in air pollution exposures across the United States using home and work addresses

Priyanka deSouza[1,2,3,*], Susan Anenberg[4], Carrie Makariewicz[1], Manish Shirgaokar[1], Fabio Duarte[3], Carlo Ratti[3], John Durant[5,^], Patrick Kinney[6,^], Deb Niemeier[7,^]

1: Department of Urban and Regional Planning, University of Colorado Denver, CO USA
2: CU Population Center, University of Colorado Boulder, CO, USA
3: Senseable City Lab, Massachusetts Institute of Technology, MA, USA
4: Milken Institute School of Public Health, George Washington University, Washington, DC, USA
5: Department of Civil and Environmental Engineering, Tufts University, Medford, MA, USA
6: Boston University School of Public Health, MA, USA
7: Department of Civil and Environmental Engineering, University of Maryland, College Park, MD 20742

*: Corresponding author (priyanka.desouza@ucdenver.edu)
^: Contributed equally

## Abstract

While human mobility plays a crucial role in determining air pollution exposures and health risks, research to-date has assessed risks based solely on residential location. Here we leveraged a database of ~ 130 million workers in the US and published $PM_{2.5}$ data between 2011-2018 to explore how incorporating information on both workplace and residential location changes our understanding of disparities in air pollution exposure. In general, we observed higher workplace exposures (W) relative to home exposures (H), as well as increasing exposures for non-white and less educated workers relative to the national average. Workplace exposure disparities were higher among racial and ethnic groups and job-types than by income, education, age, and sex. Not considering workplace exposures can lead to systematic underestimations in disparities to exposure among these subpopulations. We also quantified the error in assigning workers H, instead of a weighted home-and-work (HW) exposure. We observed that biases in associations between $PM_{2.5}$ and health impacts by using H instead of HW were highest among urban, younger populations.

**Synopsis**: Research that has evaluated disparities in exposure to $PM_{2.5}$ has not considered mobility patterns. This research finds that considering workplace in addition to residential location results in larger disparities in exposure to $PM_{2.5}$ than previously reported.

# Introduction

Epidemiologic evidence has linked ambient mass concentrations of fine particulate matter (particles with an aerodynamic diameter < 2.5 µm: $PM_{2.5}$) with premature mortality and a multitude of morbidities [1–4]. Accurately capturing $PM_{2.5}$ exposure can be a challenge for large cohort epidemiologic studies, especially given peoples' tendency to move over time through different environments under changing pollutant concentrations [5,6]. According to the 2017 national household transport survey (NHTS), residents in the United States (US) tend to make between 4-5 trips a day, and spend ~ 20% of a calendar year at their workplace [7]. However, given the complexity of personal monitoring, researchers evaluating the impacts of $PM_{2.5}$ in large nation-wide cohorts typically assign exposure estimates on the basis of residential location [8,9]. Depending on the circumstances, this exposure assignment can introduce significant bias in health effect estimates [10,11], which diminishes the power of epidemiologic studies to correctly evaluate associations between ambient pollution concentrations and health effects.

Research has also shown that despite declines in overall $PM_{2.5}$ levels in the US, relative racial disparities in exposures at residential locations have been *increasing* over time [12,13]. It is unknown whether such disparities may be amplified or reduced when exposures at workplaces are also considered. Research has attempted to account for variability in daily exposure by using daily travel surveys, GPS surveys, mobile phone traces, and modeling tools to account for mobility patterns [6,11,14–16]. All these studies are limited by either a small sample due to the effort required to collect mobility data (travel surveys and daily diaries), or a lack of data on the socioeconomic status (SES) of individuals (e.g., when using cell phone traces).

We expand on prior analyses by evaluating home-based exposures (which we term as 'H') and workplace-based exposures (which we term as 'W') for ~130 million workers across the US. This paper represents the nation-wide first effort to evaluate disparities in H *and* W by race, ethnicity, state, urban/rural status, sex, income, education, age and job-type for the years 2011-2018. We also report disparities in H and W relative to policy thresholds for acceptable $PM_{2.5}$ concentrations. By reporting disparities in W as well as H, we show that housing segregation, and facility siting are not the only forces of structural racism at work in generating inequalities in the US. Something more systemic is generating inequalities in exposure experienced by different groups across residential and workplaces. Our work serves as an impetus to better characterize these forces to address environmental injustices in a more holistic manner.

In addition, we quantify the error in exposure when using H compared to using $PM_{2.5}$ levels considering both home and workplaces (which we term "HW"), which we evaluate as the difference between H and HW (H - HW). We identify locations where the error is high and can influence the results from epidemiologic research. We explicitly evaluate biases in associations between H as opposed to HW and various health effects derived from epidemiologic research.

Finally, we report the variation of the calculated error and bias by income, age, state, urban/rural status, and job-type over the years 2011-2018. The error and bias we derive for various subpopulations can be used by epidemiologic researchers to evaluate more robust associations in cohorts, considered.

# Methods

## Population

We used the LODES (version 7.5) dataset (https://lehd.ces.census.gov/data/lodes/LODES7/, Last accessed December 8, 2022) from the US Census Bureau for every year between 2011 and 2018. The LODES data derive from administrative records (e.g., state employment insurance reporting and federal worker earnings records) of home and work addresses of individual workers (aged 14 and over) and are aggregated to the home and work census blocks for a representative sample of workers (n ~ 128 million living in 72,534 tracts and working in 72,032 tracts in 2011; n ~ 144 million living in 72,242 tracts and working in 72,115 tracts in 2011) for every state (excluding Hawaii and Alaska and territories). The data cover some 95% of jobs in the United States (**Figure S1.1**). Information on race, ethnicity, age groups, sex, education, income, and job-type was also acquired from LODES (see section **S1** in the *SI*). LODES consists of three primary datasets at the census block level from 2011-2018: (1) Residence Area Characteristics (RAC), which provides information on workers by census block of residence, (2) Workplace Area Characteristics (WAC), which provides information on workers by census block of employment, and (3) Origin-Destination (OD), which provides information on individual commuting links between home and work census blocks. In this analysis we only considered primary jobs, alone, and not secondary or tertiary jobs.

## Exposure Assessment

We used a well-validated public dataset of annual-averaged ground-level $PM_{2.5}$ concentrations for the years 2011-2018 available at a 1.1 km x 1.1 km resolution across the US[17]. These data comprise estimates of ground-level $PM_{2.5}$ concentrations over North America derived by combining aerosol optical depth (AOD) retrievals from different satellite instruments with the GEOS-Chem chemical transport model. The $PM_{2.5}$ concentration estimates were subsequently adjusted to regional ground-based observations of $PM_{2.5}$ using geographically weighted regressions (GWR). The estimated $PM_{2.5}$ concentrations were generally consistent with ground-based measurements ($R^2$ varying between 0.6 and 0.8). The accuracy of the model was similar for low and high-levels of exposure signifying no large differences in performance between more polluted grid cells (e.g. in urban areas) and less polluted cells (e.g., in rural areas)[17].

We aggregated the LODES OD, RAC and WAC LODES data to the census-tract level to be consistent with the resolution of the pollution data we use. The $PM_{2.5}$ exposure is a reasonable spatial match (at 1.1 $km^2$) to that of census tracts[18]. We then estimated the census tract-level exposures (using the 2010 census tract boundaries) to annual-average PM$_{2.5}$ by using a spatially weighted mean of the grid cells grid points within a census tract using the exact_extract package in R[19]. A map of census-tract PM$_{2.5}$ concentrations is displayed for the year 2011 in **Figure S2.1**. We did not have data for 22 census-tracts in the Florida Keys (islands), and 1 census-tract (also an island) in California; therefore, we excluded these from the analysis.

## Three exposure metrics (H, W, HW, error)

For our home-based exposure (H), we assumed all workers resided in their designated home census tract as extracted from the LODES RAC file. For these, we assign the corresponding PM$_{2.5}$ exposure. For the workplace exposure (W), we assume workers' primary employment locale is within their designated work census tract (these are extracted from the LODES WAC file), and we assign the corresponding PM$_{2.5}$ exposure.

We calculated an annual average population-weighted exposure (H) for each year using **Equation 1**:

$$\overline{PM_{2.5}} = \frac{\sum_h PM_{2.5,h} p_h}{\sum_h p_h} \quad (1)$$

where PM$_{2.5,h}$ denotes the PM$_{2.5}$ concentrations for residential census tract $h$ (H); p$_h$ signifies the number of workers residing in home census tract $h$. When evaluating population-weighted exposure using workplaces (W), PM$_{2.5,w}$ concentrations for work census tract $w$ were used (W) in **Equation 1.** p$_h$ is replaced by p$_w$ corresponding to the number of workers working in census tract $w$. The residential population-weighted PM$_{2.5}$ concentrations in each group $k$ is given by **Equation 2**:

$$\overline{PM_{2.5,k}} = \frac{\sum_h PM_{2.5,h} p_{k,h}}{\sum_h p_{k,h}} \quad (2)$$

where p$_{k,h}$ represents the number of workers in group $k$ living in census tract $h$. PM$_{2.5,h}$ is the PM$_{2.5}$ level in census tract $h$. We estimated these disaggregated exposures for W as well. We also report population-weighted H and W for workers in groups disaggregated by race/ethnicity, sex, education, income, job-type (information in the RAC and WAC files), and urban/rural designation for the years 2011 - 2018. In addition to reporting the population-weighted mean exposures (H, W) for the total population and different subgroups every year we also report the 10th and 90th percentile H and W population-weighted concentrations in supplementary analyses.

The LODES OD files, which have information on home and the corresponding work census tract for workers, was used to evaluate PM$_{2.5}$ exposures for workers considering both home and work census tracts. For the residence-workplace exposure metric (HW), we assumed that individuals were in their workplace census tract for 1,801 hours per year[20] (based on an 8 hour work day, 5 days a week, 45 weeks per year) out of a total of 8760 hours per year (20.6%), and we used their home census tract for the remaining time. We thus evaluate their HW exposure as 79.4% of H + 20.6% of W. We calculate population-weighted HW exposure using **Equation 3**:

$$\overline{PM_{2.5,HW}} = 0.794 \times H + 0.206 \times W \quad (3)$$

We took the difference between H and HW (H minus HW) as the error (i.e., exposure misclassification) caused by using H instead of HW. We report the error and percent error: (100×(H-HW)/H) for all workers for each year between 2011 and 2018, as well as for workers disaggregated by age, income, job-type, and urban/rural census tract location. We also report the variation in the overall error and percent error experienced by workers in each census tract by the fraction of workers belonging to different racial/ethnic groups, income, age, education, job-type, sex, and urban/rural residence in a tract.

To distinguish urban from rural exposures, we used the U.S. Census Bureau's definition of 'urban areas' which includes both "urbanized areas," (population of at least 50,000,) and "urban clusters," (population of greater than 2,500 but less than 50,000). We downloaded urban areas for the year 2014 (the midpoint of the years considered in our study) using the tigris package in R (**Figure S3.1**). Note that from 2011 onwards, the TIGER shapefiles provided by the U.S. Census Bureau incorporate minor tract numbering corrections for 24 census tracts in New York, Arizona and California. In our urban/rural analysis, we exclude these 24 census tracts from our analysis.

# Evaluation of Disparities in H and W

## Absolute and Relative Differences

Our primary disparity metric for quantifying disparities is the absolute difference[21] in population-weighted average $PM_{2.5}$ concentrations (**Equation 3**) between the subpopulation with the highest mean national exposure (most exposed group) and the group with the lowest mean national exposure (least exposed group). In addition, we derived the percent difference relative to the national mean exposure level: {(population-weighted mean of most exposed group) - (population-weighted mean of least exposed group)}x100/national mean exposure. We also include a relative exposure disparity metric, defined as the ratio of population-weighted mean of the most-exposed group/population minus the weighted mean of the least exposed group. In supplementary analyses, as a sensitivity test for conclusions based on comparisons of mean values' rank order for exposure between groups, we evaluated disparities using different metrics of the exposure distribution (i.e., 10th and 90th percentiles).

We also evaluated exposure disparities based on the percentage of workers belonging to different subpopulations living (or working) in a tract. Here, we rank-ordered all census tract bins based on the percentage of a given group (e.g., percentage of white workers). For example, the first census tract bin was the first percentile, and consisted of residents with the smallest percentage of workers living (working) belonging to the group. The last census tract bin comprised tracts with the largest percentage of workers belonging to the group under consideration. The annual exposure ($PM_{2.5,ig}$) for group *i* for the *gth* percentile census tract bin (i.e the average exposure across all census tracts in the *gth* percentile for proportion of residents that belong to a specific group) was calculated for a given year using **Equation 4**.

$$\overline{PM_{2.5,\iota g}} = \frac{\sum_{j=1}^{n_g} PM_{2.5j} p_{ij}}{\sum_{j=1}^{n_g} p_{ij}} \quad (4)$$

where $PM_{2.5j}$ is the exposure for census tract *j*. $p_{ij}$ is the population of group *i* in census tract *j*, and $n_g$ is the number of census tracts in the *gth* percentile tract bin. To quantify disparities at the highest $PM_{2.5}$ concentrations, we estimated and compared the proportion of all workers and workers belonging to different subgroups (by race, sex, income, education, job-type) exposed to the highest versus lowest decile of $PM_{2.5}$.

As reported previously[21], an important limitation to these metrics (based on differences in mean exposure) is a lack of information about disparities across the full exposure distribution. To address this limitation, we used the Atkinson Index, as another metric to quantify exposure inequalities.

## Atkinson Index

We use the Atkinson Index (AI) to quantify H and W exposure inequality. The AI has since been widely used in the environmental inequality literature, and is defined in **Equation 5**:

$$Between - Group\ AI = 1 - \left(\sum_{j=1}^{n} f_j \left[\frac{y_j}{\bar{y}}\right]^{1-\varepsilon}\right)^{\frac{1}{1-\varepsilon}} \quad (5)$$

where *n* represents the number of individuals in the population, $f_j$ represents the fraction of the total population in each subgroup, $y_j$ represents mean exposure of each subgroup, *y* represents the mean exposure over the full population within a given geographic boundary (national, state, rural or urban), and $\epsilon$ represents an explicit inequality aversion parameter.

The AI is a relative measure of inequality. By comparing each subgroup's weighted exposure to the overall population average exposure within a defined geography, the between-group AI represents the magnitude, in relative terms, of exposure disparities between population subgroups. This is an overall measure of inequality between defined subgroups- it does not explicitly provide information about which of those subgroups are most inequitably exposed.

The inequality aversion parameter ($\epsilon$) is an expression of societal concern about inequality. It allocates relative weights across the exposure distribution. The parameter ranges from zero to infinity, with increasing values allocating greater weight to the bottom of the distribution. Unlike income, environmental exposures are worse at higher levels of pollution, so we perform all AI calculations with the inverse of the pollution concentrations to allow for interpretable results. We apply an aversion parameter of 0.75, consistent with the literature[22]. As a sensitivity analysis, we report the AI for alternative inequality aversion parameters (0.25, 0.50, 1.00, 1.25, 1.50, 1.75, 2.00).

## Home and Work Exposure Disparities by State

We also investigated patterns of disparities among the 48 states of the contiguous United States plus the District of Columbia (DC) (hereafter, "states" refers to 48 states and DC, a total of 49 geographic units in state-level related calculations). For each state and subpopulation, we calculated the normalized population-weighted disparity as the difference between the annual exposure for a subpopulation *i* living/working in a state ($PM_{2.5,i}$), and the annual exposure for the total population living/working in the state as a whole ($PM_{2.5,state}$) relative to the annual exposure of workers living/working across the entire United States ($PM_{2.5,national}$) using **Equation 6**:

$$disparity_i = \frac{PM_{2.5,i} - PM_{2.5,state}}{PM_{2.5,national}} \quad (6)$$

We evaluated disparities in H and W of different subpopulations, relative to the overall state average for the years 2011 and 2018.

## Disparities relative to policy-standards

We used a previously established methodology to quantify disparities relative to policy standards[13]. Equality was defined as a state in which equal proportions have H or W exposure higher than a defined threshold. We choose this threshold in relation to the U.S Environmental Protection Agency (EPA) annual-averaged standard (*T*) for $PM_{2.5}$: 12 μg/m$^3$. In supplementary analyses we considered other policy thresholds (*T*) of 10 μg/m$^3$ (World Health Organization (WHO) guideline for annual averaged $PM_{2.5}$ before 2021 and a policy standard that the U.S EPA is considering) and 5 μg/m$^3$ (WHO guideline for annual average $PM_{2.5}$ after 2021).

We define an additional $PM_{2.5}$-related variable (*q*) to quantify disparities in exposures among different population groups. The variable *q* is defined as the percentage of a population with H or W levels above threshold *T*. We calculated a separate *q* for home and work population subgroups of interest. We compute the coefficient of variation (*CoV*) of q, referred to as the 'between group' variance as a metric of disparities relative to the policy standard or T using **Equation 7**:

$$CoV = \frac{\sqrt{Var(q)}}{\mu(q)} \quad (7)$$

where *Var(q)* is the variance of *q*, and μ is the mean of *q*.

## Statistical comparison of H, W, and HW and calculation of Bias

Finally, we statistically compared the distribution of H and HW $PM_{2.5}$ exposures for the workers in our dataset using Wilcoxon Rank Sum Tests.

General classical error theory suggests that *Z = X + E*, where *Z* is the value of the surrogate exposure, *X* is the true value, and *E* is the error in measuring *X*[10]. Although we acknowledge that personalized air pollution exposure measurements would be the best measure of exposure, we consider our 'HW' as the reference exposure in this study. We assume that the 'H' is analogous to *Z*. Mobility-based exposure estimates, 'HW' are closer to *X*. For the classical error model, *E*, defined in **Equation 8** is assumed to be independent of *X*.

$$E = H(Z) - (0.794H + 0.206W)(X) = 0.206H + 0.206W \quad (8)$$

In our dataset E shows varying degrees of correlation with X. Therefore, in the presence of the correlation between E and X, we calculate the bias factor expected in regression coefficients of simple linear models evaluating associations between PM$_{2.5}$ and health effects according to the **Equation 9**:

$$Bias = \frac{\sigma^2 + \phi}{\sigma^2 + 2\phi + \omega^2} \quad (9)$$

Where $\sigma^2$ is the variance of X, $\phi$ is the covariance of (*X,E*) and $\omega^2$ is the variance of *E*. **Equation 9** provides a multiplier that would apply to the relative risk assessment produced using the home exposure estimate H or W instead of HW. For this analysis, we use the OD LODES files which have information on both residential and the corresponding workplace census tract for workers. The OD also contains information on age, income, and job-type but not race. We thus evaluate bias for the total study population for each year 2011 - 2018, as well as for the study population disaggregated by age, income and job-type, and urban/rural.

# Results

## Trends in home and work exposures

The trends in overall H and W, and H and W by SES are displayed in **Figure 1 (**and by job-type in **Figure S4.1** in *Supplementary Information (SI)*). For all subpopulations except for agricultural workers, W is greater than H across the entire study period (**Figure S4.1**). In general, H and W decrease until 2016 and then show a small increase in 2017 and 2018, likely in part, due to increases in wildfire-related PM$_{2.5}$ concentrations [23]. Exposure disparities at the home *and* workplace are visible in our dataset across all years with white workers having the lowest H and W in comparison to all other racial and ethnic groups, except for Native American workers. Although male and female workers experienced the same H, male workers were exposed to slightly higher W concentrations, overall. Men travel further distances than women to work according to the 2017 NHTS [7]. The larger separation between home and work for men likely explains the larger difference observed between H and W. Interestingly, workers in the highest income bracket (> USD $3,333/month) experienced the lowest H but highest W compared to workers who earned less, a possible indicator of greater distance between home and workplaces for wealthier workers (**Figure 1**). Transportation and warehouse workers experienced the highest W across all job-types (**Figure S4.1**).

When evaluating H and W trends disaggregated by urban/rural designation, we observed that rural H and W are lower than the corresponding urban exposures across all subpopulations (**Figure S5.1-S5.2**). In general, there is also a smaller difference between H and W in rural areas likely due to the larger separation between home and workplaces and the greater variation in PM$_{2.5}$ concentrations in urban areas. An important difference was that Native Americans have the lowest exposures in rural areas but have higher exposures than white populations in urban areas (**Figure S5.1**; see **section S5** in the *SI* for more details). A full description of H and W for workers belonging to different subpopulations, disaggregated by urban/rural designation is provided in section **S5** in the *SI*.

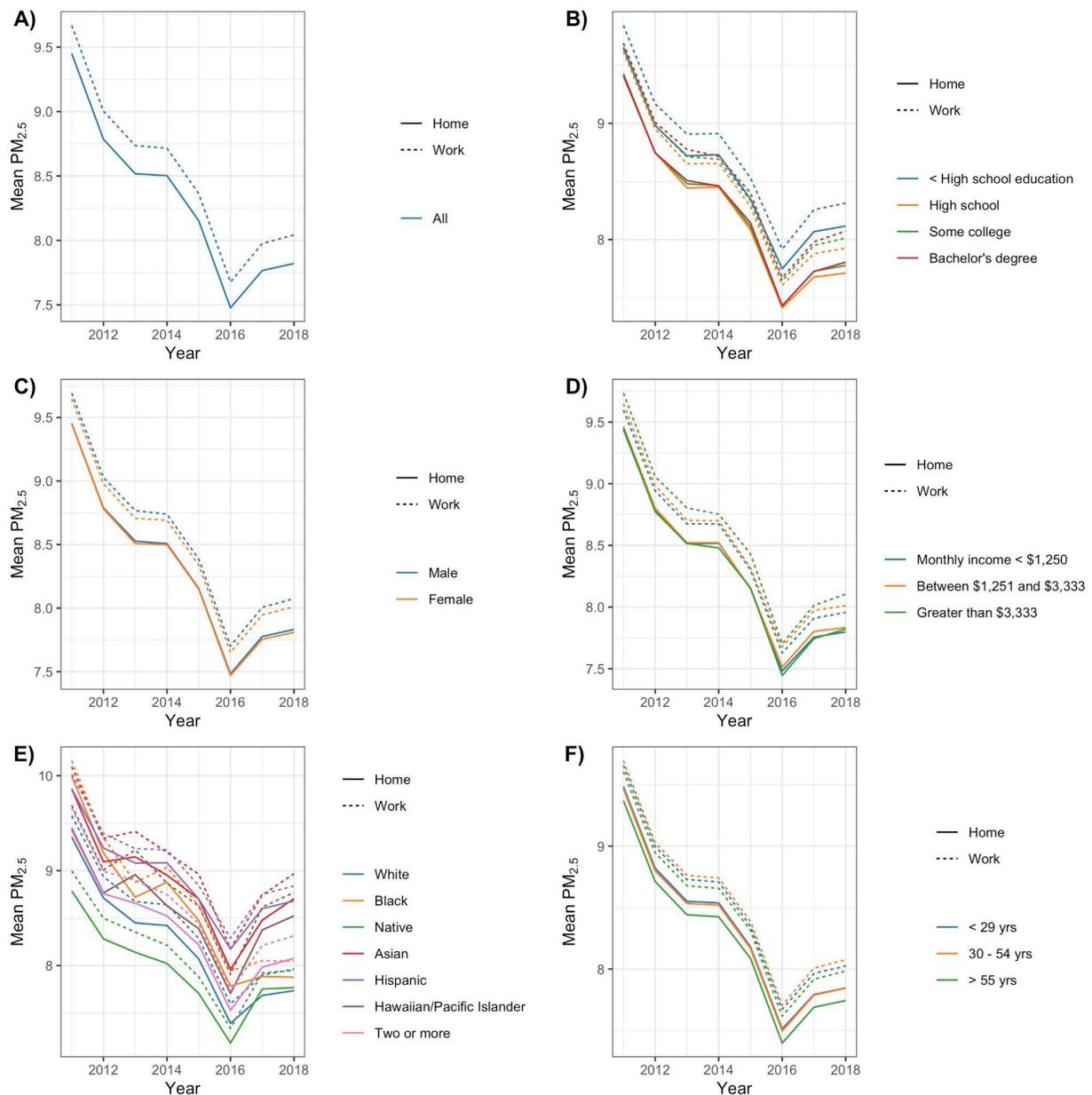

***Figure 1***: *Weighted-mean exposure to PM$_{2.5}$ (µg/m$^3$) experienced at home (H) and work (W) locations disaggregated by race, sex, income, and education*

We next examined the ratio of H (and W) for workers in different SES categories (**Figure 2**) and job-types (**Figure S4.2**) to the H (and W) experienced by all workers. Although overall H and W are decreasing over time, these decreases are not experienced equally over the respective subpopulations. Instead, relative to the national average, H and W for several subpopulations are *increasing* over time. The steepest relative increases were observed in the H and W for the least formally educated workers (to 3.8% and 3.4% higher than national H and W, respectively, in 2018), all non-white workers except Black workers (to 11.3%, 9.0%, and 11.0% higher than the national H for Asian, Hawaiian and Pacific Islander, and Hispanic workers in 2018; and to 11.5%, 9.0%, and 9.9% higher than the national W for Asian, Hawaiian and Pacific Islander, and Hispanic workers in 2018), male workers (to 0.1% higher than the national H for male worker, and 0.4% higher than the national W in 2018), and agricultural workers (to 6.6% higher than national H and 2.4% higher than the national W in 2018). Black worker's H and W are higher than the national average but decreasing over time. Native Americans had lower H and W than the national average until 2016, although H and W relative to the national exposures are increasing. The latter finding is in keeping with past research [24]. Asian workers are disproportionately located in California and Hispanic workers are disproportionately located in the south-west (**Figure S1.2**) where wildfires are increasing $PM_{2.5}$ levels [25], which could explain the increases observed among these groups.

We observe very similar results when evaluating trends in H and W exposures experienced by subgroups relative to national averages disaggregated by urban/rural designation (**Figure S5.3** in the *SI;* full description in section **S5**). We note however, widening of disparities on the basis of race/ethnicity and education over time is more pronounced in urban than rural areas. Exposure levels for workers in the agriculture, forestry, fishing and hunting sector are increasing fastest in urban areas (to 12.5% higher than the national urban H and W in 2018) (**Figure S5.4**). When using population-weighted 10th and 90th percentile $PM_{2.5}$ concentrations instead of population-weighted mean concentrations we observe similar results, indicating that disparities in exposure are observed across the range of $PM_{2.5}$ levels. A full description of trends in H and W using the population-weighted 10th and 90th percentile of $PM_{2.5}$ levels, instead of mean concentrations is provided in section **S6** in the *SI.*

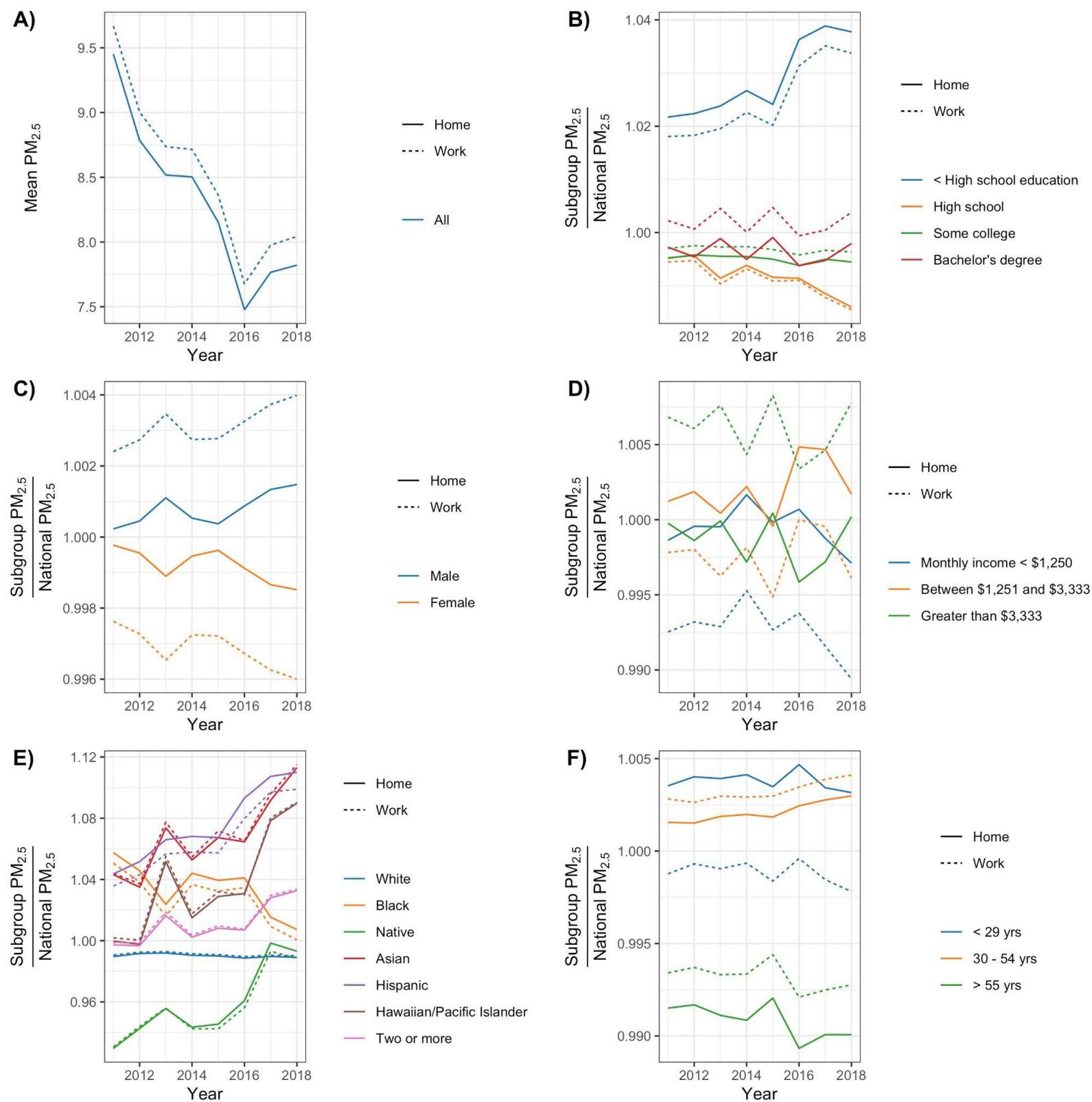

*Figure 2*: National average population-weighted exposure to PM$_{2.5}$ for A) all workers, and Ratio of population-weighted PM$_{2.5}$ exposure for B) Male and Female workers, C) white, Black,

*Hispanic, Asian, Native American workers, D) Workers with a monthly income ≤ USD 1,250/month, USD 1,251 - USD 3,333/month , ≥ USD 3,333/month, E) Workers with No high school degree, High school education, only, Some college or associate degree, Bachelor's degree or advanced degree to national average population-weighted exposure for years 2011 - 2018*

When evaluating the error and percent error in assigning workers exposure based on home, instead of home *and* workplace, we find that there are systematic variations in the error in exposure assignment by subpopulation and geography that could potentially lead to biases in evaluating associations between $PM_{2.5}$ and health (**Figure S7.1**). Overall, the magnitude of the error in assignment of $PM_{2.5}$ to the US working population (by using H instead of HW that better captures mobility patterns) has not changed much over time (overall error is - 0.05 µg/m$^3$ in 2018), while the magnitude of the percent error has generally increased between 2011 and 2018 to -0.6% in 2018 (**Figure S7.2**). The magnitude of the error and percent error is higher in rural than urban areas (**Figure S7.3**). The magnitude of the error and percent error in urban and rural locations varies by subpopulation with the largest error observed for high income populations, and workers in the goods producing sector.

## Quantifying Disparities in H and W

We present the absolute difference, percent difference, and relative difference in H and W experienced by the most exposed and least exposed subpopulations in **Figure 3** corresponding to groups disaggregated by race, education, income, sex and job-type. The magnitude of disparities in H and W by race and job-type are the highest (absolute difference ~ 0.75-1.21 µg/m$^3$, and percent absolute difference ~ 8.6-12.8%), followed by disparities across education levels (absolute difference ~ 0.21-0.39 µg/m$^3$, and percent absolute difference ~ 2.4-5.2%). Disparities based on W were higher than those based on H, overall by race, job-type, income, and sex.

Disparities by race and job-type, derived from using population-weighted 10th and 90th percentile $PM_{2.5}$ concentrations, instead of the mean, remain most pronounced (**Figures S8.1.1-2**). Disparities by race quantified using these metrics are most pronounced in rural areas (absolute difference ~ 0.70-2.1 µg/m$^3$, and percent absolute difference ~ 10.1-20.2%), while those of job-type are most pronounced in urban areas (absolute difference ~ 0.69-1.5 µg/m$^3$, and percent absolute difference ~ 7.5-18.9%). Disparities in W quantified were higher across all categories than H in rural areas. Disparities in W were higher than H by race and sex in urban areas; however, disparities in H by job-type and education were higher than disparities in W (**Figure 3**).

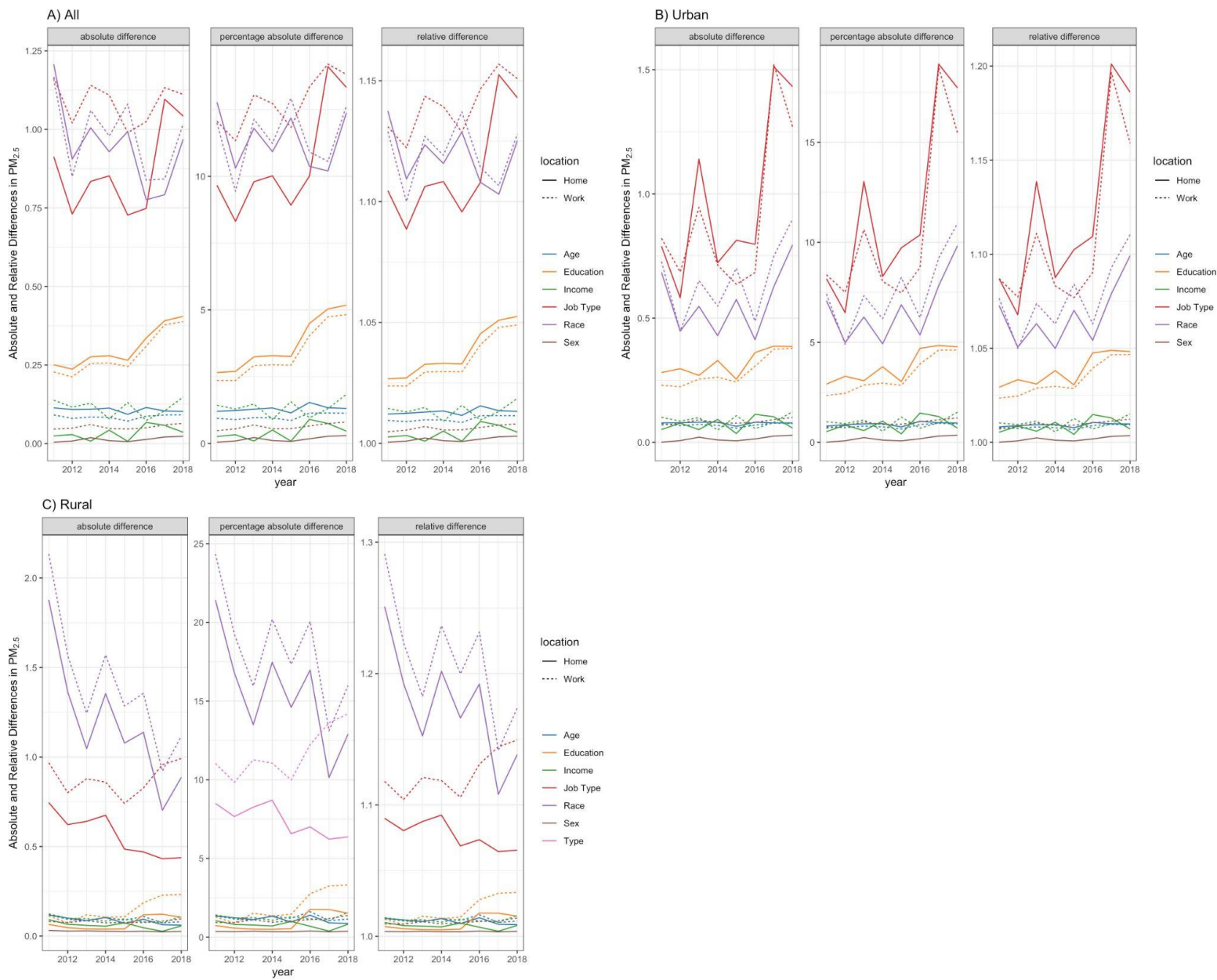

*Figure 3*: Absolute difference (μg/m$^3$), percent absolute difference, and relative difference in exposures between the most exposed and least exposed subpopulations within different groups for A) All workers, B) Workers in urban areas, C) Workers in rural areas

Plots of the mean and 95% CI of H and W experienced by workers living and working in census tracts containing different fractions of subpopulations, categorized by deciles for the years 2011 and 2018 (**Figure S8.2.1**), indicate similar disparities as that described (for more details refer to **section S8.2** in the *SI*). The trends in disparities do not change between the two years considered. We quantified the disparities observed by plotting the difference between PM$_{2.5}$ experienced by census tracts in the top and bottom decile of census tracts characterized by the fraction of different subpopulations living/working in each tract in **Figure S8.2.2** for the years 2011-2018. We observed that the difference in H and W experienced by census tracts with the highest fraction of Asian, and Hispanic residents compared to tracts with the lowest fraction of these groups was increasing over time to ΔW: 2.6 µg/m$^3$, ΔH: 2.2 µg/m$^3$, and ΔW: 2.6 µg/m$^3$, ΔH: 2.4 µg/m$^3$ in 2018, respectively. Similar increases were observed for census tracts categorized based on the prevalence of Native Americans (ΔW: 0.5 µg/m$^3$, ΔH: 0.7 µg/m$^3$ in 2018), Hawaiian and Pacific Islanders (ΔW: µg/m$^3$, ΔH: µg/m$^3$ in 2018), and Two or More Races (ΔW: 0.7 µg/m$^3$, ΔH: 1.4 µg/m$^3$ in 2018). However, for these groups ΔH was higher than ΔW. Census tracts with the highest fraction of white residents had lower H and W compared to census tracts with the lowest fraction of white residents (ΔW: -1.5 µg/m$^3$, ΔH: -1.2 µg/m$^3$ in 2018). No clear pattern was observed in the change in ΔH concentrations over time, although ΔW concentrations were decreasing. Census tracts with the highest fraction of Black residents had higher H and W concentrations than tracts with the lowest fraction of Black residents (ΔW: 1.3 µg/m$^3$, ΔH: 1.3 µg/m$^3$ in 2018). However, ΔH and ΔW levels were decreasing over time. The largest absolute ΔH observed was based on the prevalence of workers with a high school or equivalent degree (ΔH: -1.9 µg/m$^3$ in 2018), while the largest absolute ΔW was observed for tracts based on the prevalence of workers in mining, quarrying, and oil and gas extraction (ΔW: -3.1 µg/m$^3$ in 2018). When repeating this analysis in urban and rural areas, separately, we observe similar trends with some differences (**Figure S8.2.3-5**; for more details refer to **section S8.2** in the *SI*).

Conversely, we display the mean fractions of workers belonging to different subpopulations living/working in census tracts corresponding to different PM$_{2.5}$ concentrations (classified by deciles) for years 2011 and 2018 in **Figure S8.3.1**. We observe very different patterns in the mean fractions of workers belonging to a specific subpopulation exposed to different deciles of PM$_{2.5}$ concentrations for home versus workplace. For example, census tracts with the highest PM$_{2.5}$ levels have the lowest fraction of white residents residing in them (9.7%), while the highest fraction of white residents working in them (11.3%). The difference between the percent of white residents in census tracts corresponding to highest H compared to the lowest H, based on deciles is -0.5%, while for W is 2.9% in 2018 (**Figure S8.3.2**). We did not observe such trends in other subpopulations. Census tracts with the highest PM$_{2.5}$ concentrations had the workers with little formal education and the highest fraction of all non-white subpopulations (except for Native Americans) living and working in them. Specifically, in 2018 the differences between the percent of Black, Hispanic, and Asian workers with the highest H compared to the lowest H, based on deciles were 2.4%, 19.1%, 22.7%, respectively, while for W were 4.2%, 21.1%, 25.7%, respectively (**Figure S8.3.2**). The difference in trends observed could be due to more white workers living in suburbs than non-white residents who tend to live in inner-city areas closer to places of work. Inequalities in access to commuting infrastructure that allow access to certain jobs is an important contributor to these patterns [26]. Our work thus demonstrates that remedying inequalities in transportation infrastructure is also core to achieving environmental justice.

When evaluating trends, disaggregated by urban and rural, we noticed similar trends with some differences (**Figures S8.3.3-5**). The magnitude of the difference between the fraction of subpopulations in tracts with the highest H and W (top decile) compared to those with the lowest H and W (bottom decile) was lower in rural compared to urban locations. The magnitude of the difference in subpopulations in tracts with the highest H compared to those with the lowest H were lower than the difference in subpopulations in tracts with the highest W compared with the lowest W in urban areas, while we observed mixed results in rural areas.

When we evaluate the trend of the Atkinson index, we observed that disparities in race were larger than disparities by education, income, sex, age or job-type (**Figure S8.4.1**). However, in general, as observed in other research [22], the Atkinson Index values estimates are small due to small absolute differences in H and W between population groups. Small differences in H and W, however, can contribute to important health disparities[27,28]. In addition we observed that disparities in W were greater than in H. These disparities were robust to the choice of the inequality aversion parameter needed to estimate the Atkinson index (**Figure S8.4.2**). We observed similar results when disaggregating our results by urban/rural designation (**Figures S8.4.3-4**).

We investigated patterns of disparities among the 48 states of the contiguous United States plus the District of Columbia (DC) (hereafter, "states" refers to 48 states and DC, a total of 49 geographic units in state-level related calculations; **Figure 4**). For each state and subpopulation, we calculated the normalized population-weighted disparity as the difference between the annual exposure for a subpopulation living/working in a state, and the annual exposure for the total population living/working in the state as a whole relative to the annual exposure of workers living/working across the entire United States.

We observed that the white population in every state experienced lower H and W compared to the state-averages. Non-white populations experienced higher H and W relative to the state-average in almost every state. Unsurprisingly, workers belonging to certain job-types such as agriculture experienced lower H and W than the state-wide averages for most states. The patterns observed did not change very much between the years 2011 and 2018. There were substantial variations in disparities across states for some subpopulations. For example, the W for Black workers in New York and Nebraska relative to the state-wide average exposure is higher than state average by an amount of 6.4% and 8.2% of the national mean W in 2011, respectively and 5.0% and 5.8%, respectively in 2018. The H of agricultural workers in Washington state is higher than the state-wide average in 2018 by 10% of the national mean likely due to wildfires [25]. Our state-specific analysis also revealed new emerging sources of disparities in pollution. For example, W experienced by transportation and warehouse workers and workers in wholesale trade in California and Arizona was higher than the state-average by 12.3% and 11.2% of the national average in 2018, up from 10.0 and 8.6%, respectively in 2011.

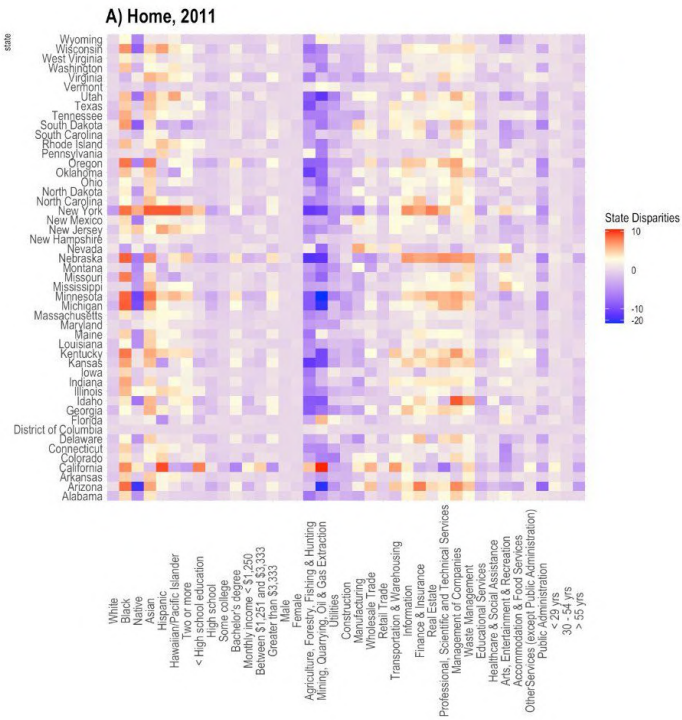
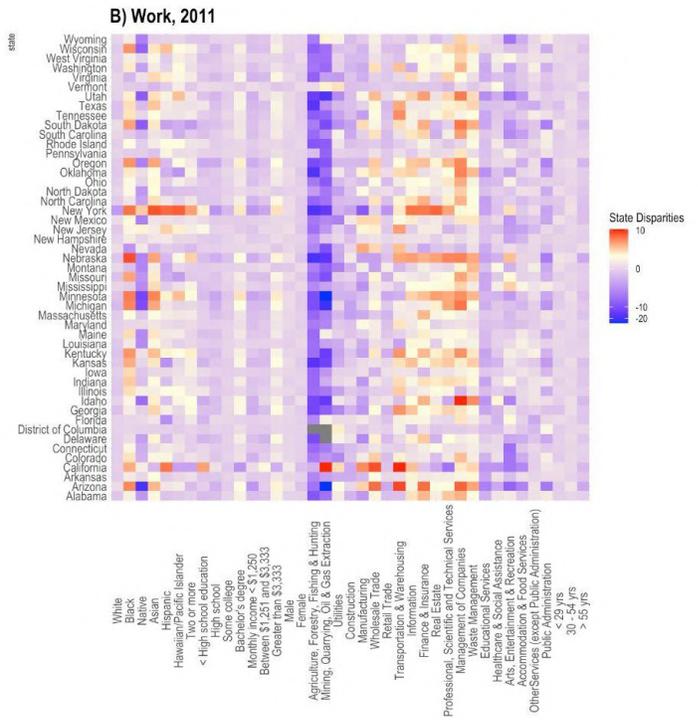
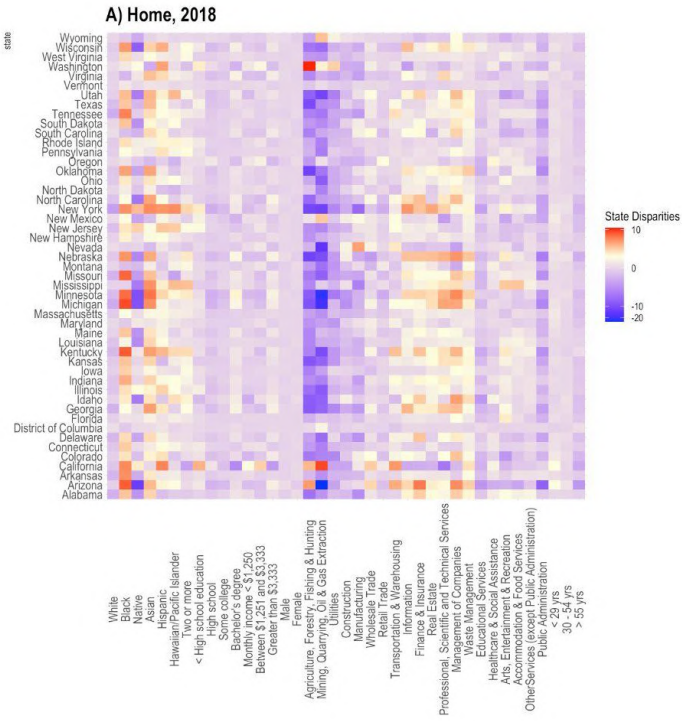
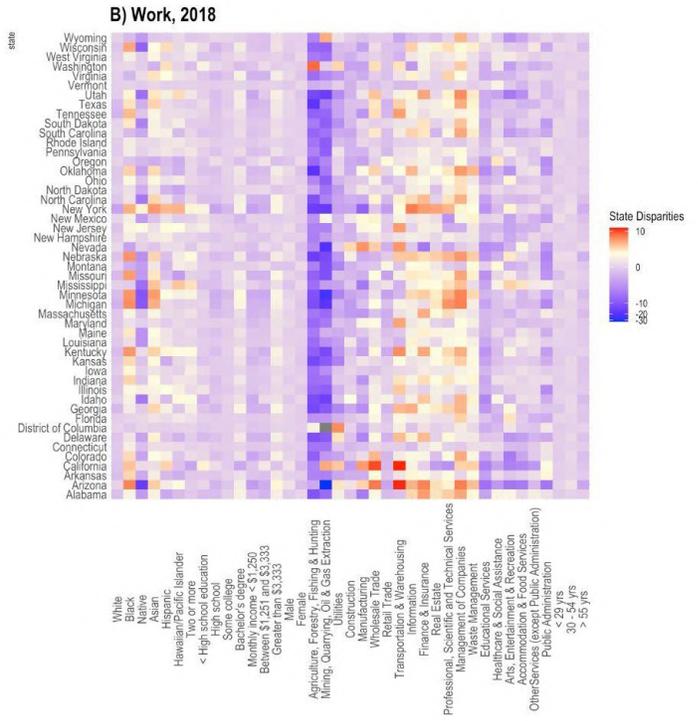

*Figure 4*: Disparities in H and W by state for the years 2011 and 2018. Gray indicates missing data

## Disparities Relative to Policy Thresholds

Finally, we evaluated disparities in H and W, relative to the annual average National Ambient Air Quality Standard of 12 µg/m$^3$ using a previously established methodology[13]. Briefly, we estimated across the study period the proportion of every subpopulation that was exposed to PM$_{2.5}$ levels higher than 12 µg/m$^3$. A state of equality (or lack of relative disparities) among various populations was defined as a state of equal proportions above the chosen safety standard across groups. We provide numerical summaries of disparities in the proportions of different subpopulations exposed to PM$_{2.5}$ levels above the chosen standard by using the coefficient of variation (CoV) described in *Methods* in **Table S8.5.1**.

We observe that the relative disparities in the proportion of workers belonging to different racial groups and working in different sectors exposed to PM$_{2.5}$ concentrations above the current standard are higher than by other categorizations (**Figure S8.5.1**). We find these results are robust to using standard levels of 10 and 5 µg/m$^3$, with the only difference being that when we used the latter threshold, disparities on the basis of education were much smaller, while that by race and education continued to be the largest in magnitude (**Figures S8.5.2-3**).

## Statistical comparison of H, W, and HW and calculation of Bias

Wilcoxon tests revealed that H, W and HW varied significantly across all years, in both urban and rural areas. We display the estimates of bias in associations between pollution and health impacts derived from epidemiologic studies when using H instead of HW over time (**Figure 5**). Overall, we find that the bias is decreasing over time. Differences in bias are observed for different subpopulations. Overall and especially in urban areas, associations between health and pollution levels for young workers experience the greatest bias. In rural areas, it is for workers in the trade, transportation and utility sector that we will obtain the greatest bias in health impact calculations. We did not have information on the home and corresponding work census tracts, for workers disaggregated by race and ethnicity, and therefore could not evaluate biases by race.

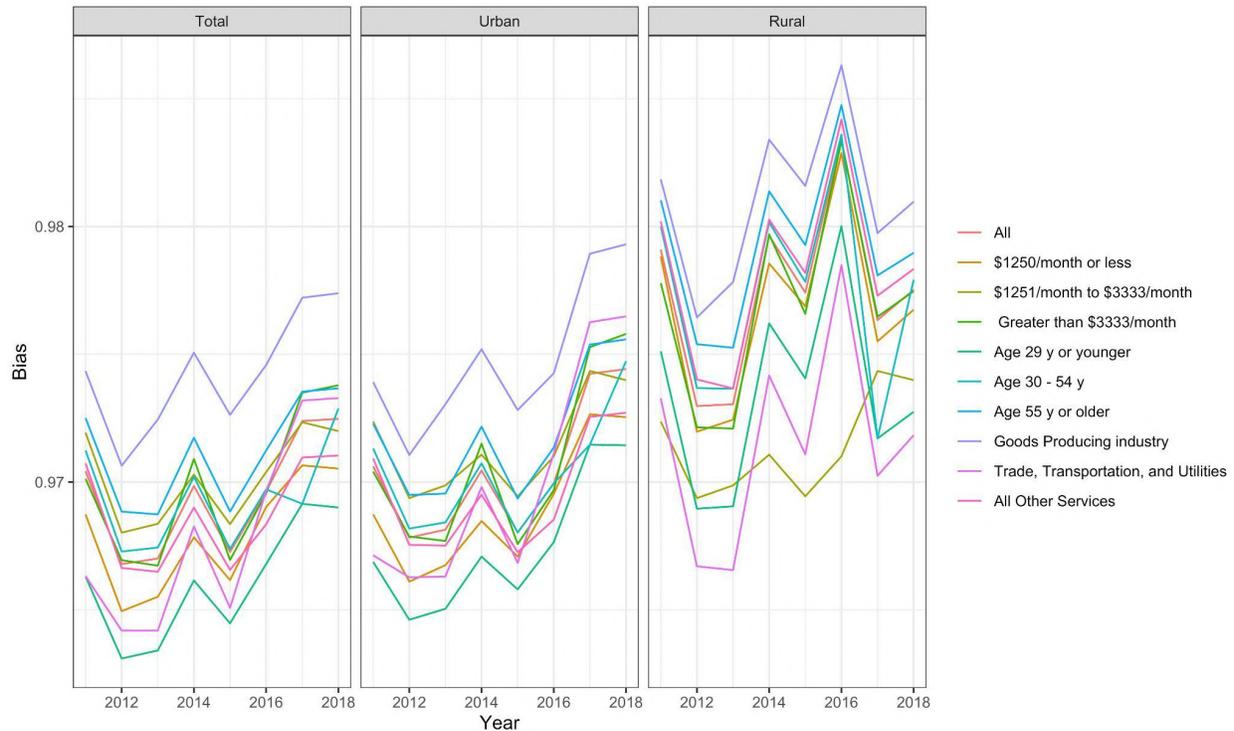

*Figure 5*: Bias in associations derived between pollution and health outcomes when using H instead of HW between 2011 - 2018 for different subpopulations, as well as disaggregated by urban and rural areas.

## Discussion

Using a rich dataset of home and workplaces for ~ 130 million workers in the United States from 2011-2018, we present the first evaluation of disparities in $PM_{2.5}$ exposures at residential *and* workplace census tracts by race, ethnicity, education status, income, age, and job-type using a wide range of metrics. Importantly, we observed that disparities estimates when using home locations, alone, instead of home and work locations could result in systematically underestimating exposure to $PM_{2.5}$ concentrations and disparities in exposure for racial and ethnic minority groups, in particular. We also identified that there were significant disparities in exposure to pollution by sector of employment, rivaling that by racial and ethnic groups, suggesting that job-type is an important axis to evaluate environmental justice concerns. Disaggregating our data by urban/rural designations accentuated some of the disparities, identified. Our work also uncovered new forms of environmental injustice that have emerged over the last decade and which demand our urgent attention. Specifically, we observed that the workplace-based exposure of warehouse and transportation workers in California and Arizona were increasing dramatically in recent years compared to the national average. In addition, we evaluated biases in associations derived in epidemiologic research between $PM_{2.5}$ and health outcomes by assigning workers' exposure on the basis of home location alone, instead of considering home and workplace. Our results can allow epidemiologists to more robustly evaluate associations between pollution and health outcomes. We tested the robustness of our results using many sensitivity analyses.

Our work could be strengthened by addressing some caveats. First, the $PM_{2.5}$ exposures used in our study rely on satellite-derived datasets and are therefore subject to error. Nonetheless, the performance of this data has been validated [17] finding that $PM_{2.5}$ values estimated are generally consistent with direct ground-based $PM_{2.5}$ values. Still it is important to interpret this dataset with caution. Second, when assigning 'true' exposures in our study, we only consider home and work locations, and do not consider exposures during commute, detailed mobility patterns, the built environment which can impact indoor and outdoor levels of pollution experienced, which can impact our results. Relatedly, we only consider the primary place of employment of all workers in this study, and not secondary of tertiary jobs. We also assume that all workers in our dataset spend the same amount of time at work. Finally, the

Nonetheless, we demonstrate the importance of considering primary workplace locations in addition to home locations when evaluating disparities in exposure to pollution. Further research is needed to explore the drivers of disparities in home and workplace locations.

# Acknowledgements

None

# Funding

PD is grateful for funding from the Presidential Initiative that supported this work.

# Author Contributions

Conceptualization: PD
Methodology: PD
Formal analysis: PD
Investigation: All authors
Writing-original draft: PD
Writing- review and editing: All authors

# Competing Interests

None

# Data and Material Availability

All the data used in this article is publicly available.

# References

(1) deSouza, P.; Braun, D.; Parks, R. M.; Schwartz, J.; Dominici, F.; Kioumourtzoglou, M.-A. Nationwide Study of Short-Term Exposure to Fine Particulate Matter and Cardiovascular Hospitalizations Among Medicaid Enrollees. *Epidemiology* **2020**, *32* (1), 6–13. https://doi.org/10.1097/EDE.0000000000001265.

(2) deSouza, P. N.; Hammer, M.; Anthamatten, P.; Kinney, P. L.; Kim, R.; Subramanian, S. V.; Bell, M. L.; Mwenda, K. M. Impact of Air Pollution on Stunting among Children in Africa. *Environ Health* **2022**, *21* (1), 128. https://doi.org/10.1186/s12940-022-00943-y.

(3) deSouza, P. N.; Dey, S.; Mwenda, K. M.; Kim, R.; Subramanian, S. V.; Kinney, P. L. Robust Relationship between Ambient Air Pollution and Infant Mortality in India. *Science of The Total Environment* **2022**, *815*, 152755. https://doi.org/10.1016/j.scitotenv.2021.152755.

(4) Boing, A. F.; deSouza, P.; Boing, A. C.; Kim, R.; Subramanian, S. V. Air Pollution, Socioeconomic Status, and Age-Specific Mortality Risk in the United States. *JAMA Network Open* **2022**, *5* (5), e2213540. https://doi.org/10.1001/jamanetworkopen.2022.13540.

(5) Larkin, A.; Hystad, P. Towards Personal Exposures: How Technology Is Changing Air Pollution and Health Research. *Curr Envir Health Rpt* **2017**, *4* (4), 463–471. https://doi.org/10.1007/s40572-017-0163-y.

(6) Nyhan, M.; Grauwin, S.; Britter, R.; Misstear, B.; McNabola, A.; Laden, F.; Barrett, S. R. H.; Ratti, C. "Exposure Track"—The Impact of Mobile-Device-Based Mobility Patterns on Quantifying Population Exposure to Air Pollution. *Environ. Sci. Technol.* **2016**, *50* (17), 9671–9681. https://doi.org/10.1021/acs.est.6b02385.

(7) Guckin, N.; Fucci, A. Summary of Travel Trends: 2017 National Household Travel Survey. *US Dept. Transp. Federal Highway Administration, Washington, DC, USA, Tech. Rep. FHWA-PL-18-019* **2018**.

(8) Josey, K. P.; deSouza, P.; Wu, X.; Braun, D.; Nethery, R. Estimating a Causal Exposure Response Function with a Continuous Error-Prone Exposure: A Study of Fine Particulate Matter and All-Cause Mortality. *JABES* **2022**. https://doi.org/10.1007/s13253-022-00508-z.

(9) Krewski, D.; Burnett, R.; Jerrett, M.; Pope, C. A.; Rainham, D.; Calle, E.; Thurston, G.; Thun, M. Mortality and Long-Term Exposure to Ambient Air Pollution: Ongoing Analyses Based on the American Cancer Society Cohort. *Journal of Toxicology and Environmental Health, Part A* **2005**, *68* (13–14), 1093–1109. https://doi.org/10.1080/15287390590935941.

(10) Setton, E.; Marshall, J. D.; Brauer, M.; Lundquist, K. R.; Hystad, P.; Keller, P.; Cloutier-Fisher, D. The Impact of Daily Mobility on Exposure to Traffic-Related Air Pollution and Health Effect Estimates. *J Expo Sci Environ Epidemiol* **2011**, *21* (1), 42–48. https://doi.org/10.1038/jes.2010.14.

(11) Nyhan, M. M.; Kloog, I.; Britter, R.; Ratti, C.; Koutrakis, P. Quantifying Population Exposure to Air Pollution Using Individual Mobility Patterns Inferred from Mobile Phone Data. *J Expo Sci Environ Epidemiol* **2019**, *29* (2), 238–247. https://doi.org/10.1038/s41370-018-0038-9.

(12) Colmer, J.; Hardman, I.; Shimshack, J.; Voorheis, J. Disparities in PM2.5 Air Pollution in the United States. *SCIENCE*, 2020, *369*, 575+. https://doi.org/10.1126/science.aaz9353.

(13) Jbaily, A.; Zhou, X.; Liu, J.; Lee, T.-H.; Kamareddine, L.; Verguet, S.; Dominici, F. Air

Pollution Exposure Disparities across US Population and Income Groups. *Nature* **2022**, *601* (7892), 228–233. https://doi.org/10.1038/s41586-021-04190-y.

(14) Shekarrizfard, M.; Minet, L.; Miller, E.; Yusuf, B.; Weichenthal, S.; Hatzopoulou, M. Influence of Travel Behaviour and Daily Mobility on Exposure to Traffic-Related Air Pollution. *Environmental Research* **2020**, *184*, 109326. https://doi.org/10.1016/j.envres.2020.109326.

(15) Gulliver, J.; Briggs, D. J. Personal Exposure to Particulate Air Pollution in Transport Microenvironments. *Atmospheric Environment* **2004**, *38* (1), 1–8. https://doi.org/10.1016/j.atmosenv.2003.09.036.

(16) Yu, X.; Stuart, A. L.; Liu, Y.; Ivey, C. E.; Russell, A. G.; Kan, H.; Henneman, L. R. F.; Sarnat, S. E.; Hasan, S.; Sadmani, A.; Yang, X.; Yu, H. On the Accuracy and Potential of Google Maps Location History Data to Characterize Individual Mobility for Air Pollution Health Studies. *Environmental Pollution* **2019**, *252*, 924–930. https://doi.org/10.1016/j.envpol.2019.05.081.

(17) Hammer, M. S.; van Donkelaar, A.; Li, C.; Lyapustin, A.; Sayer, A. M.; Hsu, N. C.; Levy, R. C.; Garay, M. J.; Kalashnikova, O. V.; Kahn, R. A.; Brauer, M.; Apte, J. S.; Henze, D. K.; Zhang, L.; Zhang, Q.; Ford, B.; Pierce, J. R.; Martin, R. V. Global Estimates and Long-Term Trends of Fine Particulate Matter Concentrations (1998–2018). *Environ. Sci. Technol.* **2020**. https://doi.org/10.1021/acs.est.0c01764.

(18) Kerr, G. H.; Goldberg, D. L.; Anenberg, S. C. COVID-19 Pandemic Reveals Persistent Disparities in Nitrogen Dioxide Pollution. *Proceedings of the National Academy of Sciences* **2021**, *118* (30), e2022409118. https://doi.org/10.1073/pnas.2022409118.

(19) Baston, D.; ISciences, L.; Baston, M. D. Package 'Exactextractr.' *terra* **2022**, *1*, 17.

(20) Inc, C. *Working hours by country and industry*. Clockify. https://clockify.me/working-hours (accessed 2023-02-26).

(21) Liu, J.; Clark, L. P.; Bechle, M. J.; Hajat, A.; Kim, S.-Y.; Robinson, A. L.; Sheppard, L.; Szpiro, A. A.; Marshall, J. D. Disparities in Air Pollution Exposure in the United States by Race/Ethnicity and Income, 1990–2010. *Environmental Health Perspectives 129* (12), 127005. https://doi.org/10.1289/EHP8584.

(22) Rosofsky, A.; Levy, J. I.; Zanobetti, A.; Janulewicz, P.; Fabian, M. P. Temporal Trends in Air Pollution Exposure Inequality in Massachusetts. *Environmental Research* **2018**, *161*, 76–86. https://doi.org/10.1016/j.envres.2017.10.028.

(23) Burke, M.; Childs, M. L.; De la Cuesta, B.; Qiu, M.; Li, J.; Gould, C. F.; Heft-Neal, S.; Wara, M. Wildfire Influence on Recent US Pollution Trends. National Bureau of Economic Research January 2023. https://doi.org/10.3386/w30882.

(24) Li, M.; Hilpert, M.; Goldsmith, J.; Brooks, J. L.; Shearston, J. A.; Chillrud, S. N.; Ali, T.; Umans, J. G.; Best, L. G.; Yracheta, J.; van Donkelaar, A.; Martin, R. V.; Navas-Acien, A.; Kioumourtzoglou, M.-A. Air Pollution in American Indian Versus Non–American Indian Communities, 2000–2018. *Am J Public Health* **2022**, *112* (4), 615–623. https://doi.org/10.2105/AJPH.2021.306650.

(25) Childs, M. L.; Li, J.; Wen, J.; Heft-Neal, S.; Driscoll, A.; Wang, S.; Gould, C. F.; Qiu, M.; Burney, J.; Burke, M. Daily Local-Level Estimates of Ambient Wildfire Smoke PM2.5 for the Contiguous US. *Environ. Sci. Technol.* **2022**, *56* (19), 13607–13621. https://doi.org/10.1021/acs.est.2c02934.

(26) Bunten, D. M.; Fu, E.; Rolheiser, L.; Severen, C. The Problem Has Existed Over Endless Years: Racialized Difference in Commuting, 1980–2019. Rochester, NY April 1, 2022. https://doi.org/10.21799/frbp.wp.2022.13.

(27) Atkinson, R. W.; Kang, S.; Anderson, H. R.; Mills, I. C.; Walton, H. A. Epidemiological Time Series Studies of PM2.5 and Daily Mortality and Hospital Admissions: A Systematic Review and Meta-Analysis. *Thorax* **2014**, *69* (7), 660–665. https://doi.org/10.1136/thoraxjnl-2013-204492.

(28) Shi, L.; Zanobetti, A.; Kloog, I.; Coull, B. A.; Koutrakis, P.; Melly, S. J.; Schwartz, J. D. Low-Concentration PM2.5 and Mortality: Estimating Acute and Chronic Effects in a Population-Based Study. *Environmental Health Perspectives* **2016**, *124* (1), 46–52. https://doi.org/10.1289/ehp.1409111.

# Supplementary Information for

Quantifying population exposure to air pollution using home and work addresses

## S1 Subgroup Information in the RAC, WAC and OD files

Additional information on race (white alone, Black alone, American Indian or Alaska Native, Asian alone, Native Hawaiian or other Pacific Islander, Two or More Race Groups), ethnicity (Hispanic or Latino, Not Hispanic or Latino), age (≤ 29 y, 30 - 54 y, ≥ 55 y), sex (Male, Female), education (Less than highschool, Highschool or equivalent, Some college or Associate degree, Bachelor's degree or advanced degree), income (≤ USD $1,250/month, USD $1,251 - USD $3,333/month, and ≥ USD $3,333/month) and job-type (Agriculture/Forestry/Fishing/Hunting, Mining/Quarrying/Oil and Gas Extraction, Utilities, Construction, Manufacturing, Wholesale Trade, Retail Trade, Transportation and Warehousing, Information, Finance and Insurance, Real Estate and Rental and Leasing, Professional/Scientific and Technical Services, Management of Companies and Enterprises, Administrative and Support and Waste Management and Remediation Services, Educational Services, Healthcare and Social Assistance, Arts/Entertainment and Recreation, Accommodation and Food Services, Other Services except Public Administration), Public Administration is also available for the years considered in this analysis for RAC and WAC (**Tables S1.1-1.2**). Maps of the % of workers living/working in census tracts across the United States belonging to different subpopulations are displayed in **Figures S1.1-S1.6** for the year 2011.

The OD files contain additional information on age (≤ 29 y, 30 - 54 y, ≥ 55 y), income (≤ USD $1,250/month, USD $1,251 - USD $3,333/month, and ≥ USD $3,333/month), and type of job (Goods Producing Industry sectors, Trade/ Transportation/Utilities sectors, All Other Services sectors) (**Table S1.3**).

The number of workers residing in the 48 states + Washington D.C from the RAC files is slightly higher (0.01%) than the total number of workers obtained from the WAC files working in the same geographic area. This is because there are a small number of workers residing in the geographic boundary of this study whose primary work location are in geographies that we have excluded (Alaska, Hawaii, US territories). Given the small difference between these numbers, we ignore this discrepancy in our analysis (**Tables S1.1**-**S1.3**).

Overall, we had information on workers who lived and worked in the 49 states in the US (including D.C.) for the years 2011 - 2018. Sample sizes by year and by population subgroups are provided in **Tables S1.1-S1.3** for the Residence Area Characteristics (RAC), Work Area Characteristics (WAC), and Origin-Destination (OD) datasets, respectively, and in **Figures S1.1-1.6**.

*Table S1.1*: *Number of total workers and workers belonging to different subpopulations disaggregated by race, sex, income, education, and job type from RAC. Note: The number of workers residing in the 48 states + Washington D.C from the RAC files is more (0.01%) than the total number of workers obtained from the WAC files working in the same geographic area. This is because there are a small number of workers residing in the geographic boundary of this study whose primary work location are in geographies that we have excluded (Alaska, Hawaii, US territories). Given the small difference between these numbers, we ignore this discrepancy in our analysis.*

|  | 2011 | 2012 | 2013 | 2014 | 2015 | 2016 | 2017 | 2018 |
|---|---|---|---|---|---|---|---|---|
|  | Home | Home | Home | Home | Home | Home | Home | Home |
| All | 128,413,842 | 129,559,623 | 131,796,807 | 134,487,437 | 137,148,401 | 140,164,755 | 141,741,355 | 143,915,044 |
| Race |  |  |  |  |  |  |  |  |
| white | 103,890,935 (80.9%) | 104,514,073 (80.7%) | 106,114,308 (80.5%) | 107,976,224 (80.3%) | 109,769,321 (80.0%) | 109,828,700 (78.4%) | 110,564,409 (78.0%) | 111,800,054 (77.7%) |
| Black | 15,291,450 (11.9%) | 15,720,593 (12.1%) | 16,164,737 (12.3%) | 16,778,674 (12.5%) | 17,423,785 (12.7%) | 18,434,474 (13.2%) | 18,852,732 (13.3%) | 19,324,879 (13.4%) |
| Asian | 6,373,883 (5.0%) | 6,397,730 (4.9%) | 6,490,714 (4.9%) | 6,590,970 (4.9%) | 6,697,936 (4.9%) | 8,019,384 (5.7%) | 8,311,128 (5.9%) | 8,616,276 (6.0%) |
| Native | 1,068,985 (0.8%) | 1,083,889 (0.8%) | 1,107,679 (0.8%) | 1,134,884 (0.8%) | 1,168,758 (0.9%) | 1,264,599 (0.9%) | 1,288,153 (0.9%) | 1,323,460 (0.9%) |
| Hawaiian or Pacific Islander | 181,487 (0.1%) | 183,249 (0.1%) | 189,769 (0.1%) | 196,454 (0.1%) | 203,712 (0.1%) | 240,094 (0.2%) | 247,737 (0.2%) | 256,343 (0.2%) |
| Two or More Races | 1,607,102 (1.3%) | 1,660,089 (1.3%) | 1,729,600 (1.3%) | 1,810,231 (1.3%) | 1,884,889 (1.4%) | 2,377,504 (1.7%) | 2,477,196 (1.7%) | 2,594,032 (1.8%) |
| Ethnicity |  |  |  |  |  |  |  |  |
| Hispanic | 16,247,537 (12.7%) | 16,496,473 (12.7%) | 16,977,604 (12.9%) | 17,495,920 (13.0%) | 17,993,282 (13.1%) | 20,465,199 (14.6%) | 21,148,149 (14.9%) | 21,956,413 (15.3%) |
| Income | | | | | | | | |
| Income 1 | 32,368,490 (25.2%) | 32,416,296 (25.0%) | 32,634,684 (24.8%) | 32,946,514 (24.5%) | 32,774,070 (23.9%) | 32,723,516 (23.3%) | 32,220,672 (22.7%) | 31,669,349 (22.0%) |
| Income 2 | 45,951,248 (35.8%) | 45,797,623 (35.3%) | 45,944,167 (34.9%) | 46,070,193 (34.3%) | 46,525,284 (33.9%) | 47,068,342 (33.6%) | 46,468,385 (32.8%) | 46,118,390 (32.0%) |
| Income 3 | 50,094,104 (39.0%) | 51,345,704 (39.6%) | 53,217,956 (40.4%) | 55,470,730 (41.2%) | 57,849,047 (42.2%) | 60,372,897 (43.1%) | 63,052,298 (44.5%) | 66,127,305 (45.9%) |
| Education | | | | | | | | |
| No highschool | 11,586,110 (9.0%) | 12,055,061 (9.3%) | 12,532,091 (9.5%) | 13,081,520 (9.7%) | 13,695,,340 (10.0%) | 13,755,591 (9.8%) | 14,306,692 (10.1%) | 14,891,593 (10.3%) |
| Highschool | 26,206,875 (20.4%) | 26,779,521 (20.7%) | 27,322,866 (20.7%) | 27,996,145 (20.8%) | 28,642,806 (20.9%) | 28,388,993 (20.3%) | 28,789,489 (20.3%) | 29,276,071 (20.3%) |
| College | 31,395,540 (24.4%) | 31,967,744 (24.7%) | 32,535,563 (24.7%) | 33,134,587 (24.6%) | 33,701,377 (24.6%) | 33,950,497 (24.2%) | 34,249,529 (24.2%) | 34,667,369 (24.1%) |
| Advanced | 29,478,944 (23.0%) | 29,707,063 (22.9%) | 29,935,448 (22.7%) | 30,090,018 (22.4%) | 30,223,570 (22.0%) | 31,600,065 (22.5%) | 31,599,353 (22.3%) | 31,860,537 (22.1%) |
| Sex | | | | | | | | |
| Male | 63,779,510 (49.7%) | 64,671,873 (49.9%) | 65,907,346 (50.0%) | 67,398,552 (50.1%) | 68,729,391 (50.1%) | 70,262,989 (50.1%) | 70,987,136 (50.1%) | 72,018,189 (50.0%) |
| Female | 64,634,332 (50.3%) | 64,887,750 (50.1%) | 65,889,461 (50.0%) | 67,088,885 (49.9%) | 68,419,010 (49.9%) | 69,901,766 (49.9%) | 70,754,219 (49.9%) | 71,896,855 (50.0%) |
| Age | | | | | | | | |
| ≤ 29 years | 29,746,373 (23.2%) | 29,050,234 (22.4%) | 29,470,839 (22.4%) | 30,0185,167 (22.4%) | 30,885,308 (22.5%) | 32,469,609 (23.2%) | 32,796,292 (23.1%) | 33,219,474 (23.1%) |
| 30 - 54 years | 72,959,236 (56.8%) | 73,129,262 (56.4%) | 73,759,425 (56.0%) | 74,545,406 (55.4%) | 75,302,743 (54.9%) | 76,030,458 (54.2%) | 76,338,706 (53.9%) | 76,997,052 (53.5%) |
| ≥ 55 yeas | 25,708,233 (20.0%) | 27,380,127 (21.1%) | 28,566,543 (21.7%) | 29,756,864 (22.1%) | 30,960,350 (22.6%) | 31,664,688 (22.6%) | 32,606,357 (23.0%) | 33,698,518 (23.4%) |
| Job-Type | | | | | | | | |
| Agriculture | 1,073,668 (0.8%) | 1,085,738 (0.8%) | 1,121,887 (0.9%) | 1,151,192 (0.9%) | 1,186,988 (0.9%) | 1,210,771 (0.9%) | 1,203,743 (0.8%) | 1,195,074 (0.8%) |
| Mining | 701,080 (0.5%) | 778,462 (0.6%) | 794,866 (0.6%) | 815,181 (0.6%) | 762,959 (0.6%) | 608,493 (0.4%) | 598,995 (0.4%) | 648,946 (0.5%) |
| Utilities | 800,858 (0.6%) | 785,982 (0.6%) | 795,782 (0.6%) | 791,262 (0.6%) | 800,585 (0.6%) | 802,499 (0.6%) | 804,273 (0.6%) | 804,685 (0.6%) |
| Construction | 5,403,947 (4.2%) | 5,458,095 (4.2%) | 5,626,515 (4.3%) | 6,132,857 (4.6%) | 6,463,089 (4.7%) | 6,843,727 (4.9%) | 7,044,977 (5.0%) | 7,327,168 (5.1%) |
| Manufacturing | 11,904,627 (9.3%) | 12,019,340 (9.3%) | 12,065,178 (9.2%) | 12,248,393 (9.1%) | 12,404,794 (9.0%) | 12,442,100 (8.9%) | 12,482,625 (8.8%) | 12,692,436 (8.8%) |
| Wholesale | 5,646,983 (4.4%) | 5,730,418 (4.4%) | 5,815,471 (4.4%) | 5,827,946 (4.3%) | 5,903,125 (4.3%) | 5,908,682 (4.2%) | 5,928,984 (4.2%) | 5,866,776 (4.1%) |
| Retail | 14,525,825 (11.3%) | 14,567,149 (11.2%) | 14,751,716 (11.2%) | 15,064,449 (11.2%) | 15,306,489 (11.2%) | 15,602,049 (11.1%) | 15,609,093 (11.0%) | 15,542,120 (10.8%) |
| Transportation & Warehousing | 4,378,755 (3.4%) | 4,461,738 (3.4%) | 4,539,588 (3.4%) | 4,668,450 (3.5%) | 4,876,728 (3.6%) | 5,073,941 (3.6%) | 5,205,209 (3.7%) | 5,477,066 (3.8%) |
| Information | 2,939,338 (2.3%) | 2,922,054 (2.3%) | 2,975,645 (2.3%) | 2,998,543 (2.2%) | 3,024,426 (2.2%) | 3,088,288 (2.2%) | 3,077,866 (2.2%) | 3,136,500 (2.2%) |
| Finance & Insurance | 5,578,977 (4.3%) | 5,615,114 (4.3%) | 5,693,823 (4.3%) | 5,678,993 (4.2%) | 5,785,165 (4.2%) | 5,875,049 (4.2%) | 5,948,716 (4.2%) | 5,977,999 (4.2%) |
| Real estate | 1,977,402 (1.5%) | 1,988,971 (1.5%) | 2,033,886 (1.5%) | 2,068,566 (1.5%) | 2,116,262 (1.5%) | 2,165,900 (1.5%) | 2,201,805 (1.6%) | 2,247,463 (1.6%) |
| Professional, Scientific, Technical Services | 7,876,132 (6.1%) | 8,065,854 (6.2%) | 8,,318,072 (6.3%) | 8,461,155 (6.3%) | 8,758,896 (6.4%) | 9,020,767 (6.4%) | 9,147,038 (6.5%) | 9,407,389 (6.5%) |
| Management of companies | 2,028,615 (1.6%) | 2,145,481 (1.7%) | 2,221,874 (1.7%) | 2,298,180 (1.7%) | 2,359,823 (1.7%) | 2,369,384 (1.7%) | 2,403,555 (1.7%) | 2,476,275 (1.7%) |
| Waste Management | 7,642,954 (6.0%) | 7,874,199 (6.1%) | 8,093,691 (6.1%) | 8,408,579 (6.3%) | 8,604,249 (6.3%) | 8,894,803 (6.3%) | 8,976,598 (6.3%) | 9,135,027 (6.3%) |
| Education | 12,910,717 (10.1%) | 12,723,369 (9.8%) | 12,737,205 (9.7%) | 12,837,027 (9.5%) | 12,907,113 (9.4%) | 12,978,505 (9.3%) | 13,054,386 (9.2%) | 13,157,030 (9.1%) |
| Healthcare | 18,415,417 (14.3%) | 18,479,592 (14.3%) | 19,253,971 (14.6%) | 19,679,003 (14.6%) | 20,134,939 (14.7%) | 20,710,903 (14.8%) | 21,123,403 (14.9%) | 21,497,127 (14.9%) |
| Arts, Entertainment & Recreation | 2,190,586 (1.7%) | 2,237,320 (1.7%) | 2,271,565 (1.7%) | 2,335,057 (1.7%) | 2,380,893 (1.7%) | 2,525,223 (1.8%) | 2,579,633 (1.8%) | 2,624,987 (1.8%) |
| Accommodation | 11,194,394 (8.7%) | 11,492,726 (8.9%) | 11,916,203 (9.0%) | 12,298,630 (9.1%) | 12,669,830 (9.2%) | 13,157,486 (9.4%) | 13,403,548 (9.5%) | 13,633,322 (9.5%) |
| Public Administration | 6,612,758 (5.1%) | 6,454,773 (5.0%) | 6,490,449 (4.9%) | 6,465,828 (4.8%) | 6,362,440 (4.6%) | 6,418,576 (4.6%) | 6,430,421 (4.5%) | 6,492,760 (4.5%) |
| Other services | 4,610,809 (3.6%) | 4,673,248 (3.6%) | 4,279,420 (3.2%) | 4,258,146 (3.2%) | 4,339,608 (3.2%) | 4,467,609 (3.2%) | 4,516,487 (3.2%) | 4,574,894 (3.2%) |

*Table S1.2*: Number of total workers and workers belonging to different subpopulations disaggregated by race, sex, income, education, and job type from WAC. Note: The number of workers residing in the 48 states + Washington D.C from the RAC files is more (0.01%) than the total number of workers obtained from the WAC files working in the same geographic area. This is because there are a small number of workers residing in the geographic boundary of this study whose primary work location are in geographies that we have excluded (Alaska, Hawaii, US territories). Given the small difference between these numbers, we ignore this discrepancy in our analysis.

| | 2011 | 2012 | 2013 | 2014 | 2015 | 2016 | 2017 | 2018 |
|---|---|---|---|---|---|---|---|---|

|  | Work | Work | Work | Work | Work | Work | Work | Work |
|---|---|---|---|---|---|---|---|---|
| All | 128,3999,118 | 129,542,393 | 131,777,555 | 134,466,180 | 137,125,124 | 140,138,326 | 141,734,619 | 143,908,458 |
| Race | | | | | | | | |
| white | 103,882,883 (80.9%) | 104,504,348 (80.7%) | 106,103,125 (80.5%) | 107,963,081 (80.3%) | 109,755,428 (80.0%) | 109,813,663 (78.4%) | 110,562,754 (78.0%) | 111,798,718 (77.7%) |
| Black | 15,290,766 (11.9%) | 15,719,787 (12.1%) | 16,163,812 (12.3%) | 16,777,699 (12.5%) | 17,422,870 (12.7%) | 18,433,311 (13.2%) | 18,852,738 (13.3%) | 19,324,928 (13.4%) |
| Asian | 6,371,777 (5.0%) | 6,395,356 (4.9%) | 6,488,010 (4.9%) | 6,588,142 (4.9%) | 6,694,591 (4.9%) | 8,014,967 (5.7%) | 8,308,905 (5.9%) | 8,614,003 (6.0%) |
| Native | 1,066,784 (0.8%) | 1,081,594 (0.8%) | 1,105,504 (0.8%) | 1,132,914 (0.8%) | 1,166,642 (0.9%) | 1,262,716 (0.9%) | 1,288,153 (0.9%) | 1,323,576 (0.9%) |
| Hawaiian or Pacific Islander | 181,030 (0.1%) | 182,687 (0.1%) | 189,150 (0.1%) | 195,797 (0.1%) | 202,887 (0.1%) | 238,864 (0.2%) | 246,706 (0.2%) | 255,257 (0.2%) |
| Two or More Races | 1,605,878 (1.3%) | 1,658,621 (1.3%) | 1,727,954 (1.3%) | 1,808,547 (1.3%) | 1,882,706 (1.4%) | 2,374,805 (1.7%) | 2,475,270 (1.7%) | 2,591,976 (1.8%) |
| Ethnicity | | | | | | | | |
| Hispanic | 16,245,520 (12.7%) | 16,494,098 (12.7%) | 16,974,829 (12.9%) | 17,493,070 (13.0%) | 17,990,410 (13.1%) | 20,462,170 (14.6%) | 21,147,233 (14.9%) | 21,955,423 (15.3%) |
| Income | | | | | | | | |
| Income 1 | 32.364.717 (25.2%) | 32,411,533 (25.0%) | 32,629,662 (24.8%) | 32,940,699 (24.5%) | 32,768,472 (23.9%) | 32,717,084 (23.3%) | 32,218,254 (22.7%) | 31,667,286 (22.0%) |
| Income 2 | 45,946,944 (35.8%) | 45,792,514 (35.3%) | 45,938,351 (34.9%) | 46,064,537 (34.3%) | 46,518,790 (33.9%) | 47,060,463 (33.6%) | 46,466,159 (32.8%) | 46,116,108 (32.0%) |
| Income 3 | 50,087,457 (39.0%) | 51,338,346 (39.6%) | 53,209,542 (40.4%) | 55,460,944 (41.2%) | 57,837,862 (42.2%) | 60,360,779 (43.1%) | 63,050,206 (44.5%) | 66,125,064 (45.9%) |
| Education | | | | | | | | |
| No highschool | 11,584,655 (9.0%) | 12,053,245 (9.3%) | 12,532,091 (9.5%) | 13,079,259 (9.7%) | 13,692,830 (10.0%) | 13,752,944 (9.8%) | 14,306,186 (10.1%) | 14,891,015 (10.3%) |
| Highschool | 26,203,748 (20.4%) | 26,775,836 (20.7%) | 27,318,869 (20.7%) | 27.991,671 (20.8%) | 28,637,692 (20.9%) | 28,383,522 (20.3%) | 28,788,286 (20.3%) | 29,274,821 (20.3%) |
| College | 31,391,901 (24.4%) | 31,963,598 (24.7%) | 32,530,860 (24.7%) | 33,129,493 (24.6%) | 33,695,431 (24.6%) | 33,943,803 (24.2%) | 34,248,106 (24.2%) | 34,665,719 (24.1%) |
| Advanced | 29,476,242 (23.0%) | 29,703,941 (22.9%) | 29,931,992 (22.7%) | 30,086,145 (22.4%) | 30,219124 (22.0%) | 31,594,836 (22.5%) | 31,598,085 (22.3%) | 31,859,237 (22.1%) |
| Sex | | | | | | | | |
| Male | 63,770,042 (49.7%) | 64,660,663 (49.9%) | 65,894,918 (50.0%) | 67,384,816 (50.1%) | 68,714,925 (50.1%) | 70,246,539 (50.1%) | 70,983,768 (50.1%) | 72,014,791 (50.0%) |
| Female | 64,629,076 (50.3%) | 64,881,730 (50.1%) | 65,882,637 (50.0%) | 67,081,364 (49.9%) | 68,410,199 (49.9%) | 69,891,787 (49.9%) | 70,750,851 (49.9%) | 71,893,667 (50.0%) |
| Age | | | | | | | | |
| ≤ 29 years | 29,742,572 (23.2%) | 29,045,773 (22.4%) | 29,465,869 (22.4%) | 30,179,612 (22.4%) | 30,880,047 (22.5%) | 32,463,221 (23.2%) | 32,793,956 (23.1%) | 33,217,666 (23.1%) |
| 30 - 54 years | 72,951,387 (56.8%) | 73,120,133 (56.4%) | 73,749,360 (56.0%) | 74,534,078 (55.4%) | 75,290,281 (54.9%) | 76,016,631 (54.2%) | 76,335,615 (53.9%) | 76,993,627 (53.5%) |
| ≥ 55 yeas | 25,705,159 (20.0%) | 27,376,487 (21.1%) | 28,562,326 (21.7%) | 29,752,490 (22.1%) | 30,954,796 (22.6%) | 31,658,474 (22.6%) | 32,605,048 (23.0%) | 33,697,165 (23.4%) |
| Job-Type | | | | | | | | |
| Agriculture | 1,073708 (0.8%) | 1,085,739 (0.8%) | 1,121,901 (0.9%) | 1,151,286 (0.9%) | 1,186,968 (0.9%) | 1,210,755 (0.9%) | 1,203,783 (0.8%) | 1,195,129 (0.8%) |
| Mining | 699,197 (0.5%) | 776,410 (0.6%) | 792,557 (0.6%) | 812,660 (0.6%) | 760,280 (0.6%) | 606,133 (0.4%) | 599,023 (0.4%) | 648,975 (0.5%) |
| Utilities | 800,709 (0.6%) | 785,809 (0.6%) | 795,598 (0.6%) | 791,093 (0.6%) | 800,426 (0.6%) | 802,302 (0.6%) | 804,191 (0.6%) | 804,608 (0.6%) |
| Construction | 5,403,195 (4.2%) | 5,457,183 (4.2%) | 5,625,454 (4.3%) | 6,131,656 (4.6%) | 6,461,708 (4.7%) | 6,842,226 (4.9%) | 7,044,465 (5.0%) | 7,326,655 (5.1%) |
| Manufacturing | 11,902,548 (9.3%) | 12,016,366 (9.3%) | 12,062,139 (9.2%) | 12,245,107 (9.1%) | 12,401,964 (9.0%) | 12,438,562 (8.9%) | 12,482,819 (8.8%) | 12,692,640 (8.8%) |
| Wholesale | 5,646,697 (4.4%) | 5,730,277 (4.4%) | 5,815,292 (4.4%) | 5,827,713 (4.3%) | 5,902,792 (4.3%) | 5,908,356 (4.2%) | 5,928,849 (4.2%) | 5,866,665 (4.1%) |
| Retail | 14,524,106 (11.3%) | 14,565,317 (11.2%) | 14,749,624 (11.2%) | 15,062,269 (11.2%) | 15,304,149 (11.2%) | 15,599,368 (11.1%) | 15,607,946 (11.0%) | 15,540,822 (10.8%) |
| Transportation & Warehousing | 4,377,897 (3.4%) | 4,460,644 (3.4%) | 4,538,344 (3.4%) | 4,667,058 (3.5%) | 4,875,156 (3.6%) | 5,072,101 (3.6%) | 5,204,906 (3.7%) | 5,476,716 (3.8%) |
| Information | 2,939,182 (2.3%) | 2,921,820 (2.3%) | 2,975,411 (2.3%) | 2,998,232 (2.2%) | 3,024,049 (2.2%) | 3,087,865 (2.2%) | 3,077,754 (2.2%) | 3,136,310 (2.2%) |
| Finance & Insurance | 5,578,770 (4.3%) | 5,614,869 (4.3%) | 5,693,658 (4.3%) | 5,678,794 (4.2%) | 5,784,915 (4.2%) | 5,874,773 (4.2%) | 5,948,648 (4.2%) | 5,977,926 (4.2%) |
| Real estate | 1,977,198 (1.5%) | 1,988,963 (1.5%) | 2,033,609 (1.5%) | 2,068,235 (1.5%) | 2,115,909 (1.5%) | 2,165,432 (1.5%) | 2,201,613 (1.6%) | 2,247,269 (1.6%) |
| Professional, Scientific, Technical Services | 7,875,779 (6.1%) | 8,065,439 (6.2%) | 8,317,564 (6.3%) | 8,460,529 (6.3%) | 8,758,163 (6.4%) | 9,019,937 (6.4%) | 9,147,010 (6.5%) | 9,407,261 (6.6%) |
| Management of companies | 2,028,555 (1.6%) | 2,145,407 (1.7%) | 2,221,747 (1.7%) | 2,297,948 (1.7%) | 2,359,611 (1.7%) | 2,369,160 (1.7%) | 2,403,512 (1.7%) | 2,476,233 (1.7%) |
| Waste Management | 7,642,247 (6.0%) | 7,873,340 (6.1%) | 8,092,662 (6.1%) | 8,407,351 (6.3%) | 8,602,761 (6.3%) | 8,892,986 (6.3%) | 8,975,644 (6.3%) | 9,133,997 (6.3%) |
| Education | 12,909,806 (10.1%) | 12,722,536 (9.8%) | 12,736,223 (9.7%) | 12,836,013 (9.5%) | 12,905,876 (9.4%) | 12,977,074 (9.3%) | 13,054,113 (9.2%) | 13,157,306 (9.1%) |
| Healthcare | 18,415,417 (14.3%) | 18.478.184 (14.3%) | 19,252,415 (14.6%) | 19,677,376 (14.6%) | 20,132,952 (14.7%) | 20,708,648 (14.8%) | 21,122,915 (14.9%) | 21,496,555 (14.9%) |
| Arts, Entertainment | 2,190,321 | 2,237,037 | 2,271,188 | 2,334,641 | 2,380,413 | 2,524,564 | 2,579,163 | 2,624,700 |
| & Recreation | (1.7%) | (1.7%) | (1.7%) | (1.7%) | (1.7%) | (1.8%) | (1.8%) | (1.8%) |
| Accommodation | 11,193,206 (8.7%) | 11,491,284 (8.9%) | 11,914,465 (9.0%) | 12,296,641 (9.1%) | 12,667,487 (9.2%) | 13,154,665 (9.4%) | 13,401,741 (9.5%) | 13,631,417 (9.5%) |
| Public Administration | 6,611,250 (5.1%) | 6,453,114 (5.0%) | 6,488,774 (4.9%) | 6,464,008 (4.8%) | 6,360,538 (4.6%) | 6,416,468 (4.6%) | 6,430,338 (4.5%) | 6,492,624 (4.5%) |
| Other services | 4,610,487 (3.6%) | 4,672,855 (3.6%) | 4,278,930 (3.2%) | 4,257,570 (3.2%) | 4,339,007 (3.2%) | 4,466,951 (3.2%) | 4,516,186 (3.2%) | 4,574,650 (3.2%) |

*Table S1.3: Number of total workers and workers belonging to different subpopulations disaggregated by race, sex, income, education, and job type from OD*

| | 2011 | 2012 | 2013 | 2014 | 2015 | 2016 | 2017 | 2018 |
|---|---|---|---|---|---|---|---|---|
| | OD | OD | OD | OD | OD | OD | OD | OD |
| All | 128,344,714 | 129,487,909 | 131,721,325 | 134,416,016 | 137,077,151 | 140,088,892 | 141,684,644 | 143,857,731 |
| Income | | | | | | | | |
| Income 1 | 32,349,592 (25.2%) | 32,396,831 (25.0%) | 32,614,331 (24.8%) | 32,927,795 (24.5%) | 32,756,568 (23.9%) | 32,705,208 (23.3%) | 32,206,599 (22.7%) | 31,655,757 (22.0%) |
| Income 2 | 45,926,360 (35.8%) | 45,772,157 (35.3%) | 45,917,459 (34.9%) | 46,045,874 (34.3%) | 46,500,876 (33.9%) | 47,042,153 (33.6%) | 46,448,276 (32.8%) | 46,098,205 (32.0%) |
| Income 3 | 50,068,762 (39.0%) | 51,318,921 (39.6%) | 53,189,535 (40.4%) | 55,442,347 (41.2%) | 57,819,707 (42.2%) | 60,341,531 (43.1%) | 63,029,769 (44.5%) | 66,103,769 (46.0%) |
| Age | | | | | | | | |
| ≤ 29 years | 29,729,382 (23.2%) | 29,033,423 (22.4%) | 29,452,922 (22.4%) | 30,168,495 (22.4%) | 30,869,484 (22.5%) | 32,452,201 (23.2%) | 32,782,838 (23.1%) | 33,206,011 (23.1%) |
| 30 - 54 years | 72,921,585 (56.8%) | 73,090,426 (56.4%) | 73,719,033 (56.0%) | 74,507,245 (55.4%) | 75,264,722 (54.9%) | 75,990,390 (54.2%) | 76,309,262 (53.9%) | 76,967,327 (53.5%) |
| ≥ 55 yeas | 25,693,747 (20.0%) | 27,364,060 (21.1%) | 28,549,370 (21.7%) | 29,740,276 (22.1%) | 30,942,945 (22.6%) | 31,646,301 (22.6%) | 32,592,544 (23.0%) | 33,684,393 (23.4%) |
| Job-Type | | | | | | | | |
| Goods Producing | 19,070,518 (14.9%) | 19,327,679 (14.9%) | 19,594,043 (14.9%) | 20,333,103 (15.1%) | 20,803,341 (15.2%) | 21,090,000 (15.1%) | 21,322,457 (15.0%) | 21,855,587 (15.2%) |
| Trade, Transportation & Utility | 25,338,272 (19.7%) | 25,530,882 (19.7%) | 25,887,623 (19.7%) | 26,338,624 (19.6%) | 26,873,322 (19.6%) | 27,372,980 (19.5%) | 27,536,754 (19.4%) | 27,679,300 (19.2%) |
| Other Services | 83,935,924 (65.4%) | 84,629,348 (65.4%) | 86,239,659 (65.5%) | 87,744,289 (65.3%) | 89,400,488 (65.2%) | 91,625,912 (65.4%) | 92,825,433 (65.5%) | 94,322,844 (65.6%) |

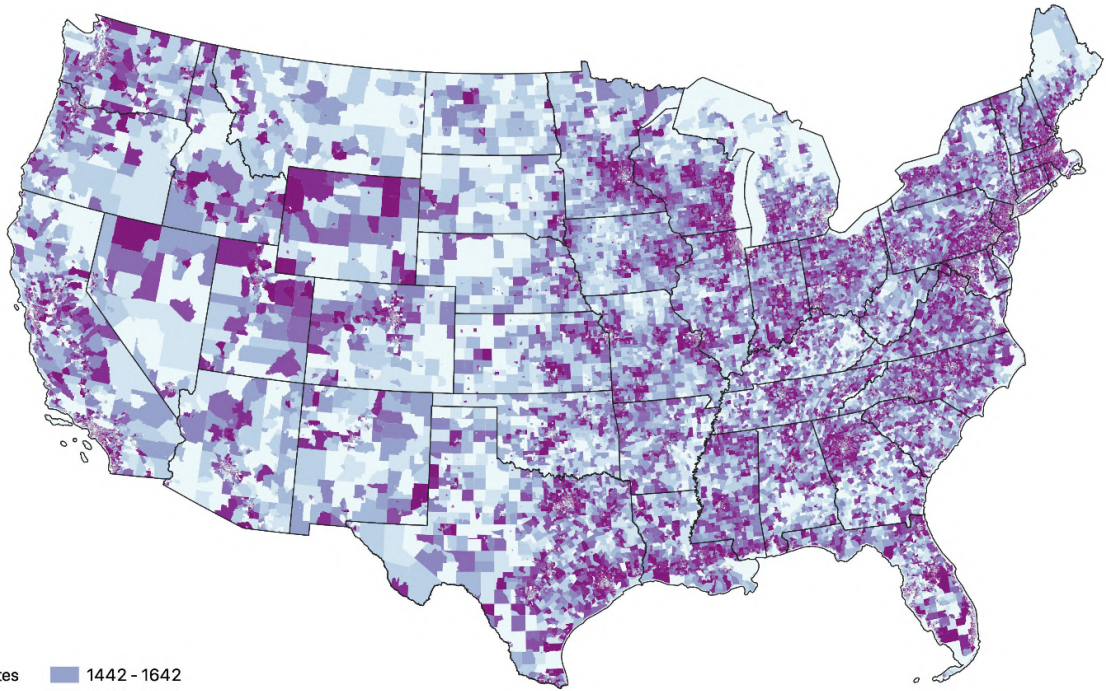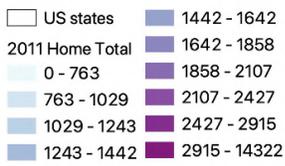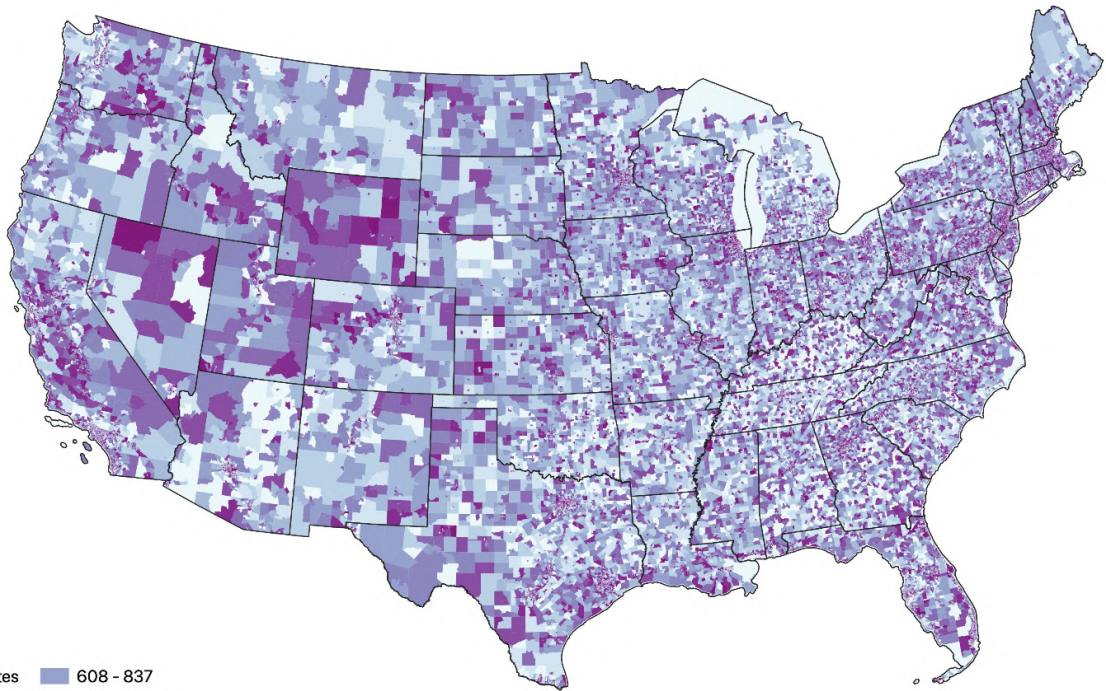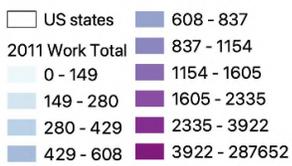

*Figure S1.1*: Deciles of the total Number of Workers A) Living and B) Working in census tracts across the United States in 2011. Note that in 2011, the LODES data recorded a small number of workers living in census tracts corresponding to the Great Lakes region in Michigan (likely on house boats). Post 2016, the LODES data indicated no residents in these areas.

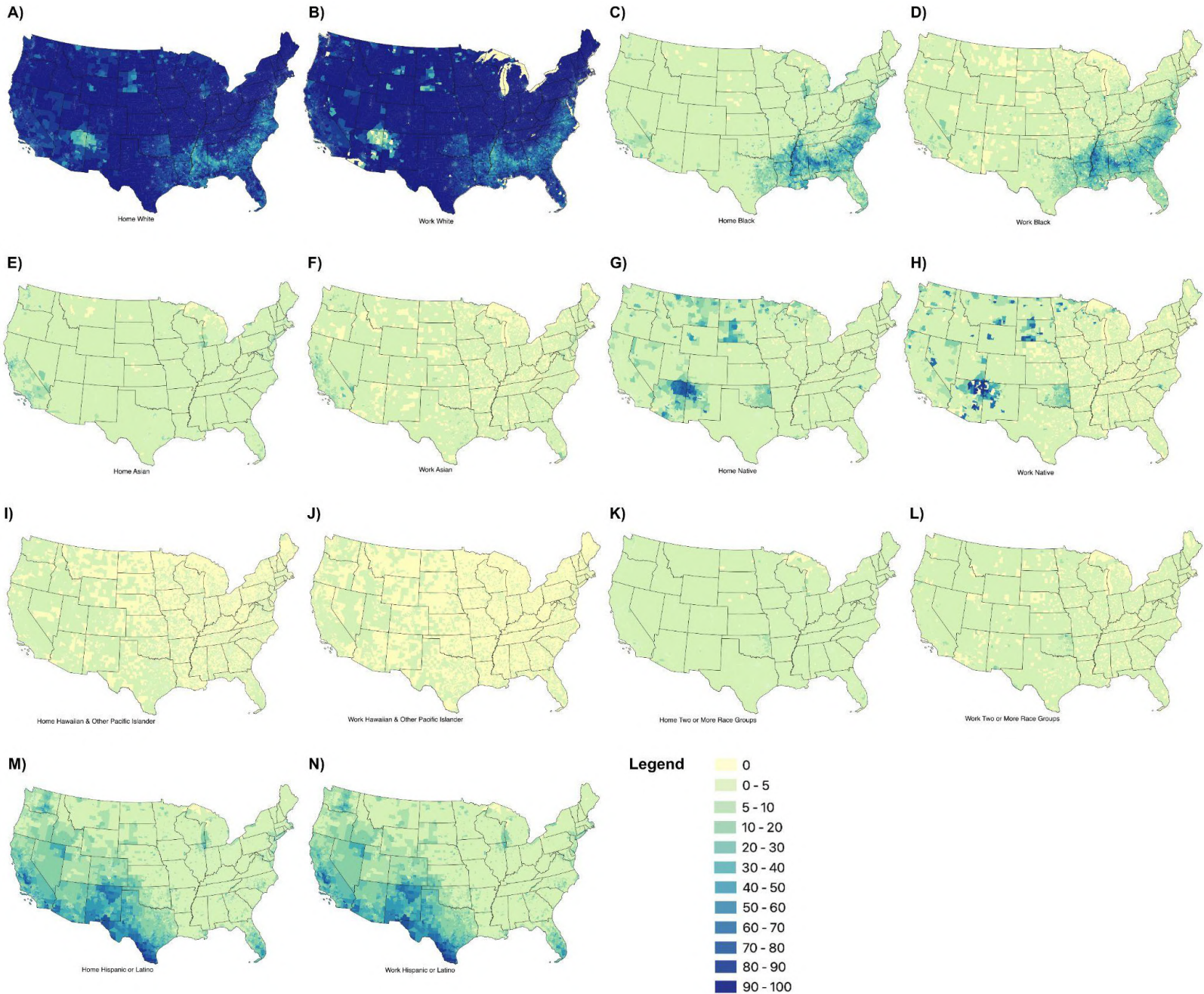

*Figure S1.2*: % of Workers living and working in census tracts across the United States in 2011 disaggregated by race and ethnicity: A) Home white, B) Work white, C) Home Black, D) Work Black, E) Home Asian, F) Work Asian, G) Home Native, H) Work Native, I) Home Hawaiian or other Pacific Islander, J) Work Hawaiian or other Pacific Islander, K) Home Two or More Races, L) Work Two or More Races, M) Home Hispanic or Latino, N) Work Hispanic or Latino. All maps have the same scale to allow for comparison. Note that in 2011, the LODES data recorded a small number of workers living in census tracts corresponding to the Great Lakes region in Michigan (likely on house boats). Post 2016, the LODES data indicated no residents in these areas.

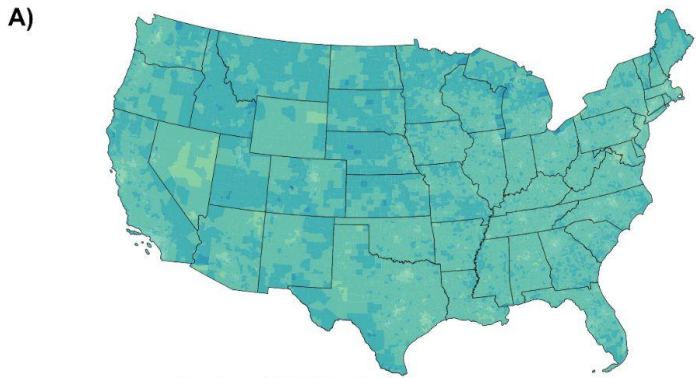
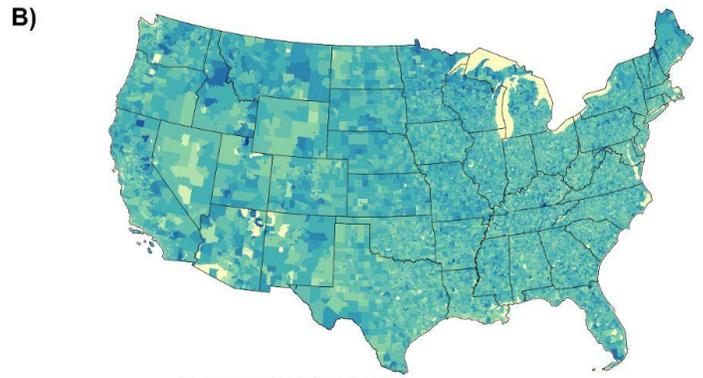
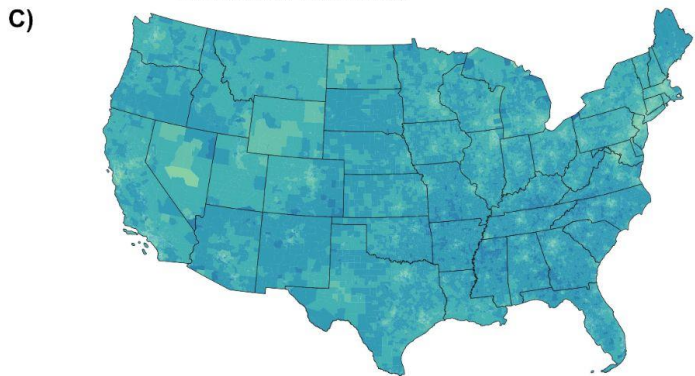
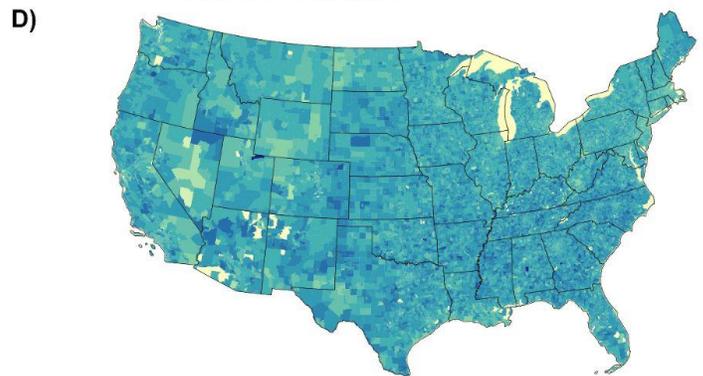
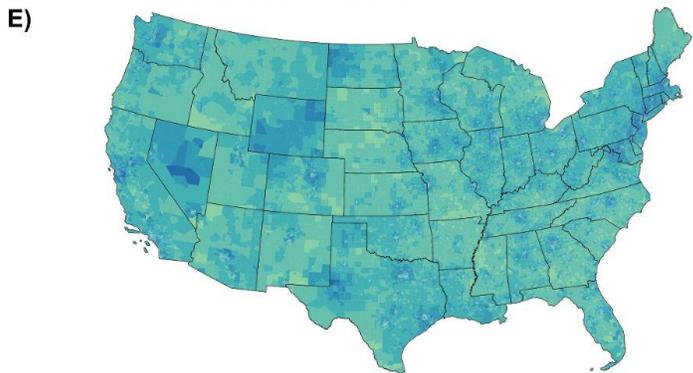
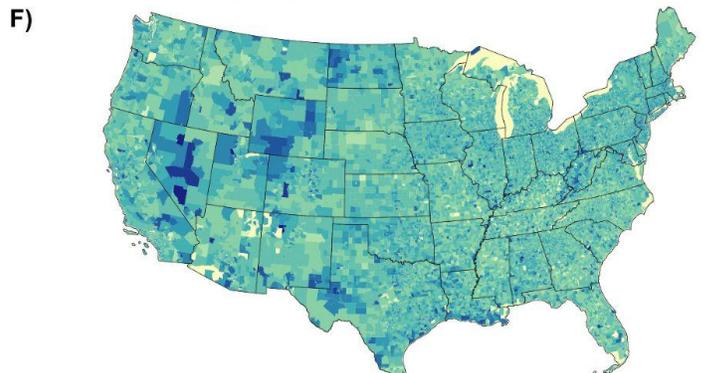
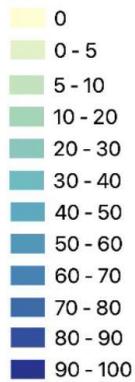

*Figure S1.3*: % of Workers living and working in census tracts across the United States in 2011 disaggregated by income: A) Home Income USD $1,250/month or less, B) Work Income USD

*$1,250/month or less, C) Home Income USD $1,251 - USD $3,333/month, D) Work Income USD $1,251 - USD $3,333/month, E) Home Income greater than USD $3,333/month, F) Work Income greater than USD $3,333/month. All maps have the same scale to allow for comparison. Note that in 2011, the LODES data recorded a small number of workers living in census tracts corresponding to the Great Lakes region in Michigan (likely on house boats). Post 2016, the LODES data indicated no residents in these areas.*

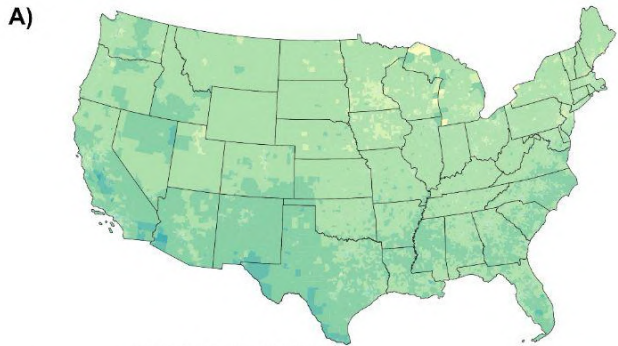
Home Less than High School Education

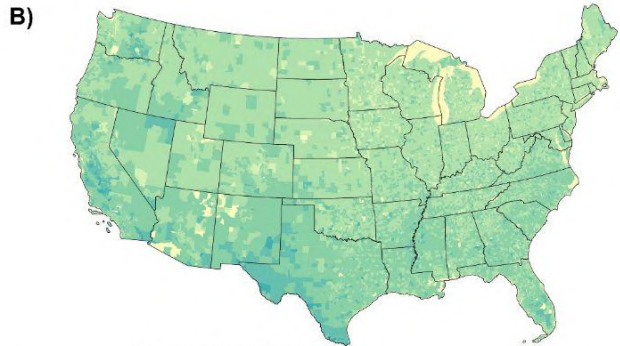
Work Less than High School Education

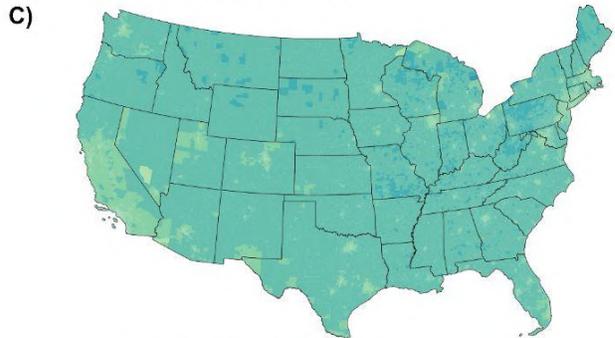
Home High School or Equivalent

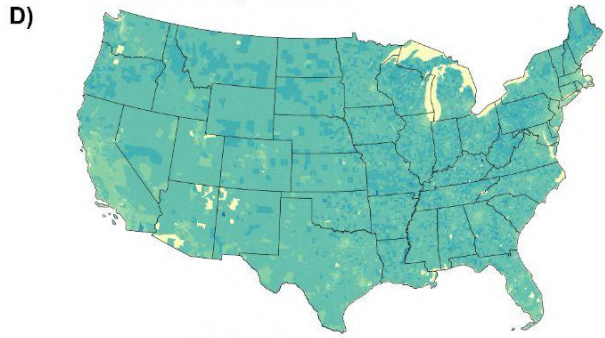
Work High School or Equivalent

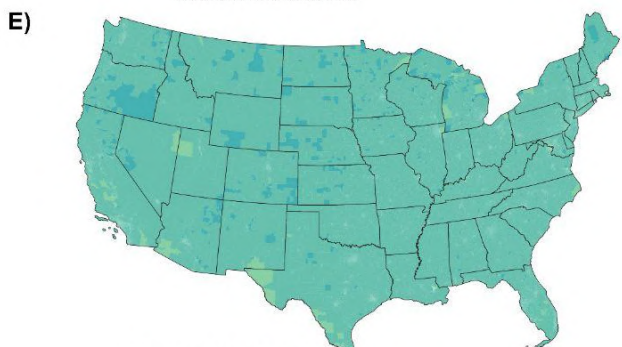
Home Some college or Associate degree

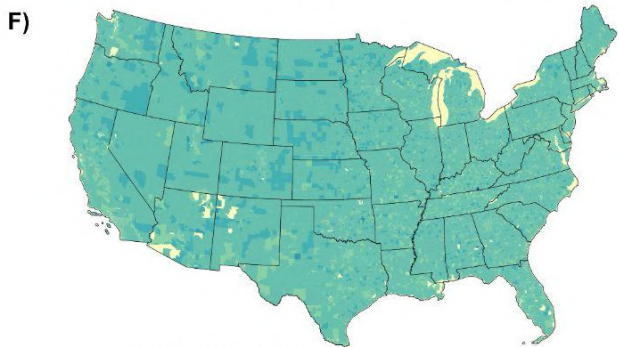
Work Some college or Associate degree

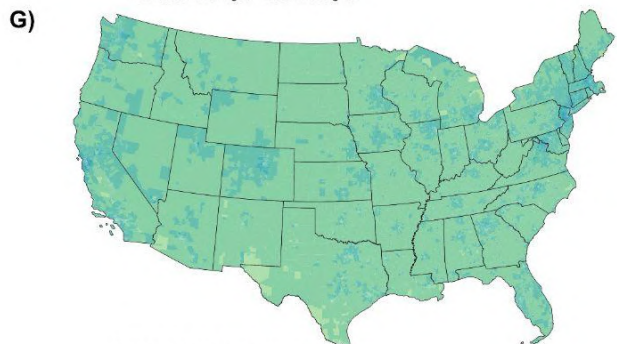
Home Bachelor's degree or advanced degree

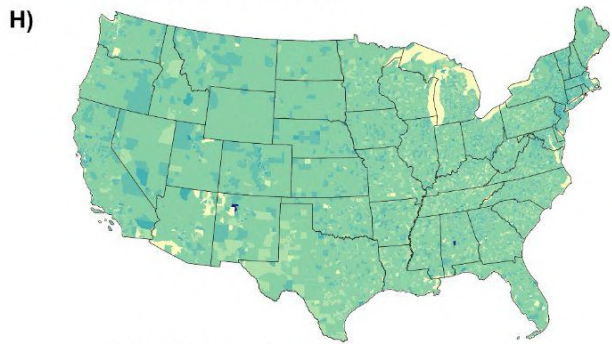
Work Bachelor's degree or advanced degree

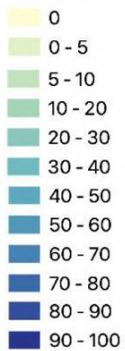

Legend

- 0
- 0 - 5
- 5 - 10
- 10 - 20
- 20 - 30
- 30 - 40
- 40 - 50
- 50 - 60
- 60 - 70
- 70 - 80
- 80 - 90
- 90 - 100

*Figure S1.4*: % of Workers living and working in census tracts across the United States in 2011 disaggregated by education: A) Home Less than High School, B) Work Less than High School, C) Home High School or Equivalent, D) Work High School or Equivalent, E) Home Some College or Associate degree, F) Work Some College or Associate degree, G) Home College degree or higher, H) Work College degree or higher. All maps have the same scale to allow for comparison. Note that in 2011, the LODES data recorded a small number of workers living in census tracts corresponding to the Great Lakes region in Michigan (likely on house boats). Post 2016, the LODES data indicated no residents in these areas.

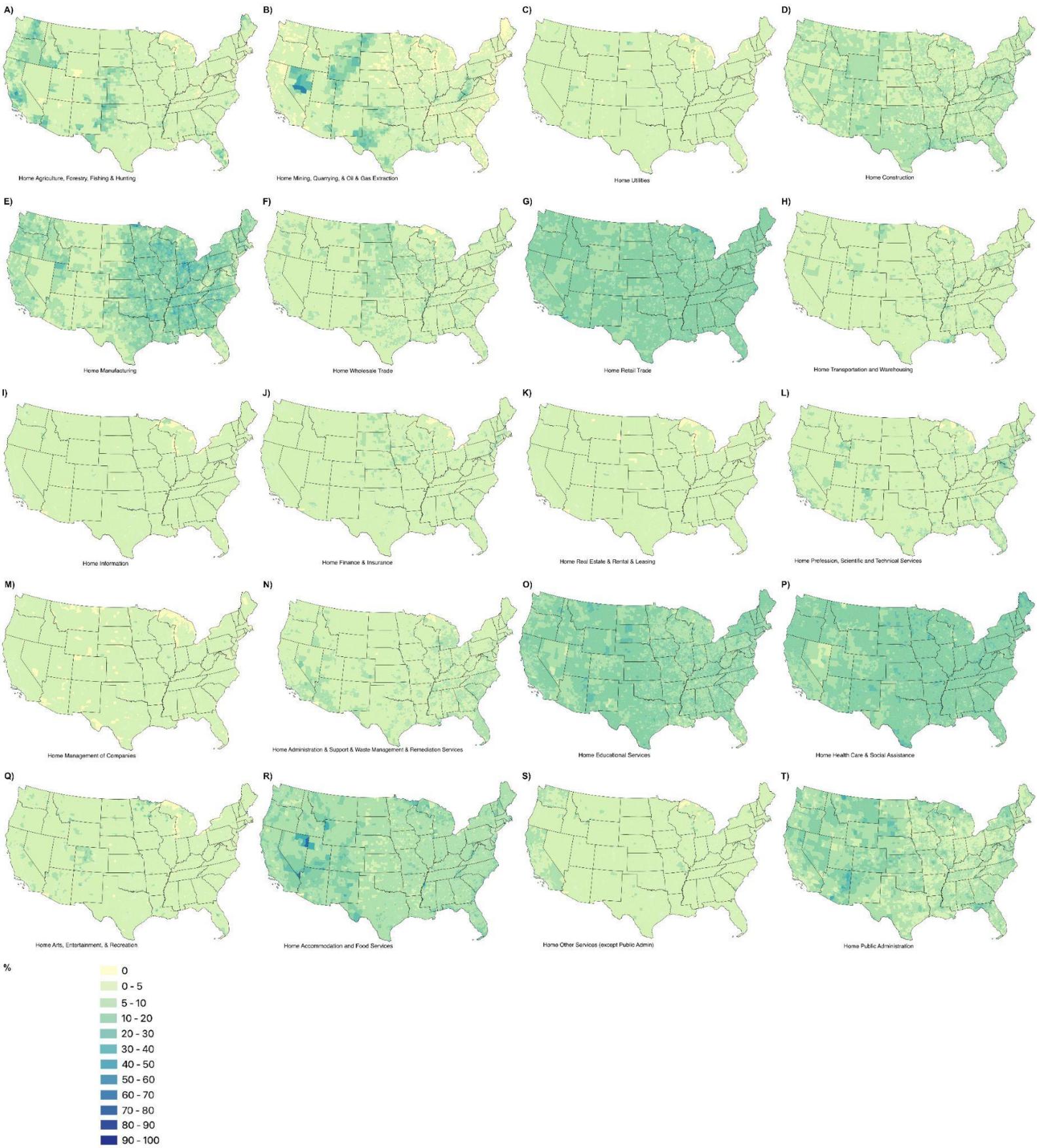

*Figure S1.5*: % of Workers living in census tracts across the United States with different job types: A) Agriculture/Forestry/Fishing/Hunting, B) Mining/Quarrying/Oil and Gas Extraction, C) Utilities, D) Construction, E) Manufacturing, F) Wholesale Trade, G) Retail Trade, H) Transportation and Warehousing, I) Information, J) Finance and Insurance, K) Real Estate and Rental and Leasing, L) Professional/Scientific and Technical Services, M) Management of Companies and Enterprises, N) Administrative and Support and Waste Management and Remediation Services, O) Educational Services, P) Healthcare and Social Assistance, Q) Arts/Entertainment and Recreation, R) Accommodation and Food Services, S) Other Services except Public Administration, T) Public Administration. All maps have the same scale to allow for comparison. Note that in 2011, the LODES data recorded a small number of workers living in census tracts corresponding to the Great Lakes region in Michigan (likely on house boats). Post 2016, the LODES data indicated no residents in these areas.

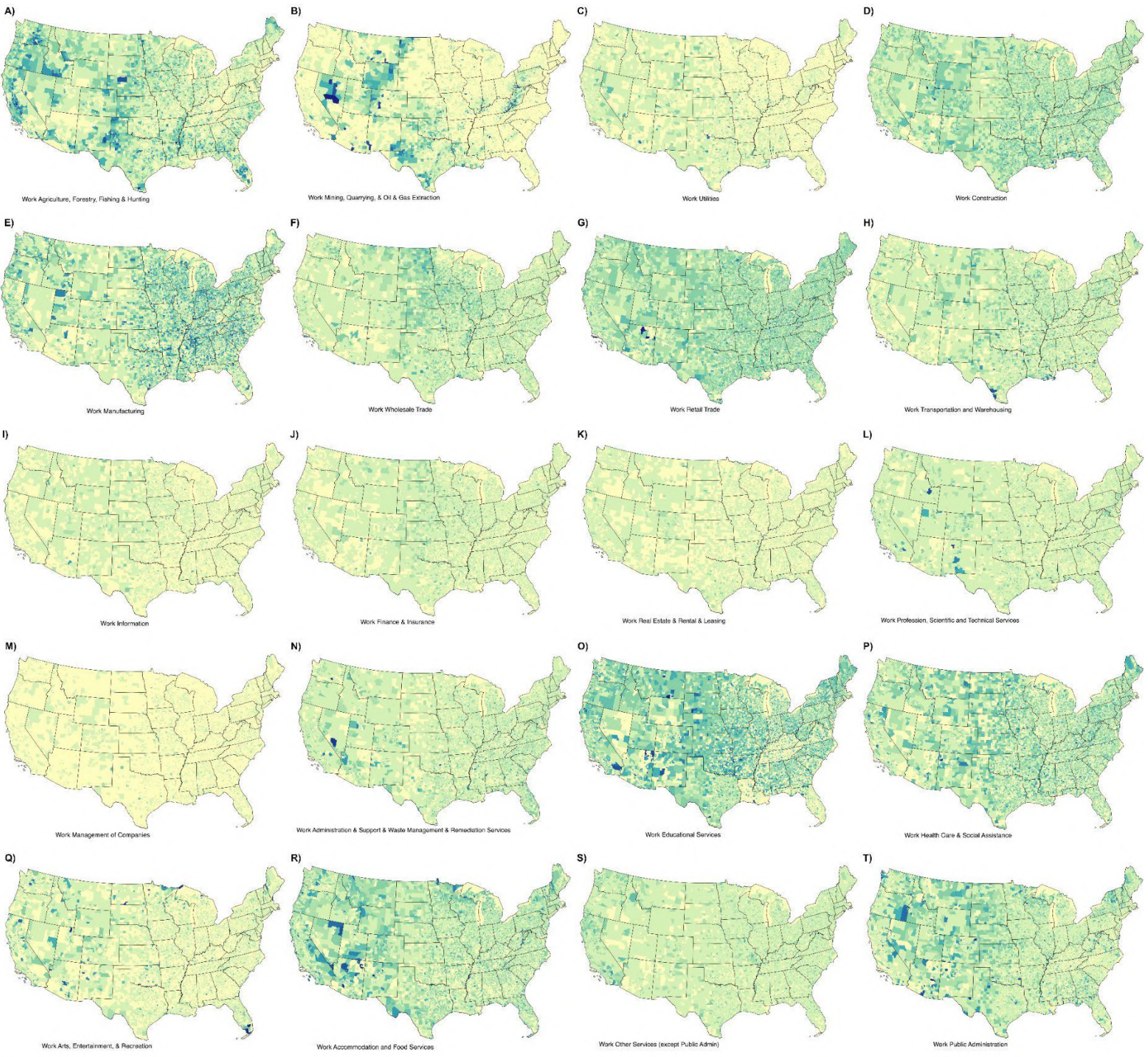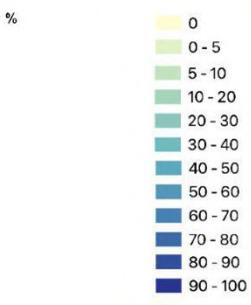

*Figure S1.6*: % of Workers working in census tracts across the United States with different job types: A) Agriculture/Forestry/Fishing/Hunting, B) Mining/Quarrying/Oil and Gas Extraction, C) Utilities, D) Construction, E) Manufacturing, F) Wholesale Trade, G) Retail Trade, H) Transportation and Warehousing, I) Information, J) Finance and Insurance, K) Real Estate and Rental and Leasing, L) Professional/Scientific and Technical Services, M) Management of Companies and Enterprises, N) Administrative and Support and Waste Management and Remediation Services, O) Educational Services, P) Healthcare and Social Assistance, Q) Arts/Entertainment and Recreation, R) Accommodation and Food Services, S) Other Services except Public Administration, T) Public Administration. All maps have the same scale to allow for comparison. Note that in 2011, the LODES data recorded a small number of workers living in census tracts corresponding to the Great Lakes region in Michigan (likely on house boats). Post 2016, the LODES data indicated no residents in these areas.

## S2: Population-weighted PM$_{2.5}$ exposures

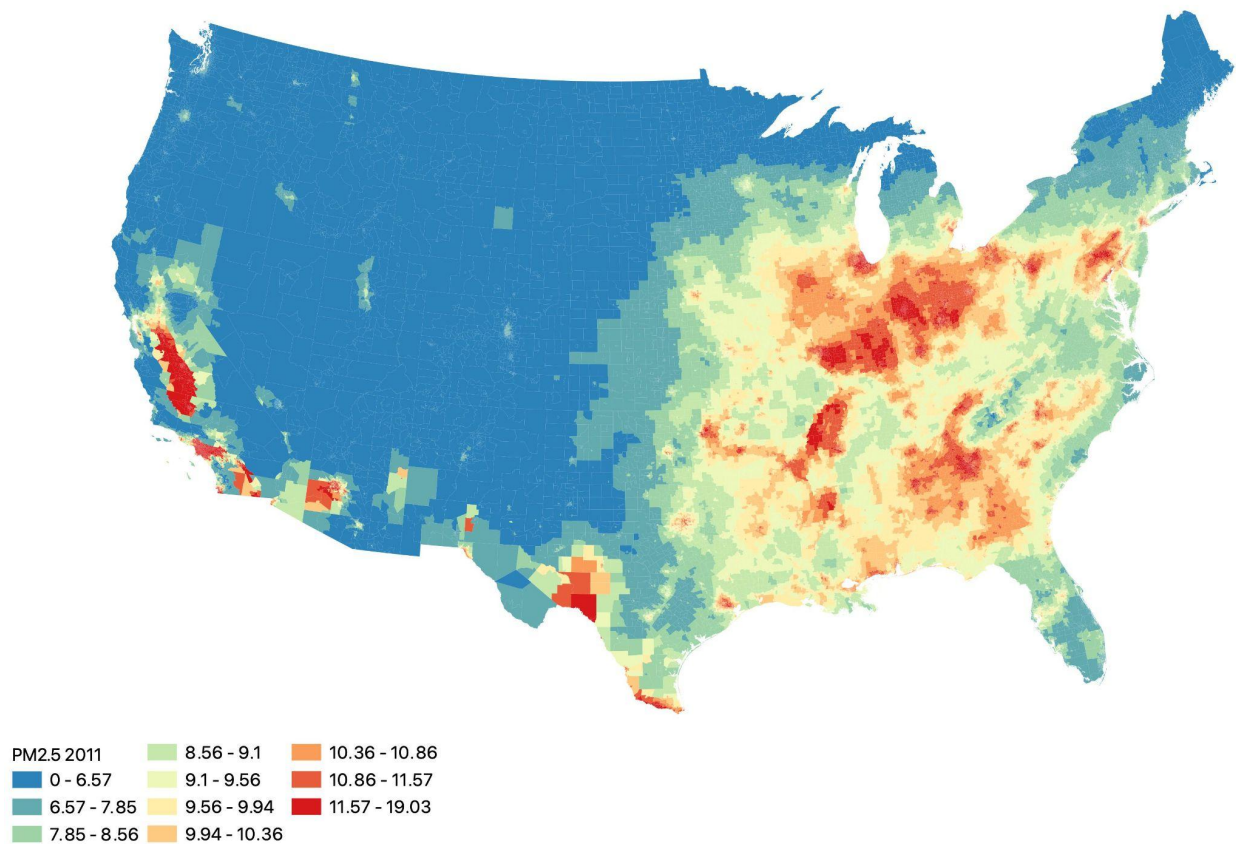

*Figure S2.1*: Deciles of annual-averaged PM$_{2.5}$ in 2011 across census tracts

# S3: Urban Areas

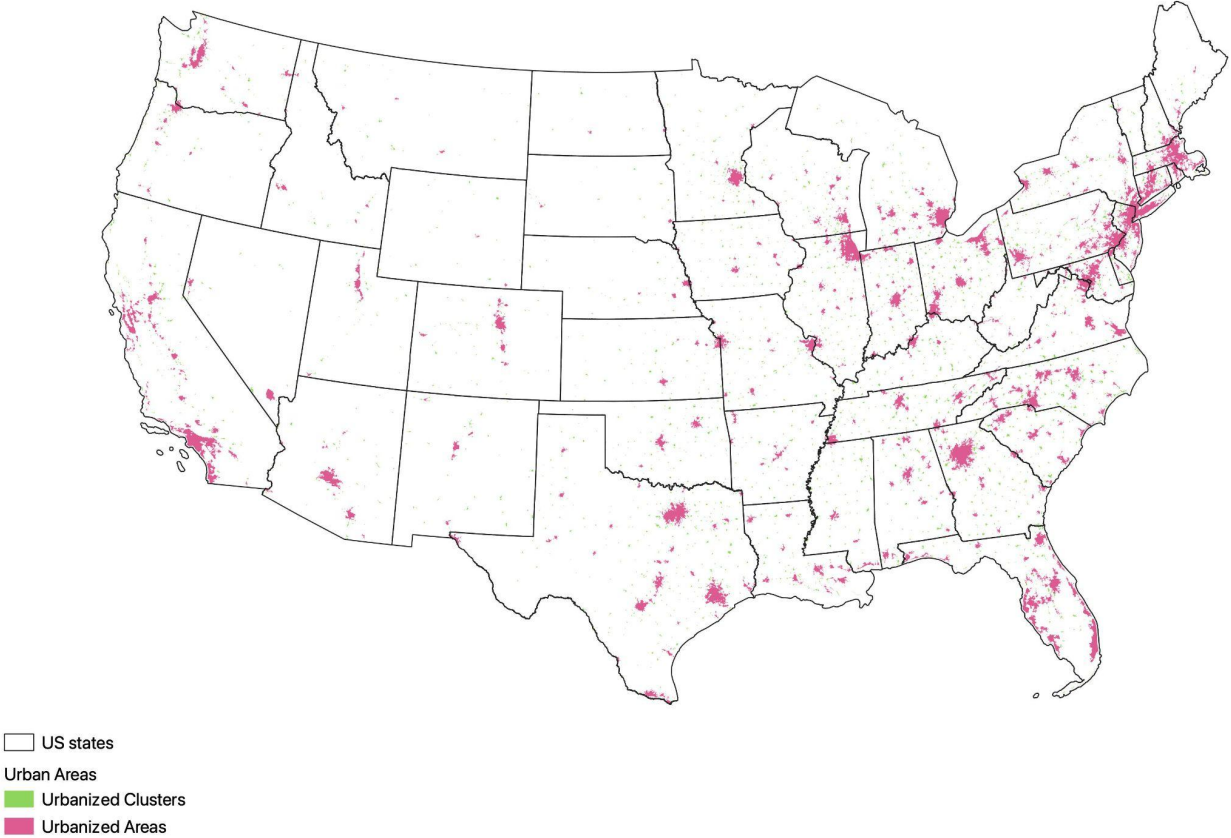

*Figure S3.1*: Urban areas

# S4: Trends in home and work exposures by job-type

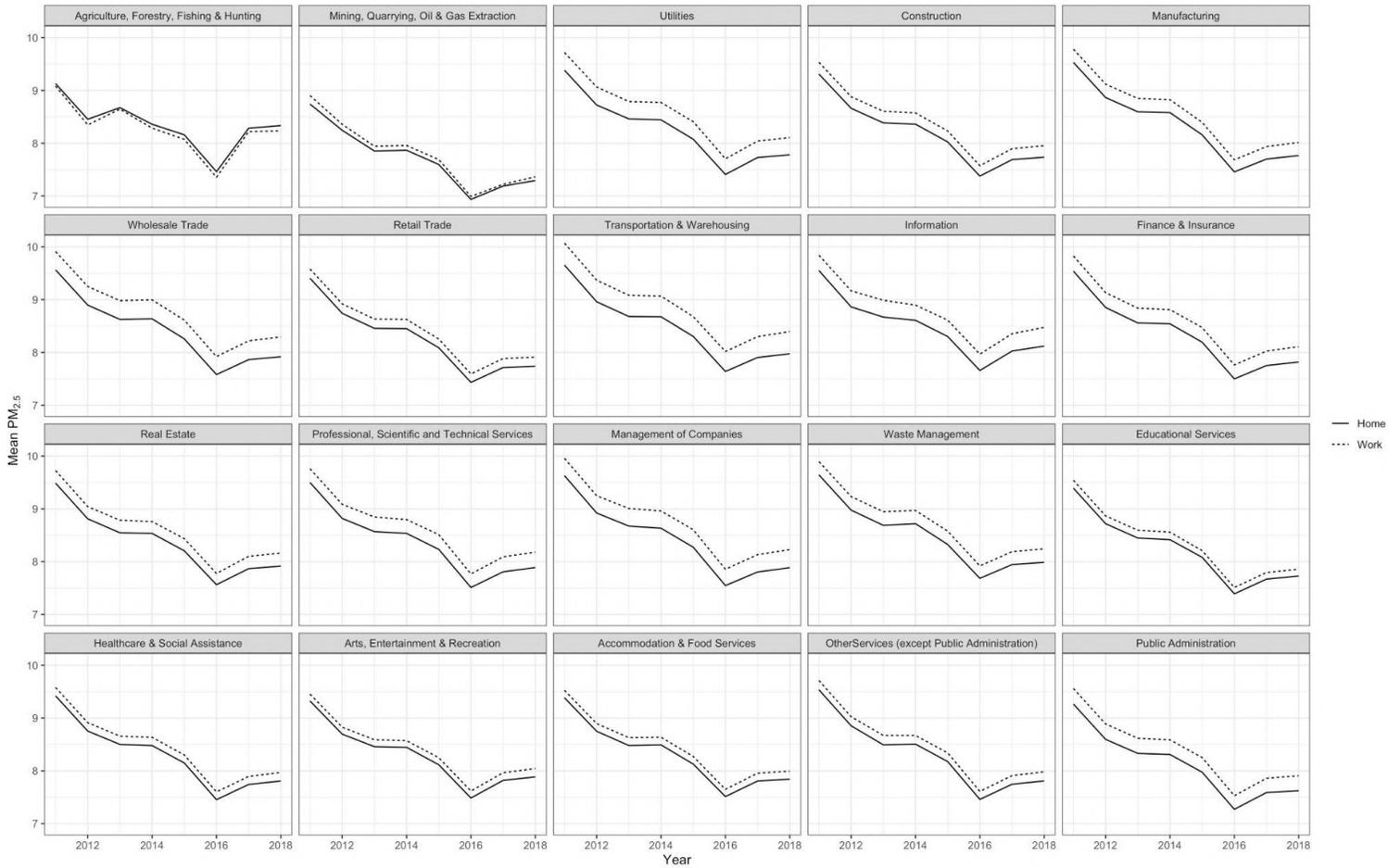

***Figure S4.1****: Weighted-mean exposure to PM$_{2.5}$ (μg/m$^3$) experienced at home (H) and work (W) locations disaggregated by job-type*

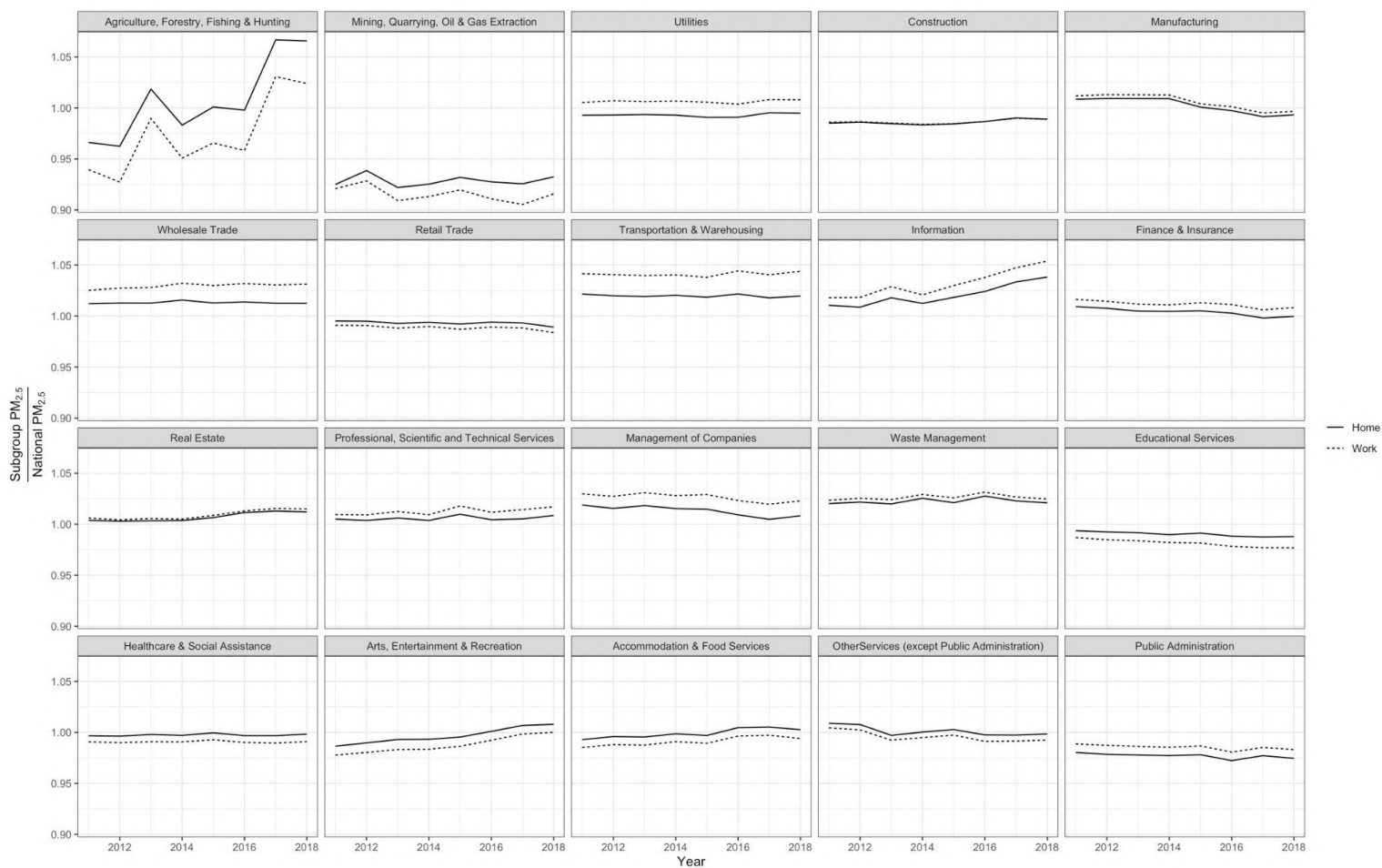

*Figure S4.2*: Ratio of population-weighted PM$_{2.5}$ exposure for workers with different job-types: A) Agriculture, Forestry, Fishing and Hunting, C) Utilities, D) Construction, E) Manufacturing, F) Wholesale Trade, G) Retail Trade, H) Transportation and Warehousing, I) Information, J) Finance and Insurance, K) Real Estate, L) Professional, Scientific and Technical Services, M) Management, N) Waste management, O) Educational Serives, P) Health Care and Social Assistance, Q) Arts, Entertainment and Recreation, R) Accommodation and Food Services, S) Public Administration, T) Other Services to national average population-weighted exposure for years 2011 - 2018

*Table S4.1*: Weighted-mean exposure to PM$_{2.5}$ (μg/m$^3$) experienced at H and W for the total population and subpopulations disaggregated by race, sex, income, education, and job type.

|  | 2011 | | 2012 | | 2013 | | 2014 | | 2015 | | 2016 | | 2017 | | 2018 | |
|---|---|---|---|---|---|---|---|---|---|---|---|---|---|---|---|---|
| Weighted PM$_{2.5}$ Mean Exposure | Home | Work | Home | Work | Home | Work | Home | Work | Home | Work | Home | Work | Home | Work | Home | Work |
| All | 9.5 | 9.7 | 8.8 | 9.0 | 8.5 | 8.7 | 8.5 | 8.7 | 8.2 | 8.4 | 7.5 | 7.7 | 7.8 | 8.0 | 7.8 | 8.0 |
| Race | | | | | | | | | | | | | | | | |
| white | 9.4 | 9.6 | 8.7 | 8.9 | 8.4 | 8.7 | 8.4 | 8.6 | 8.1 | 8.3 | 7.4 | 7.6 | 7.7 | 7.9 | 7.7 | 8.0 |

| Category | | | | | | | | | | | | | | | |
|---|---|---|---|---|---|---|---|---|---|---|---|---|---|---|---|
| Black | 10.0 | 10.2 | 9.2 | 9.3 | 8.7 | 8.9 | 8.9 | 9.0 | 8.5 | 8.6 | 7.8 | 7.9 | 7.9 | 8.1 | 7.9 | 8.0 |
| Asian | 9.9 | 10.1 | 9.1 | 9.3 | 9.1 | 9.4 | 9.0 | 9.2 | 8.7 | 9.0 | 8.0 | 8.2 | 8.5 | 8.7 | 8.7 | 9.0 |
| Native | 8.8 | 9.0 | 8.3 | 8.5 | 8.1 | 8.4 | 8.0 | 8.2 | 7.7 | 7.9 | 7.2 | 7.3 | 7.8 | 7.9 | 7.8 | 8.0 |
| Hawaiian or Pacific Islander | 9.4 | 9.7 | 8.8 | 9.0 | 9.0 | 9.2 | 8.6 | 8.9 | 8.4 | 8.6 | 7.7 | 7.9 | 8.4 | 8.6 | 8.5 | 8.8 |
| Two or More Races | 9.4 | 9.7 | 8.8 | 9.0 | 8.7 | 8.9 | 8.5 | 8.7 | 8.2 | 8.4 | 7.5 | 7.7 | 8.0 | 8.2 | 8.1 | 8.3 |
| Ethnicity | | | | | | | | | | | | | | | | |
| Hispanic | 9.9 | 10.0 | 9.2 | 9.4 | 9.1 | 9.2 | 9.1 | 9.2 | 8.7 | 8.8 | 8.2 | 8.3 | 8.6 | 8.8 | 8.7 | 8.8 |
| Income | | | | | | | | | | | | | | | | |
| Income 1 | 9.4 | 9.6 | 8.8 | 8.9 | 8.5 | 8.7 | 8.5 | 8.7 | 8.2 | 8.3 | 7.5 | 7.6 | 7.8 | 7.9 | 7.8 | 8.0 |
| Income 2 | 9.5 | 9.6 | 8.8 | 9.0 | 8.5 | 8.7 | 8.5 | 8.7 | 8.1 | 8.3 | 7.5 | 7.7 | 7.8 | 8.0 | 7.8 | 8.0 |
| Income 3 | 9.4 | 9.7 | 8.8 | 9.1 | 8.5 | 8.8 | 8.5 | 8.8 | 8.2 | 8.4 | 7.4 | 7.7 | 7.7 | 8.0 | 7.8 | 8.1 |
| Education | | | | | | | | | | | | | | | | |
| No highschool | 9.7 | 9.8 | 9.0 | 9.2 | 8.7 | 8.9 | 8.7 | 8.9 | 8.3 | 8.5 | 7.7 | 7.9 | 8.1 | 8.3 | 8.1 | 8.3 |
| Highschool | 9.4 | 9.6 | 8.7 | 9.0 | 8.4 | 8.7 | 8.5 | 8.7 | 8.1 | 8.3 | 7.4 | 7.6 | 7.7 | 7.9 | 7.7 | 7.9 |
| College | 9.4 | 9.6 | 8.7 | 9.0 | 8.5 | 8.7 | 8.5 | 8.7 | 8.1 | 8.3 | 7.4 | 7.6 | 7.7 | 7.9 | 7.8 | 8.0 |
| Advanced | 9.4 | 9.7 | 8.7 | 9.0 | 8.5 | 8.8 | 8.5 | 8.7 | 8.1 | 8.4 | 7.4 | 7.7 | 7.7 | 8.0 | 7.8 | 8.1 |
| Sex | | | | | | | | | | | | | | | | |
| Male | 9.5 | 9.7 | 8.8 | 9.0 | 8.5 | 8.8 | 8.5 | 8.7 | 8.2 | 8.4 | 7.5 | 7.7 | 7.8 | 8.0 | 7.8 | 8.1 |
| Female | 9.4 | 9.6 | 8.8 | 9.0 | 8.5 | 8.7 | 8.5 | 8.7 | 8.1 | 8.3 | 7.5 | 7.7 | 7.8 | 7.9 | 7.8 | 8.0 |
| Age | | | | | | | | | | | | | | | | |
| ≤ 29 years | 9.5 | 9.7 | 8.8 | 9.0 | 8.6 | 8.7 | 8.5 | 8.7 | 8.2 | 8.3 | 7.5 | 7.7 | 7.8 | 8.0 | 7.8 | 8.0 |
| 30 - 54 years | 9.5 | 9.7 | 8.8 | 9.0 | 8.5 | 8.8 | 8.5 | 8.7 | 8.2 | 8.4 | 7.5 | 7.7 | 7.8 | 8.0 | 7.8 | 8.1 |
| ≥ 55 yeas | 9.4 | 9.6 | 8.7 | 8.9 | 8.4 | 8.7 | 8.4 | 8.7 | 8.1 | 8.3 | 7.4 | 7.6 | 7.7 | 7.9 | 7.7 | 8.0 |
| Job-Type | | | | | | | | | | | | | | | | |
| Agriculture | 9.1 | 9.1 | 8.5 | 8.3 | 8.7 | 8.6 | 8.4 | 8.3 | 8.2 | 8.1 | 7.5 | 7.4 | 8.3 | 8.2 | 8.3 | 8.2 |
| Mining | 8.7 | 8.9 | 8.2 | 8.4 | 7.9 | 7.9 | 7.9 | 8.0 | 7.6 | 7.7 | 6.9 | 7.0 | 7.2 | 7.2 | 7.3 | 7.4 |
| Utilities | 9.4 | 9.7 | 8.7 | 9.1 | 8.5 | 8.8 | 8.4 | 8.8 | 8.1 | 8.4 | 7.4 | 7.7 | 7.7 | 8.0 | 7.8 | 8.1 |
| Construction | 9.3 | 9.5 | 8.7 | 8.9 | 8.4 | 8.6 | 8.4 | 8.6 | 8.0 | 8.2 | 7.4 | 7.6 | 7.7 | 7.9 | 7.7 | 8.0 |
| Manufacturing | 9.5 | 9.8 | 8.9 | 9.1 | 8.6 | 8.8 | 8.6 | 8.8 | 8.2 | 8.4 | 7.5 | 7.7 | 7.7 | 7.9 | 7.8 | 8.0 |

| | | | | | | | | | | | | | | | |
|---|---|---|---|---|---|---|---|---|---|---|---|---|---|---|---|
| Wholesale | 9.6 | 9.9 | 8.9 | 9.2 | 8.6 | 9.0 | 8.6 | 9.0 | 8.3 | 8.6 | 7.6 | 7.9 | 7.9 | 8.2 | 7.9 | 8.3 |
| Retail | 9.4 | 9.6 | 8.7 | 8.9 | 8.5 | 8.6 | 8.5 | 8.6 | 8.1 | 8.3 | 7.4 | 7.6 | 7.7 | 7.9 | 7.7 | 7.9 |
| Transportation & Warehousing | 9.7 | 10.1 | 9.0 | 9.4 | 8.7 | 9.1 | 8.7 | 9.1 | 8.3 | 8.7 | 7.6 | 8.0 | 7.9 | 8.3 | 8.0 | 8.4 |
| Information | 9.6 | 9.8 | 8.9 | 9.2 | 8.7 | 9.0 | 8.6 | 8.9 | 8.3 | 8.6 | 7.7 | 8.0 | 8.0 | 8.4 | 8.1 | 8.5 |
| Finance & Insurance | 9.5 | 9.8 | 8.9 | 9.1 | 8.6 | 8.8 | 8.5 | 8.8 | 8.2 | 8.5 | 7.5 | 7.8 | 7.8 | 8.0 | 7.8 | 8.1 |
| Real estate | 9.5 | 9.7 | 8.8 | 9.0 | 8.5 | 8.8 | 8.5 | 8.8 | 8.2 | 8.4 | 7.6 | 7.8 | 7.9 | 8.1 | 7.9 | 8.2 |
| Professional, Scientific, Technical Services | 9.5 | 9.8 | 8.8 | 9.1 | 8.6 | 8.8 | 8.5 | 8.8 | 8.2 | 8.5 | 7.5 | 7.8 | 7.8 | 8.1 | 7.9 | 8.2 |
| Management of companies | 9.6 | 10.0 | 8.9 | 9.2 | 8.7 | 9.0 | 8.6 | 9.0 | 8.3 | 8.6 | 7.5 | 7.9 | 7.8 | 8.1 | 7.9 | 8.2 |
| Waste Management | 9.6 | 9.9 | 9.0 | 9.2 | 8.7 | 8.9 | 8.7 | 9.0 | 8.3 | 8.6 | 7.7 | 7.9 | 7.9 | 8.2 | 8.0 | 8.2 |
| Education | 9.4 | 9.5 | 8.7 | 8.9 | 8.4 | 8.6 | 8.4 | 8.6 | 8.1 | 8.2 | 7.4 | 7.5 | 7.7 | 7.8 | 7.7 | 7.9 |
| Healthcare | 9.4 | 9.6 | 8.8 | 8.9 | 8.5 | 8.7 | 8.5 | 8.6 | 8.1 | 8.3 | 7.5 | 7.6 | 7.7 | 7.9 | 7.8 | 8.0 |
| Arts, Entertainment & Recreation | 9.3 | 9.5 | 8.7 | 8.8 | 8.5 | 8.6 | 8.4 | 8.6 | 8.1 | 8.2 | 7.5 | 7.6 | 7.8 | 8.0 | 7.9 | 8.0 |
| Accommodation | 9.4 | 9.5 | 8.7 | 8.9 | 8.5 | 8.6 | 8.5 | 8.6 | 8.1 | 8.3 | 7.5 | 7.6 | 7.8 | 8.0 | 7.8 | 8.0 |
| Public Administration | 9.3 | 9.6 | 8.6 | 8.9 | 8.3 | 8.6 | 8.3 | 8.6 | 8.0 | 8.3 | 7.3 | 7.5 | 7.6 | 7.9 | 7.6 | 7.9 |
| Other services | 9.5 | 9.7 | 8.9 | 9.0 | 8.5 | 8.7 | 8.5 | 8.7 | 8.2 | 8.3 | 7.5 | 7.6 | 7.7 | 7.9 | 7.8 | 8.0 |

# S5: Trends in home and work exposures disaggregated by urban/rural

We display H and W for all workers, and workers belonging to different SES (**Figure S5.1**), and job-types (**Figure S5.2**) disaggregated by urban/rural locations, where we observe similar trends in PM$_{2.5}$ across subpopulations. Rural H and W are lower than the corresponding urban exposures across all subpopulations.

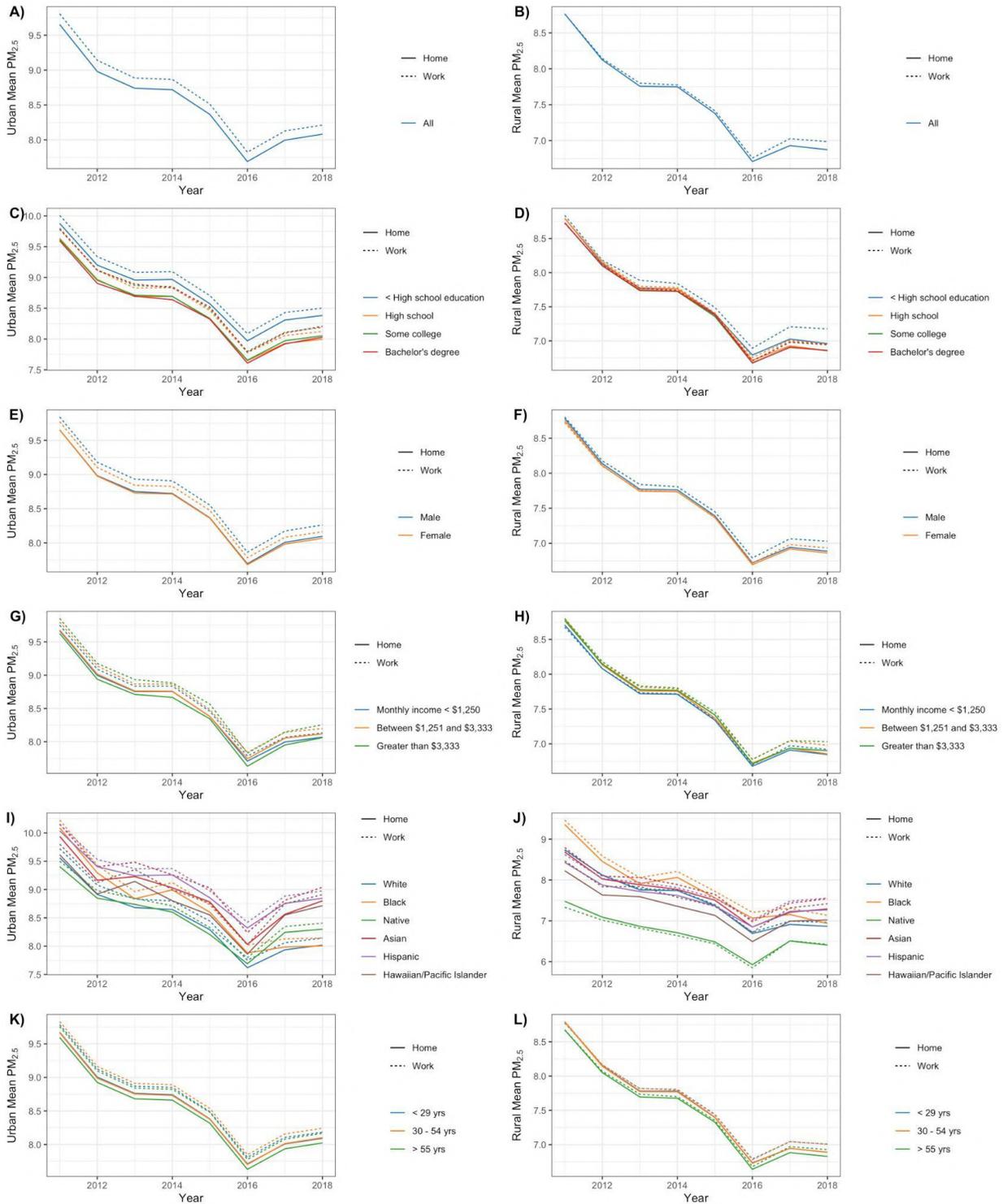

***Figure S5.1****: Weighted-mean exposure to PM$_{2.5}$ (μg/m$^3$) experienced at home (H) and work (W) locations for urban and rural census tracts disaggregated by race, sex, income, and education*

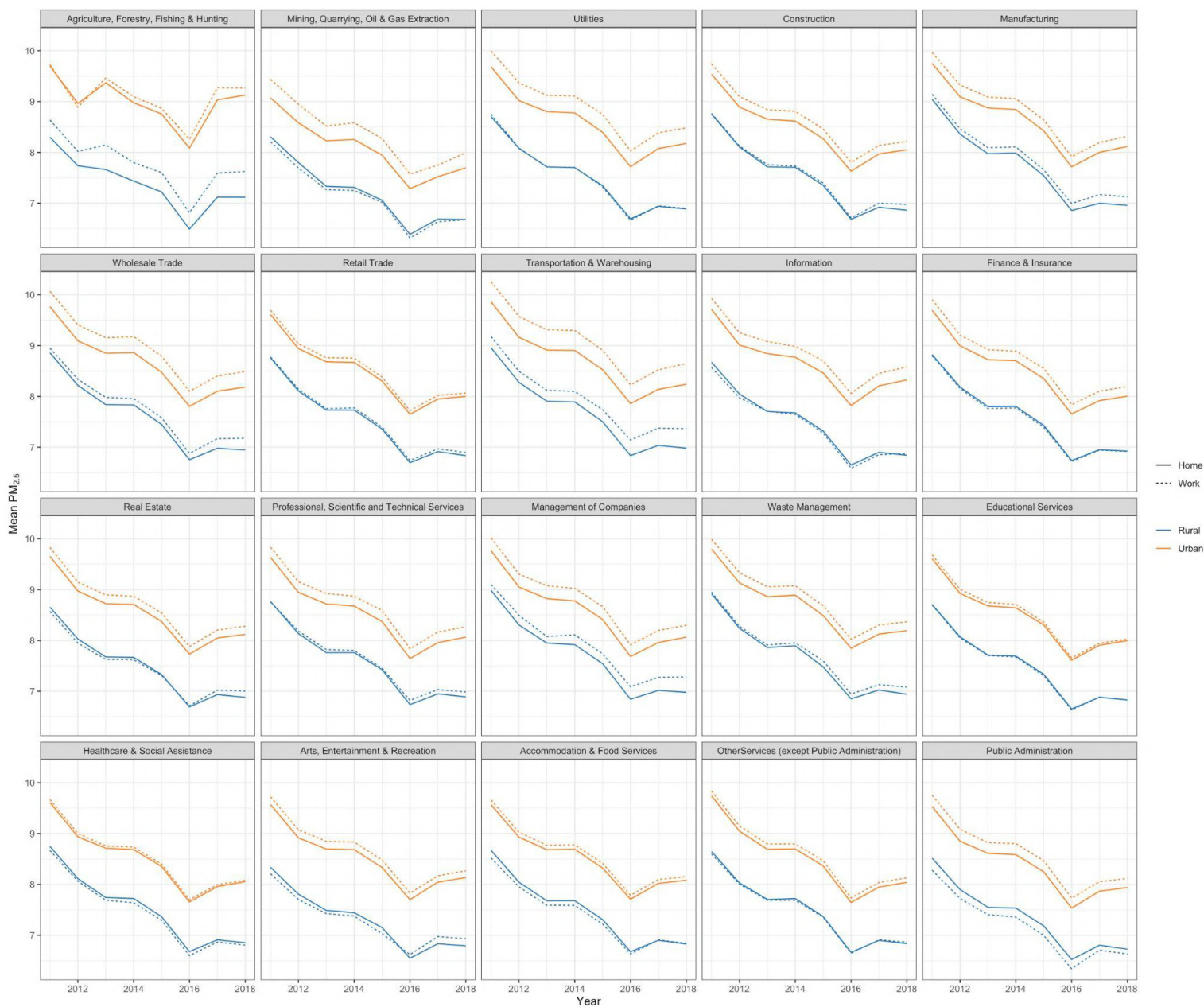

*Figure S5.2*: Population-weighted mean PM$_{2.5}$ exposure for workers with different job-types: A) Agriculture, Forestry, Fishing and Hunting, C) Utilities, D) Construction, E) Manufacturing, F) Wholesale Trade, G) Retail Trade, H) Transportation and Warehousing, I) Information, J) Finance and Insurance, K) Real Estate, L) Professional, Scientific and Technical Services, M) Management, N) Waste management, O) Educational Serives, P) Health Care and Social Assistance, Q) Arts, Entertainment and Recreation, R) Accommodation and Food Services, S) Public Administration, T) Other Services, disaggregated by urban/rural census tracts for years 2011 - 2018.

The ratio of H (and W) to the total urban and rural H (and W) experienced by different worker subpopulations over time are displayed in **Figures S5.3** and **S5.4**. Steep increases in H and W for workers with the least formal education, non-white workers (except for Black workers) relative to overall H and W were observed in both urban and rural populations, respectively (**Figure S5.3**). There are however, differences in exposures experienced by some subpopulations when disaggregated by urban/rural designation. Native Americans have the lowest exposures, relative to white exposures in rural areas, but have higher exposures than white populations in urban areas. Post 2016, in urban areas, Native Americans experience higher H and W than overall H and W urban populations, while in rural areas we observe the opposite result (**Figure S5.3**). In urban areas, workers belonging to the lowest income bracket had lower W and higher H relative to the W and H of the overall urban population. In rural areas, H and W of workers belonging to the lowest income bracket are among the lowest, relative to that of the overall rural population. We observe similar differences in the relative differences in H and W experienced by workers in urban and rural locations belonging to different job-types especially among workers in the mining sector, and health care workers, among others, over time (**Figure S5.4**).

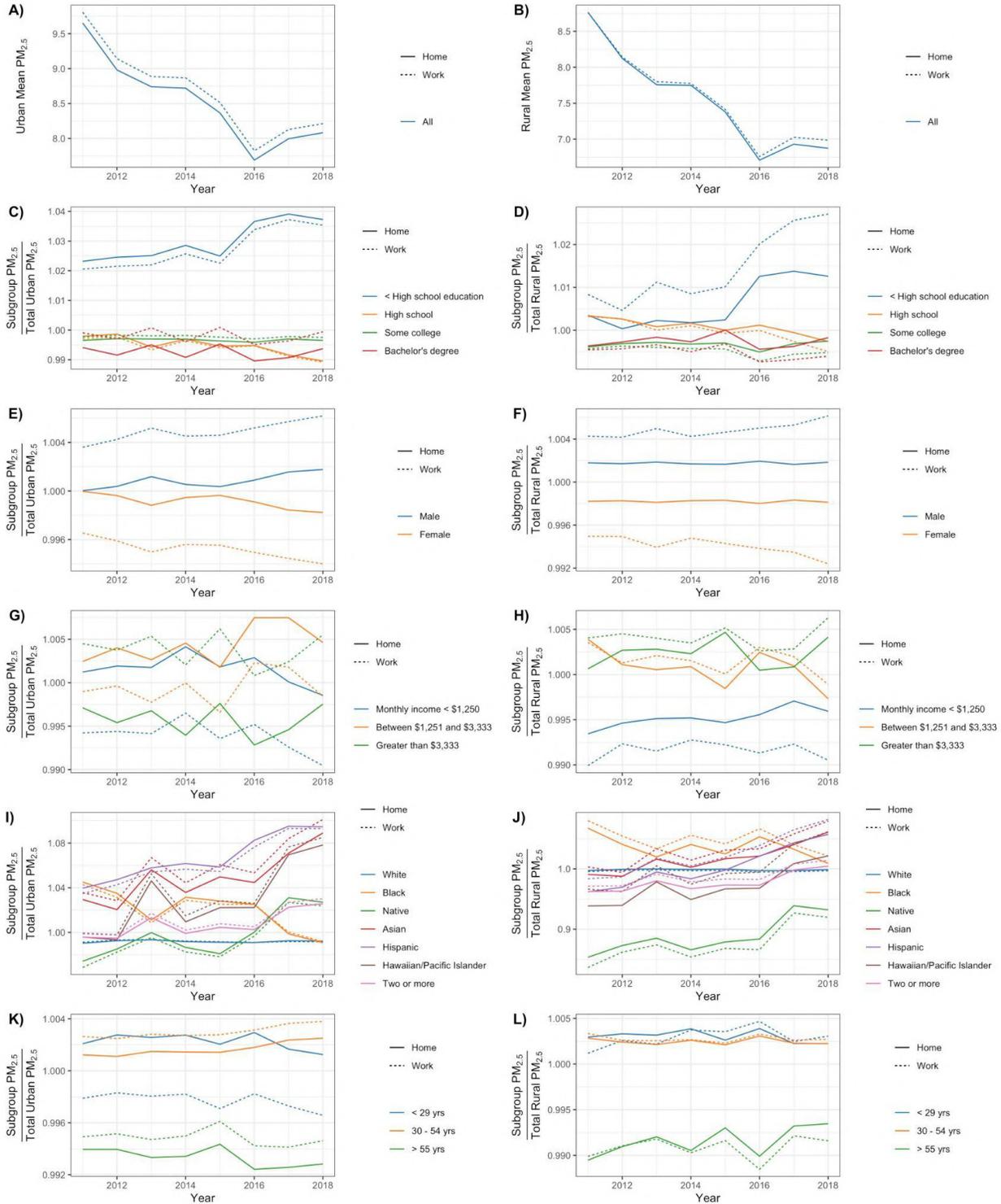

*Figure S5.3*: National average population-weighted exposure to PM$_{2.5}$ for A) All Urban workers, B) All Rural workers, and the Ratio of H and W population-weighted PM$_{2.5}$ exposure for C) Urban workers disaggregated by sex, D) Rural workers disaggregated by sex, E) Urban workers disaggregated by race/ethnicity, F) Rural workers disaggregated by race/ethnicity, G) Urban workers disaggregated by income levels, H) Rural workers disaggregated by income levels, I)

*Urban workers disaggregated by education, J) Rural workers disaggregated by education, K) Urban workers belonging to different age groups, L) Rural workers belonging to different age groups to national average urban and rural H and W exposures for years 2011 - 2018 for urban and rural locations, respectively.*

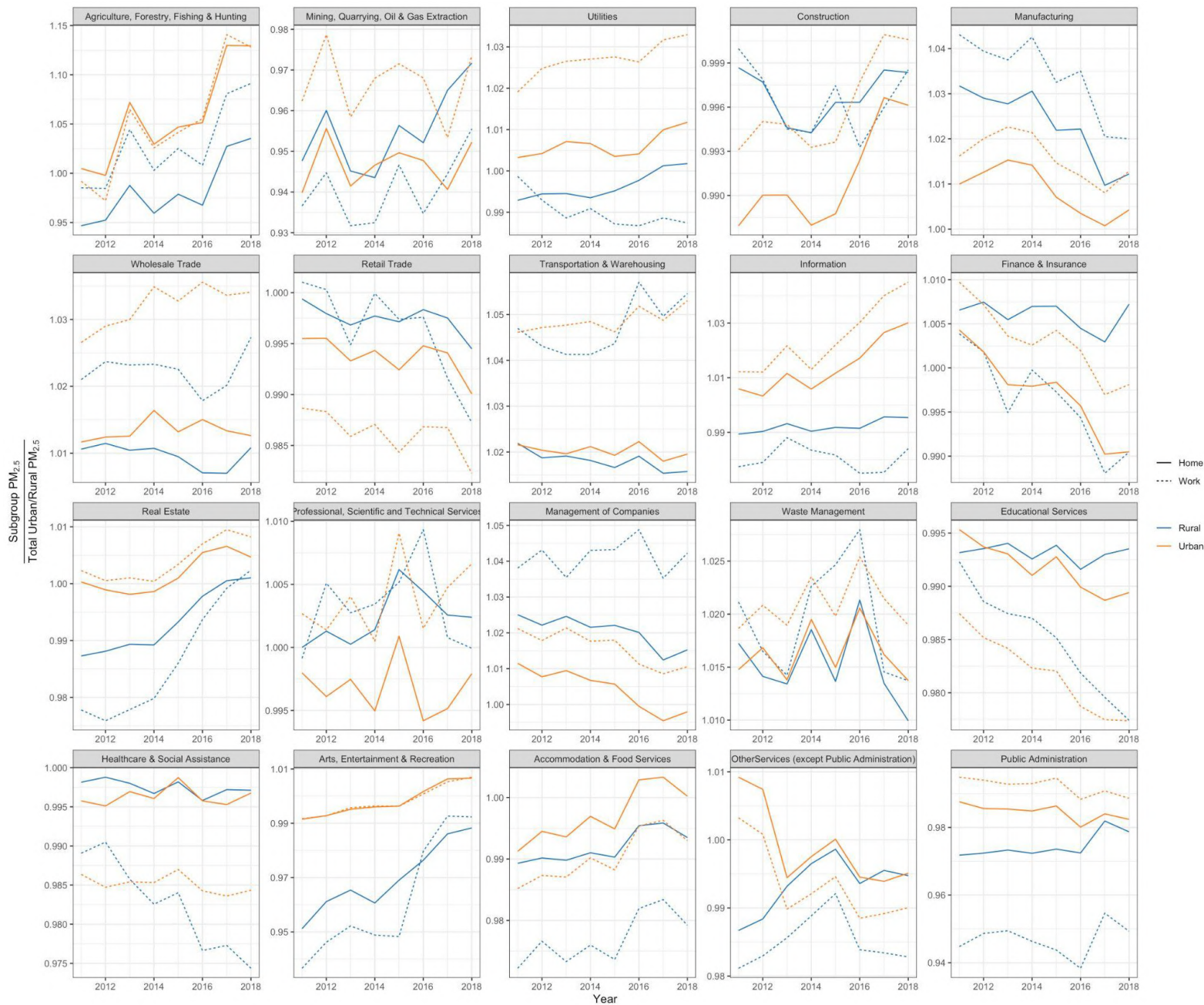

***Figure S5.4***: *Ratio of urban and rural population-weighted PM$_{2.5}$ exposures at home (H) and work locations (W) for workers with different job-types: A) Agriculture, Forestry, Fishing and Hunting, C) Utilities, D) Construction, E) Manufacturing, F) Wholesale Trade, G) Retail Trade, H) Transportation and Warehousing, I) Information, J) Finance and Insurance, K) Real Estate, L) Professional, Scientific and Technical Services, M) Management, N) Waste management, O)*

*Educational Serives, P) Health Care and Social Assistance, Q) Arts, Entertainment and Recreation, R) Accommodation and Food Services, S) Public Administration, T) Other Services for years 2011 - 2018 to overall urban and rural H and W.*

# S6: Trends in H and W using population-weighted 10th and 90th percentile of $PM_{2.5}$ concentrations, instead of population-weighted mean

Instead of calculating H and W using mean population-weighted exposure to $PM_{2.5}$ in home and work locations, we also plotted the 10th and 90th percentile population-weighted exposures to $PM_{2.5}$ for the total population as well as disaggregated by SES (**Figures S6.1**, **S6.3**) and job-type (**Figures S6.2**, **S6.4**) to visualize if differences in exposures exist across the spectrum of $PM_{2.5}$ concentrations. When evaluating trends in the 10th percentile of $PM_{2.5}$ concentrations, we still observe the same important disparities in H and W by race/ethnicity, education, and job-type, with less privileged populations experiencing higher H and W. However, when comparing trends of the 90th percentile of $PM_{2.5}$ concentrations experienced by workers, we observed post 2016 Black workers experiencing a smaller H and W than white workers. Other non-white workers experienced higher 90th percentile $PM_{2.5}$ concentrations. This surprising result is likely due to the fact that Black workers experience a smaller range of high $PM_{2.5}$ concentrations. Other workers experience a larger range of $PM_{2.5}$ concentrations, with a small number of white workers exposed to high $PM_{2.5}$ concentrations. At both the 10th and 90th percentile levels, we observe that workers with the least amount of formal education experience higher H and W.

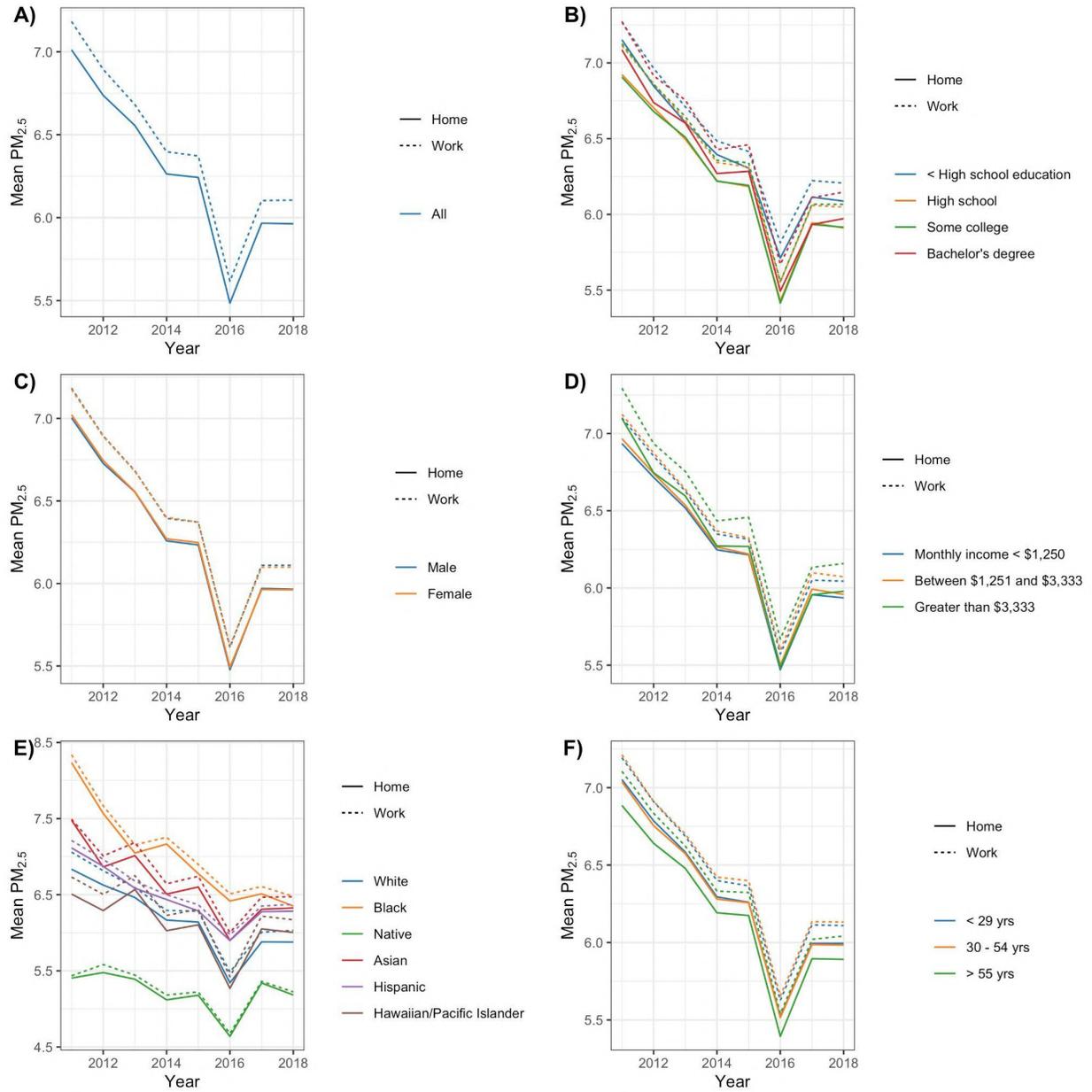

*Figure S6.1*: 10th percentile population-weighted exposure to $PM_{2.5}$ (µg/m$^3$) experienced at H and W disaggregated by race, sex, income, and education

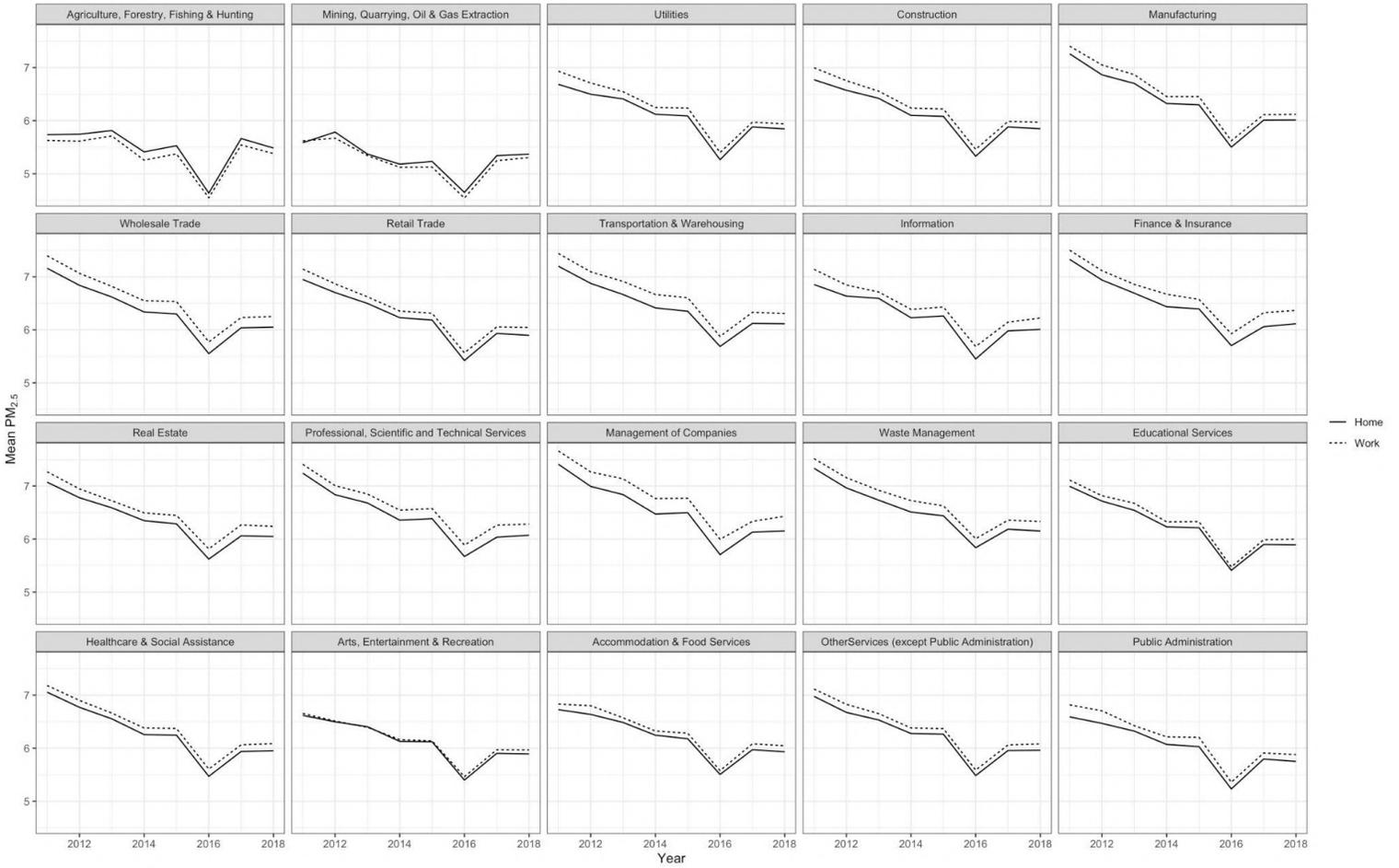

*Figure S6.2*: 10th percentile of population-weighted PM$_{2.5}$ exposure for workers with different job-types: A) Agriculture, Forestry, Fishing and Hunting, C) Utilities, D) Construction, E) Manufacturing, F) Wholesale Trade, G) Retail Trade, H) Transportation and Warehousing, I) Information, J) Finance and Insurance, K) Real Estate, L) Professional, Scientific and Technical Services, M) Management, N) Waste management, O) Educational Serives, P) Health Care and Social Assistance, Q) Arts, Entertainment and Recreation, R) Accommodation and Food Services, S) Public Administration, T) Other Services, disaggregated by urban/rural census tracts for years 2011 - 2018

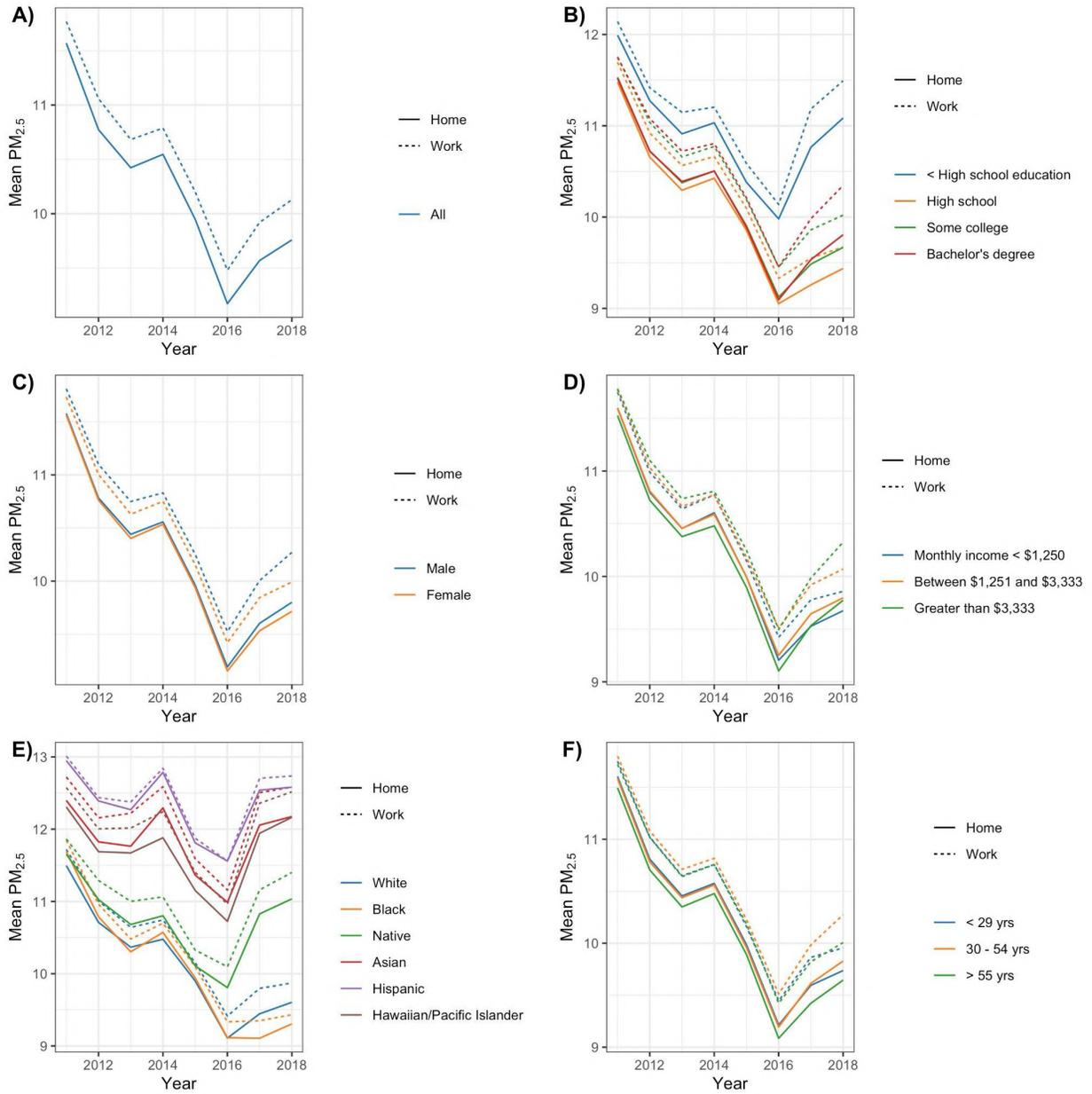

***Figure S6.3***: *90th percentile population-weighted exposure to PM$_{2.5}$ (μg/m$^3$) experienced at H and W fdisaggregated by race, sex, income, and education*

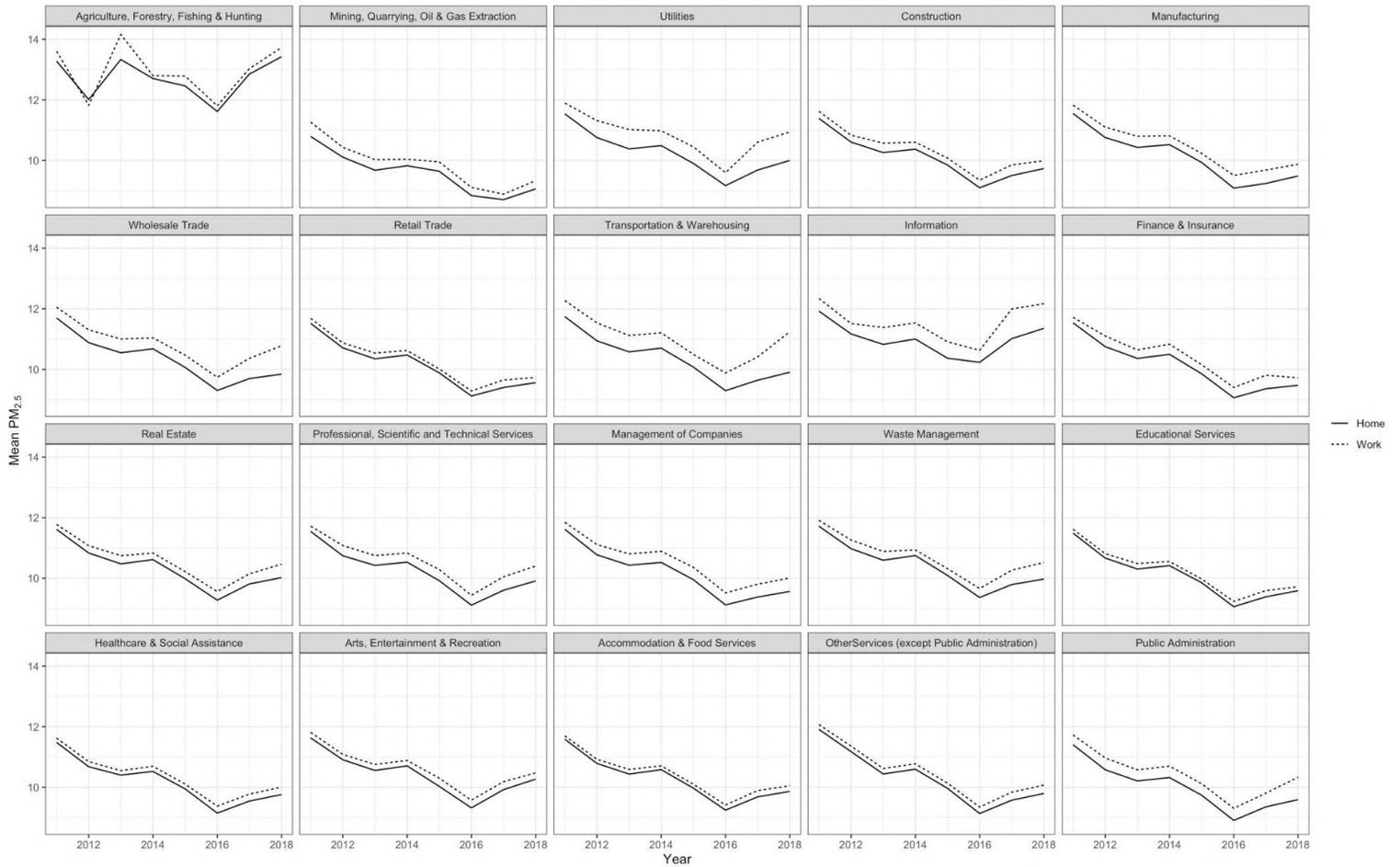

*Figure S6.4*: 90th percentile of population-weighted PM$_{2.5}$ exposure for workers with different job-types: A) Agriculture, Forestry, Fishing and Hunting, C) Utilities, D) Construction, E) Manufacturing, F) Wholesale Trade, G) Retail Trade, H) Transportation and Warehousing, I) Information, J) Finance and Insurance, K) Real Estate, L) Professional, Scientific and Technical Services, M) Management, N) Waste management, O) Educational Serives, P) Health Care and Social Assistance, Q) Arts, Entertainment and Recreation, R) Accommodation and Food Services, S) Public Administration, T) Other Services, disaggregated by urban/rural census tracts for years 2011 - 2018.

# S7: Trends in error and percent error in assigning workers exposure based on home, instead of home *and* workplace locations

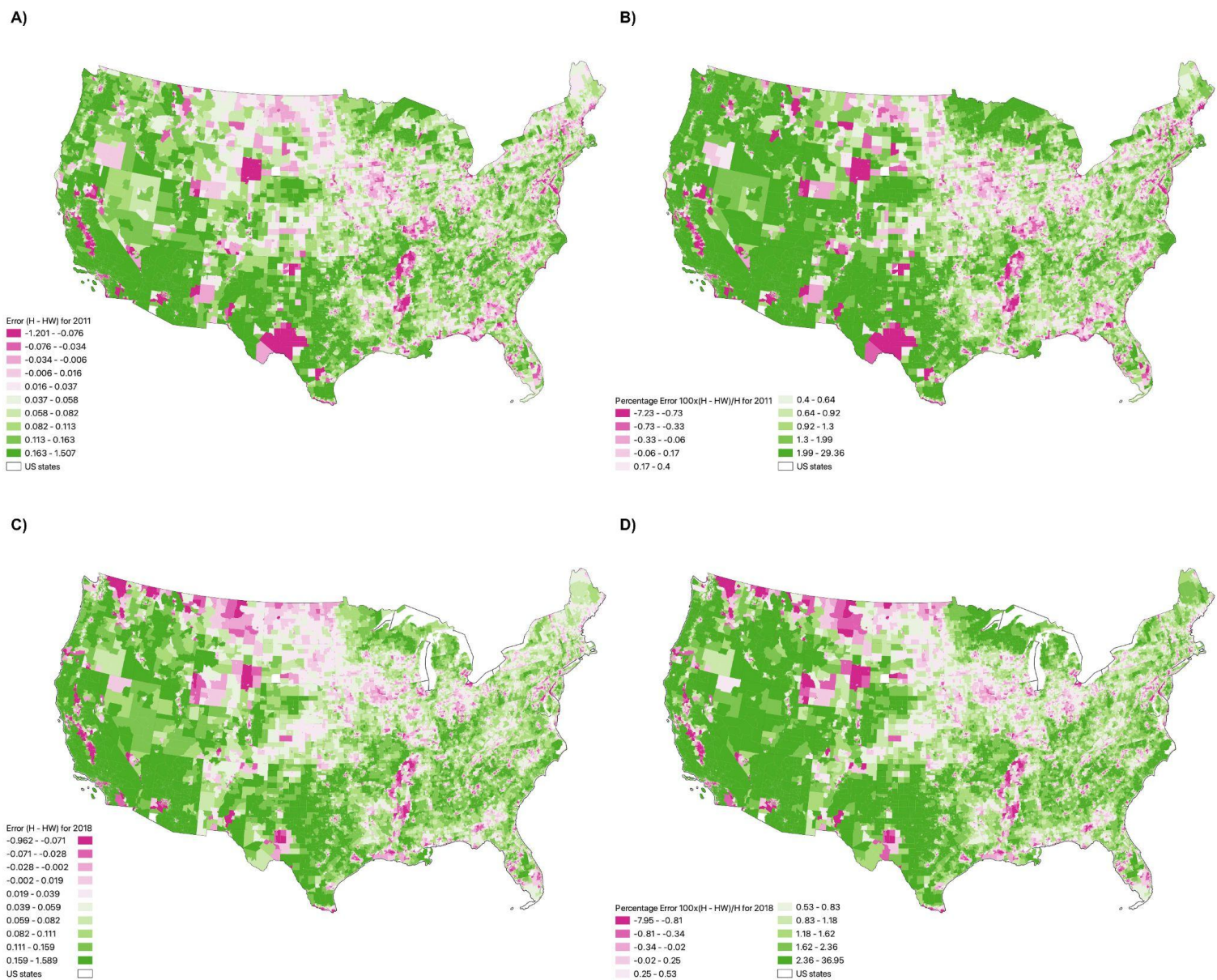

***Figure S7.1***: *Maps of Mean Error (μg/m³) in PM$_{2.5}$ concentrations (H - HW), and Percent Error: 100×(H-HW)/H classified by decile for the overall worker population for the years 2011 and 2018, respectively. Note that in 2011, the LODES data recorded a small number of workers living in census tracts corresponding to the Great Lakes region in Michigan (likely on house boats). Post 2016, the LODES data indicated no residents in these areas.*

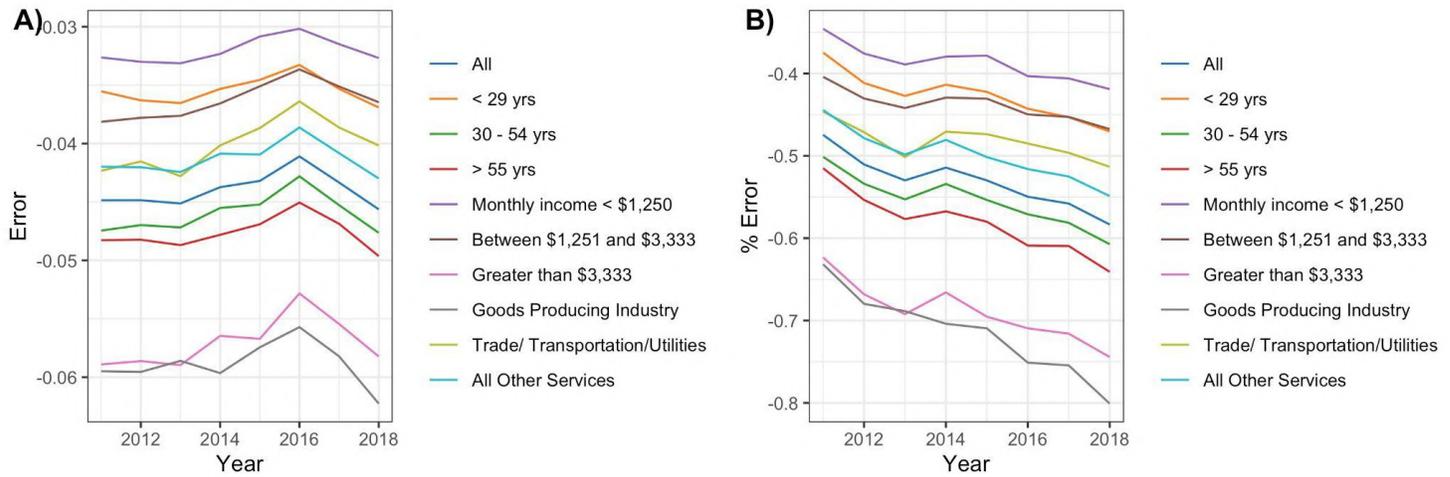

***Figure S7.2***: *Mean Error (µg/m$^3$) in PM$_{2.5}$ concentrations (H - HW), and % Error: 100×(H-HW)/H for the overall worker population for the years 2011 - 2018*

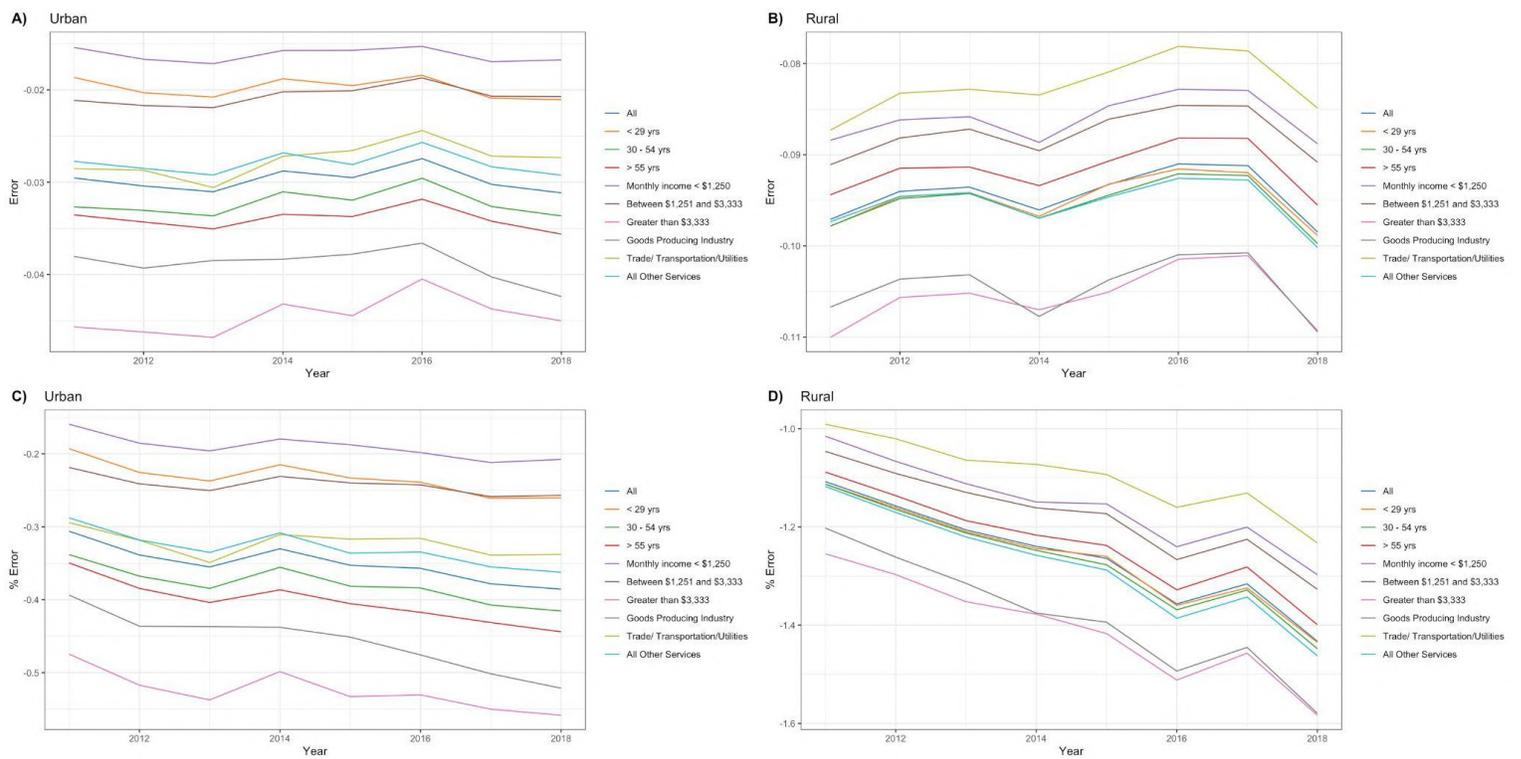

***Figure S7.3***: *Mean Error (µg/m$^3$) in PM$_{2.5}$ concentrations (H - HW), and % Error: 100×(H-HW)/H for the overall worker population as well as for different subpopulations described in the LODES OD file for the years 2011 - 2018 for urban and rural populations (assigned based on residential locations)*

# S8: Quantifying Disparities in H and W

## S8.1: Absolute and Relative Differences in H and W Between the Most Exposed and Least Exposed Subpopulatons using population-weighted 10th and 90th percentile $PM_{2.5}$ concentrations instead of the population-weighted mean

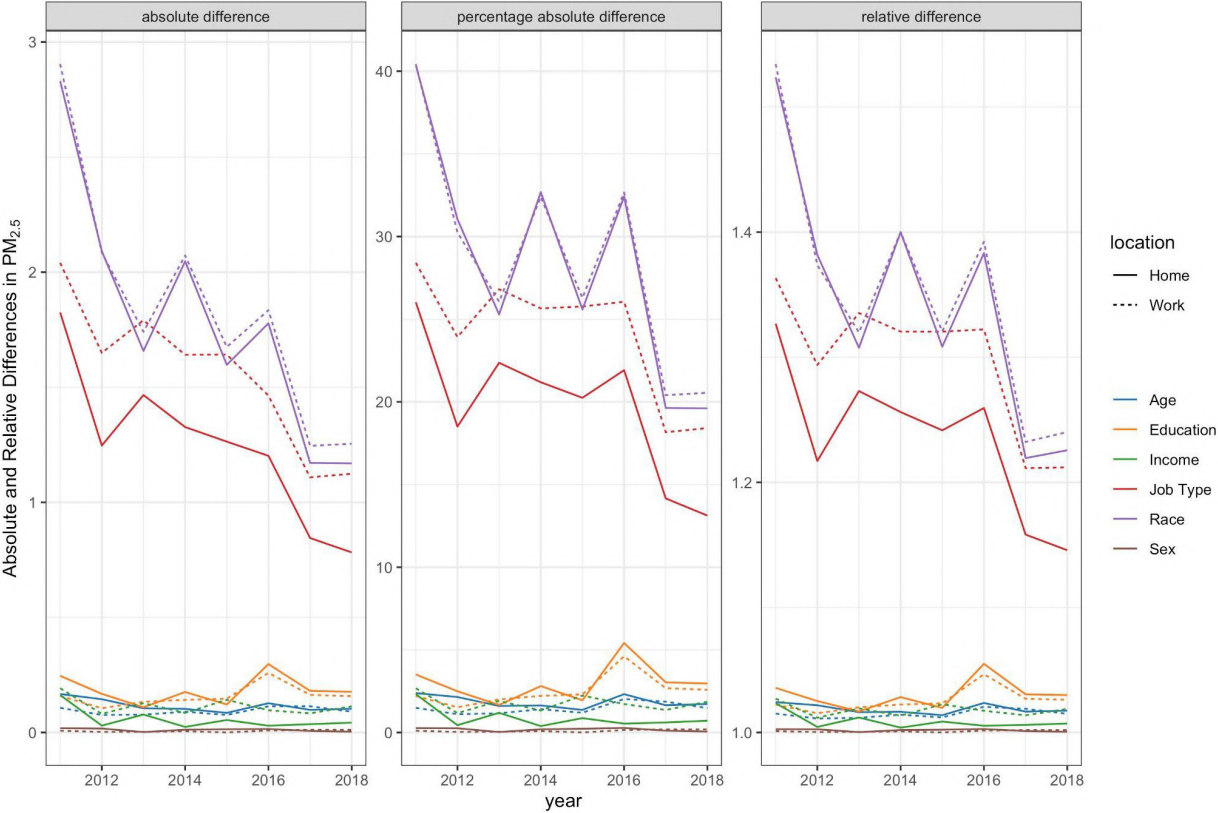

***Figure S8.1.1***: *Absolute difference (µg/m$^3$), % absolute difference, relative difference in exposures between the most exposed and least exposed population (using 10th percentile of population-weighted mean $PM_{2.5}$) within each subgroup for the years 2011-2018*

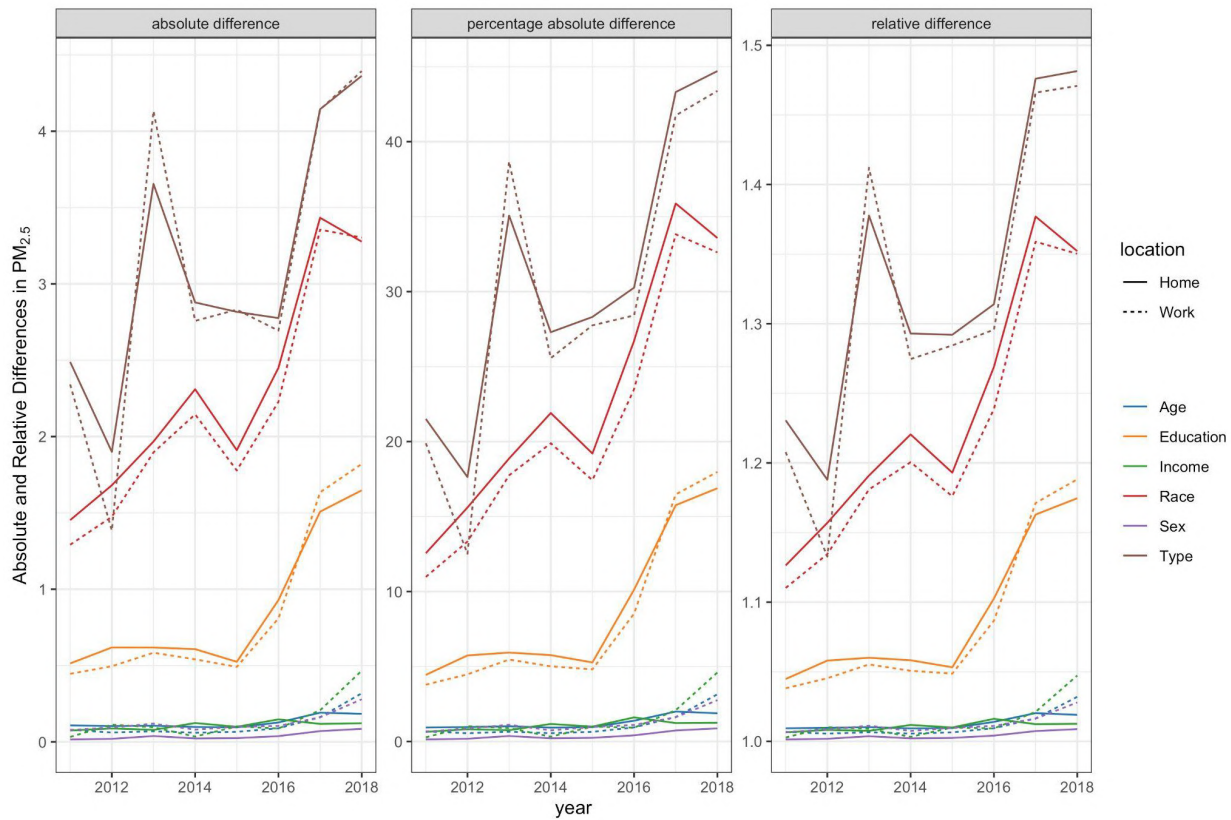

***Figure S8.1.2***: *Absolute difference (μg/m$^3$), % absolute difference, relative difference in exposures between the most exposed and least exposed population (using 90th percentile of population-weighted mean PM$_{2.5}$) within each subgroup for the years 2011-2018.*

## S8.2: Mean H and W experienced by workers in census tracts containing different fractions of subpopulations categorized by decile

Mean and 95% CI of H and W experienced by workers living and working in census tracts containing different fractions of subpopulations, categorized by deciles for the years 2011 and 2018 are displayed in **Figure S8.2.1**. We observed that overall, census tracts with the highest fraction of white workers residing/working (Decile 10) had the lowest H/W exposures for 2011 and 2018. H and W were higher in census tracts with a higher fraction of all non-white populations. H and W were substantially higher in 2018 than 2011 for most census tracts decile-groups, considered. However, census tracts with the highest fraction of Asian, Hispanic, workers belonging to two or more race groups, and Hawiian and other Pacific Islanders had similar concentrations in 2011 and 2018, indicating that these census tracts had not experienced many of the benefits of lower air pollution levels that other groups had experienced. Census tracts with the highest fraction of these subgroups also experienced the highest PM$_{2.5}$ levels compared to census groups falling in other deciles. Similar trends in H and W were observed for the years 2018 and 2011 across workers belonging to most subpopulations.

We also observed substantial increases in H and W for census tracts with a high fraction of workers with little formal education compared to census tracts with a small fraction of these workers. Opposite trends were observed for workers with a high school degree or more. We observed varying trends for workers who worked in different sectors. For example, census tracts with a higher fraction of workers living/working in them in the transportation and warehousing sector, the waste management sector, and the wholesale trade sector had the highest H and W levels, compared to tracts with a lower fraction of these workers.

Mean H and W experienced by workers living and working in census tracts containing different fractions of subpopulations categorized by decile for the years 2011 and 2018 disaggregated by urban/rural designation is displayed in **Figure S8.2.2** and **S8.2.3**. There are certain differences in trends in urban versus rural areas. For example, in rural areas, census tracts with the highest prevalence of Hispanic or Asian workers had the lowest H and W compared to census tracts with the lowest prevalence of these subpopulations in 2011. In 2018, this trend reverses. The lowest income residents in rural areas have the lowest exposures, while we observe the opposite trend in urban areas. We observe similar trends across other categories.

We quantified the differences in H and W experienced by urban and rural census tracts in the top and bottom decile when categorized by the fraction of different subpopulations in **Figure S8.2.4**. We observed that there was little difference in H and W in census tracts with the highest and lowest prevalence of white workers in rural areas, while a difference of -1 µg/m$^3$ was observed in urban areas in 2018. Census tracts with the highest and lowest prevalence of Black workers experienced similar differences in H and W in urban and rural areas (ΔW and ΔH ~ 0.3 µg/m$^3$ in 2018). Differences in H and W for tracts with the highest and lowest prevalence of Hispanic or Asian workers was higher in urban than rural areas.

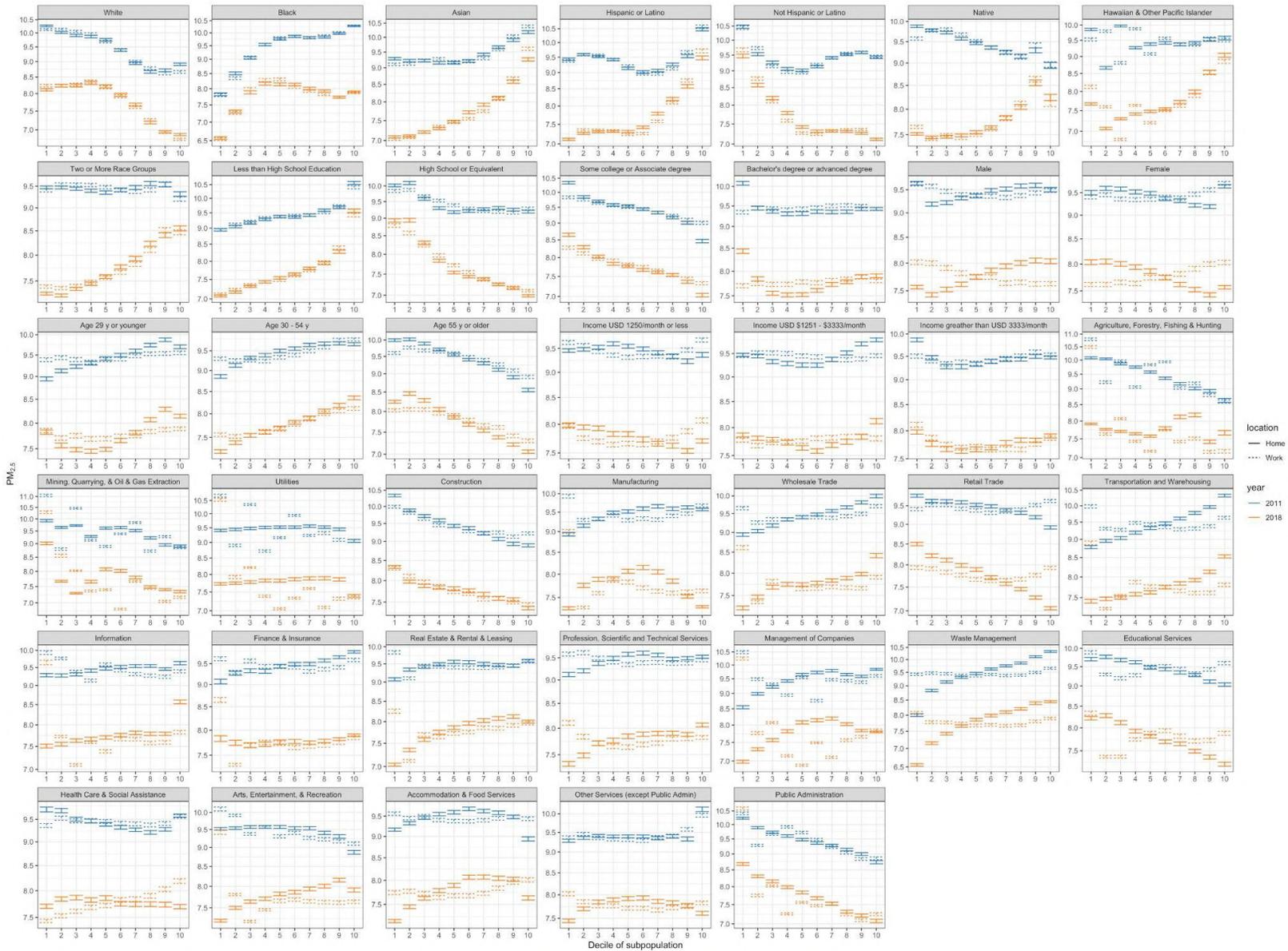

*Figure 8.2.1*: Mean and 95% CI of H and W experienced by workers living and working in census tracts containing different fractions of subpopulations, categorized by deciles for the years 2011 and 2018.

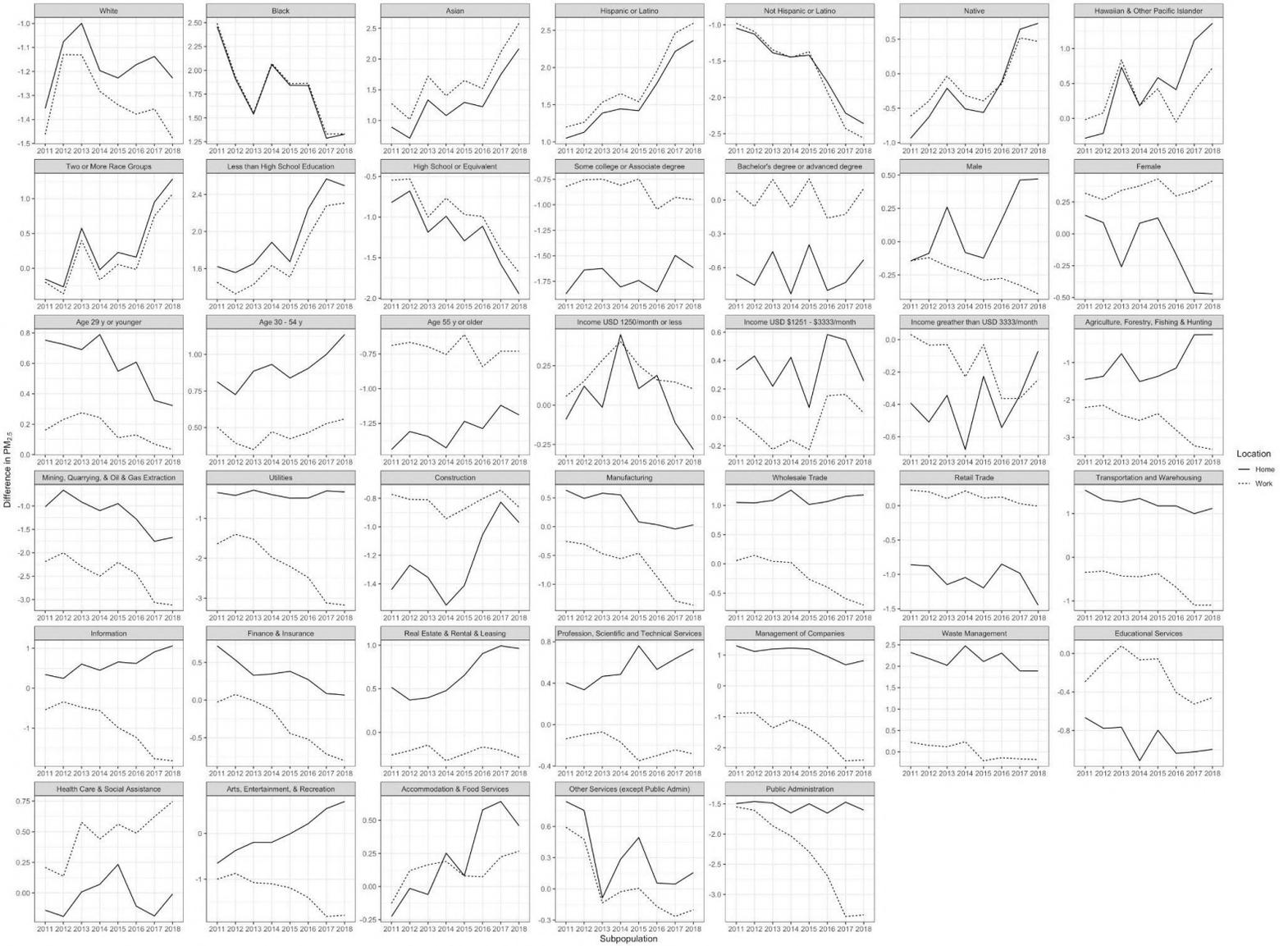

*Figure S8.2.2*: Difference in mean H and W experienced by census tracts in the top and bottom deciles based on the the fractions of subpopulations living/working in each tract, for the years 2011-2018.

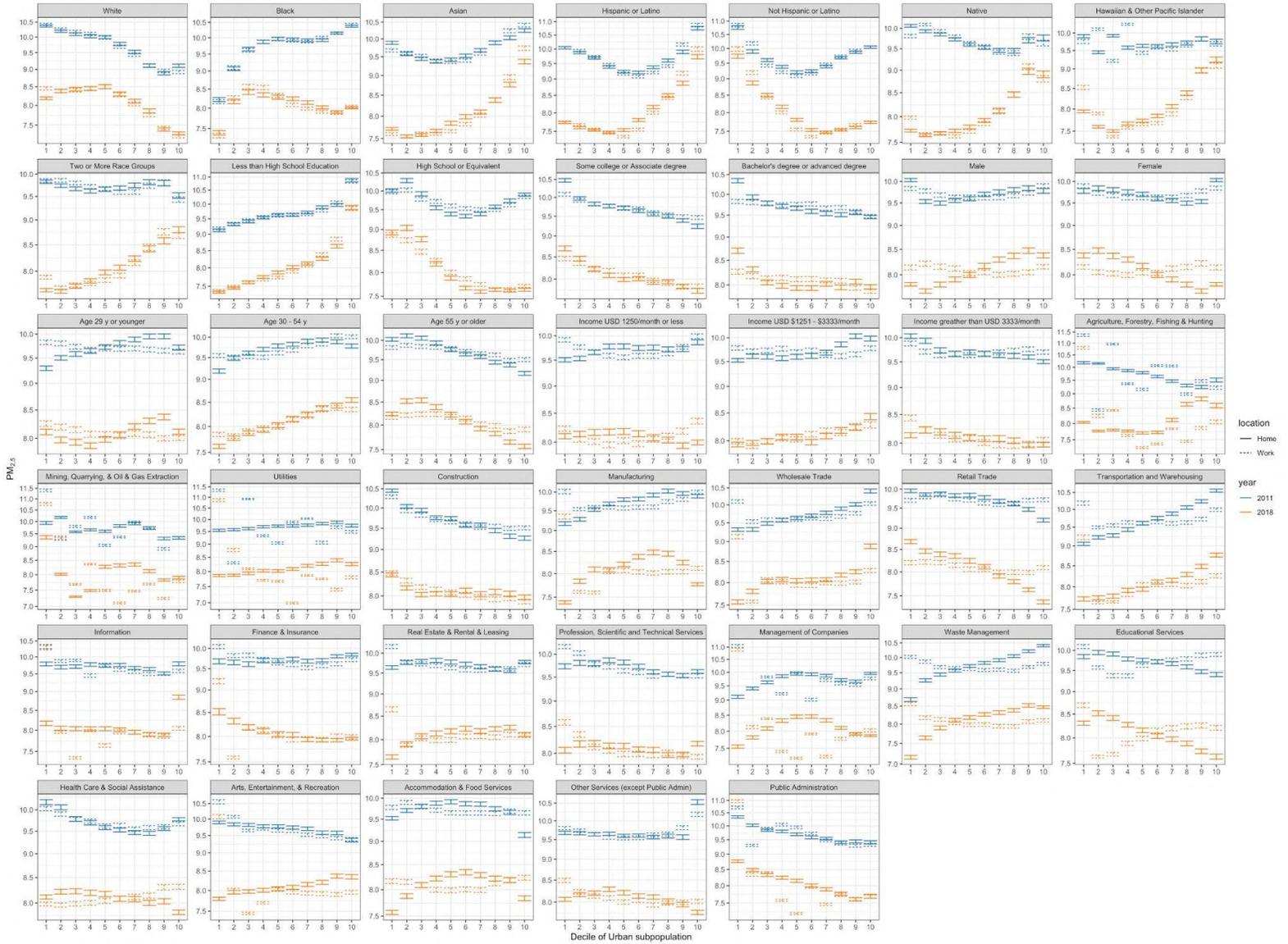

*Figure S8.2.3*: Mean and 95% CI of H and W experienced by workers living and working in urban census tracts containing different fractions of subpopulations, categorized by deciles for the years 2011 and 2018

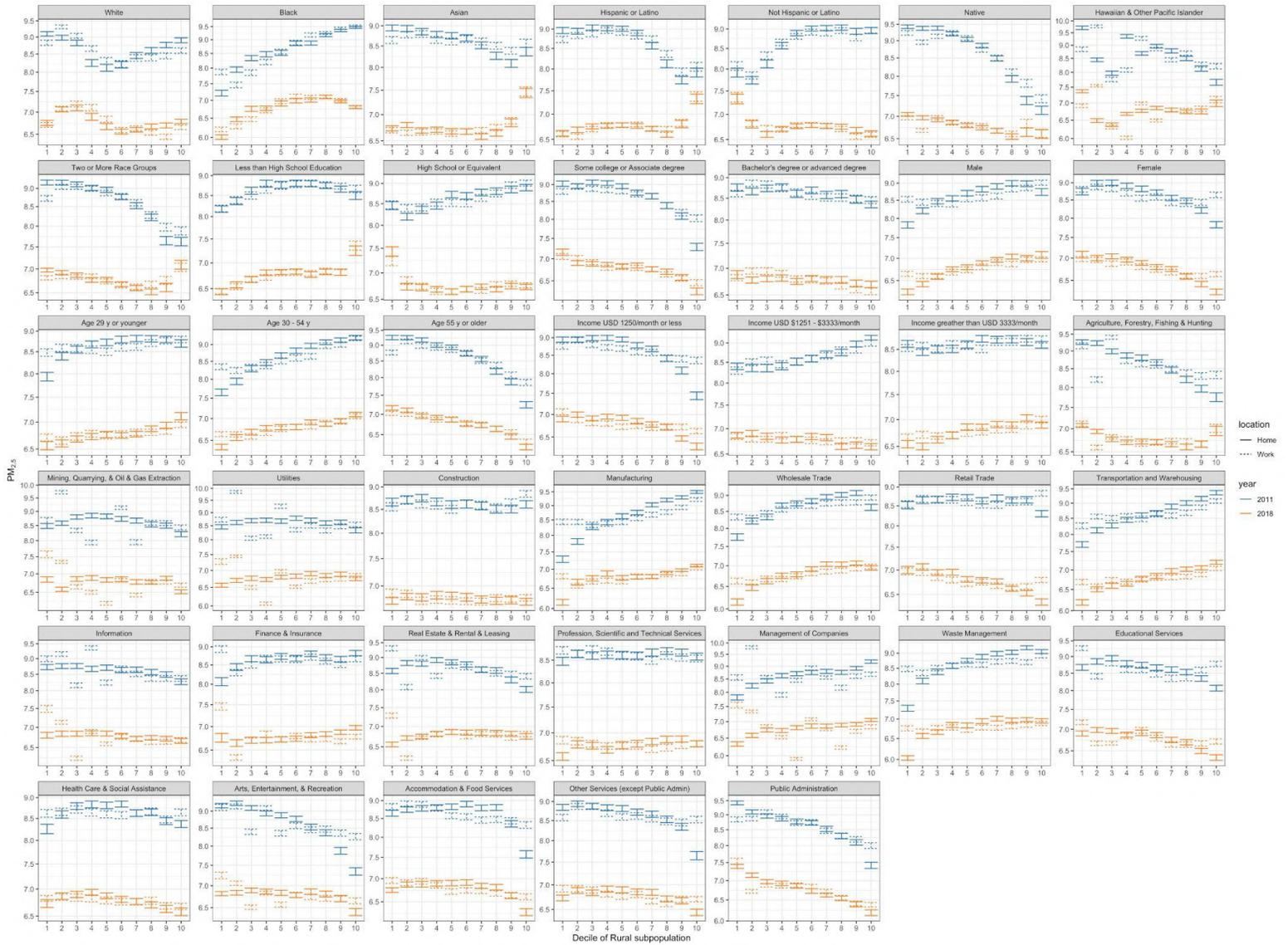

*Figure S8.2.4*: Mean and 95% CI of H and W experienced by workers living and working in rural census tracts containing different fractions of subpopulations, categorized by deciles for the years 2011 and 2018

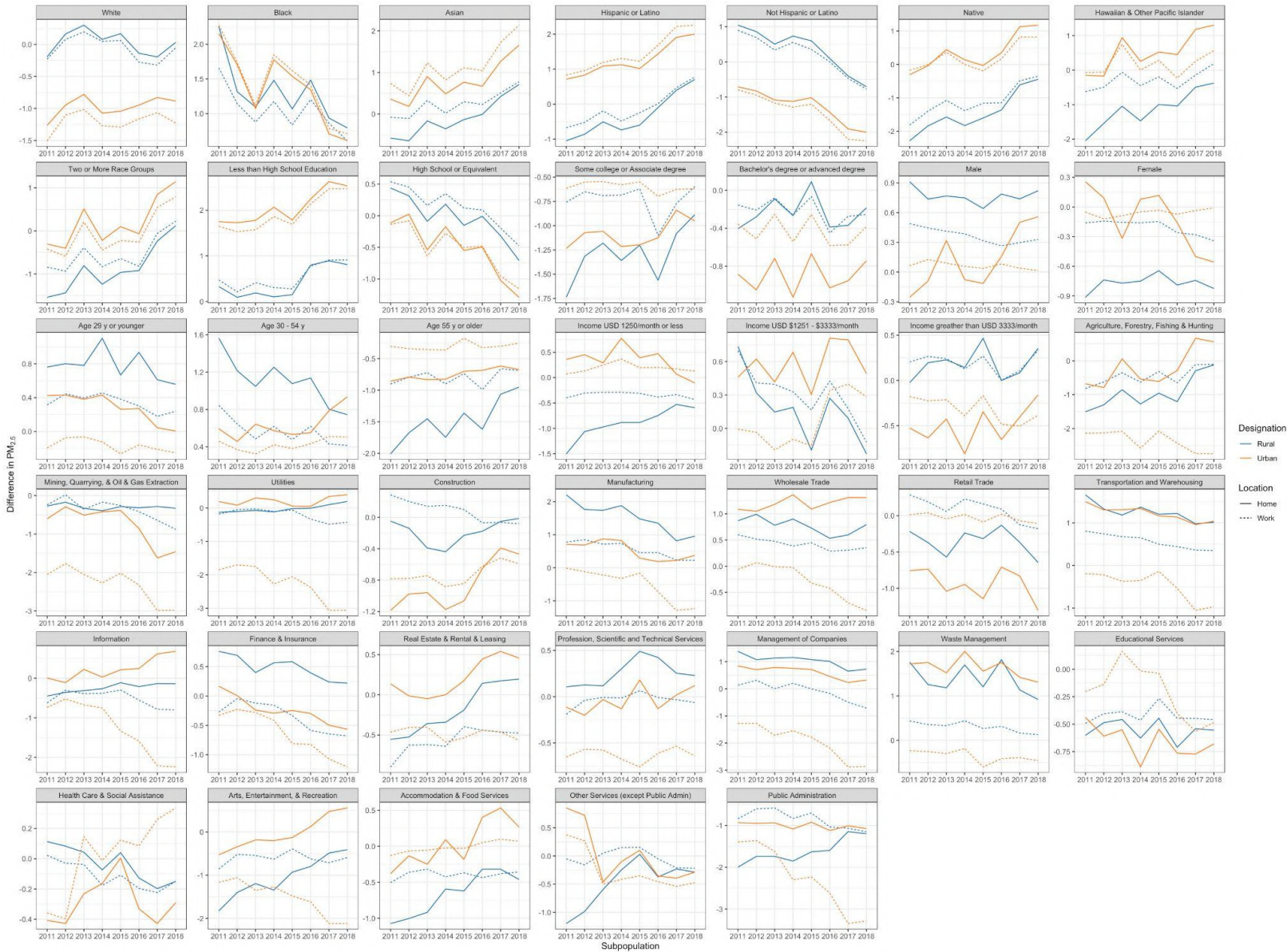

*Figure 8.2.5*: Difference in mean PM$_{2.5}$ experienced by urban and rural census tracts in the top and bottom deciles based on the the fractions of subpopulations living/working in each tract, for the years 2011-2018.

## S8.3: Mean fractions of workers belonging to different subpopulations living/working in census tracts corresponding to different H and W categorized by decile

Conversely we display the mean fractions of workers belonging to different subpopulations living/working in census tracts corresponding to different PM$_{2.5}$ concentrations (classified by deciles) for years 2011 and 2018 in **Figure S8.3.1**. We note that census tracts with the highest PM$_{2.5}$ concentrations have the lowest fraction of white residents, but the highest fraction of white workers. These census tracts had high fractions of non-white residents living and working in them. The diverging patterns observed in the fraction of each subpopulation living versus those working in census tracts by $PM_{2.5}$ decile is also observed for the more highly educated, and the oldest workers. The divergence observed was less pronounced by job-type (**Figure S8.3.1**).

We also evaluated the mean fractions of workers belonging to different subpopulations living/working in census tracts corresponding to different $PM_{2.5}$ concentrations (classified by deciles) disaggregated by urban/rural designation for years 2011 and 2018 (**Figure S8.3.2**). In urban areas, a higher fraction of workers worked, as opposed to lived in census tracts with high $PM_{2.5}$ concentrations. We observed the opposite trend in rural locations. In both rural and urban areas, we saw the same separation between the fraction of white workers living and working in census tracts with high $PM_{2.5}$ concentrations. We did not see such a separation among non-white workers.

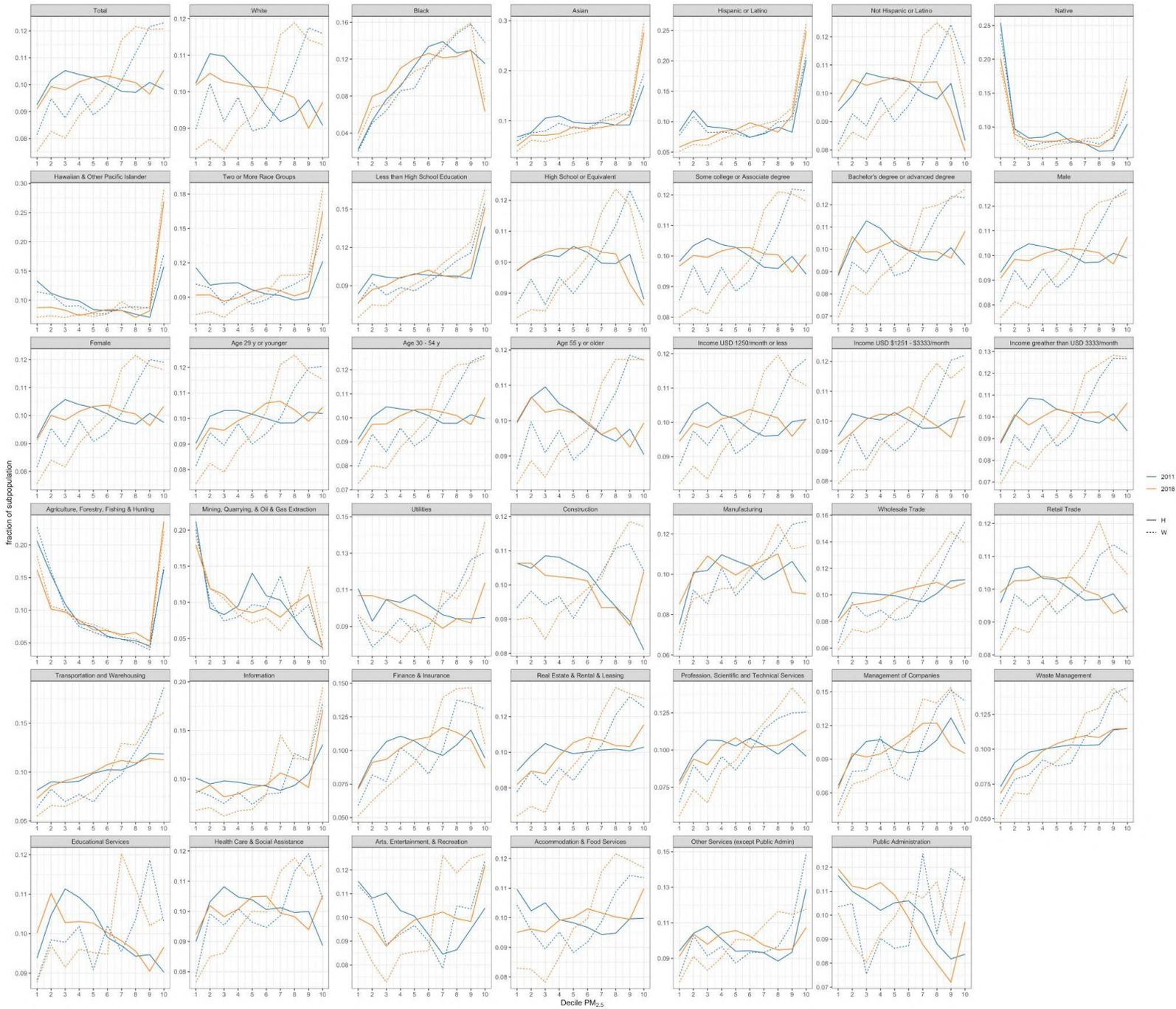

*Figure S8.3.1*: Fraction of the total population and subpopulations corresponding to different deciles of H and W $PM_{2.5}$ concentrations years 2011 and 2018

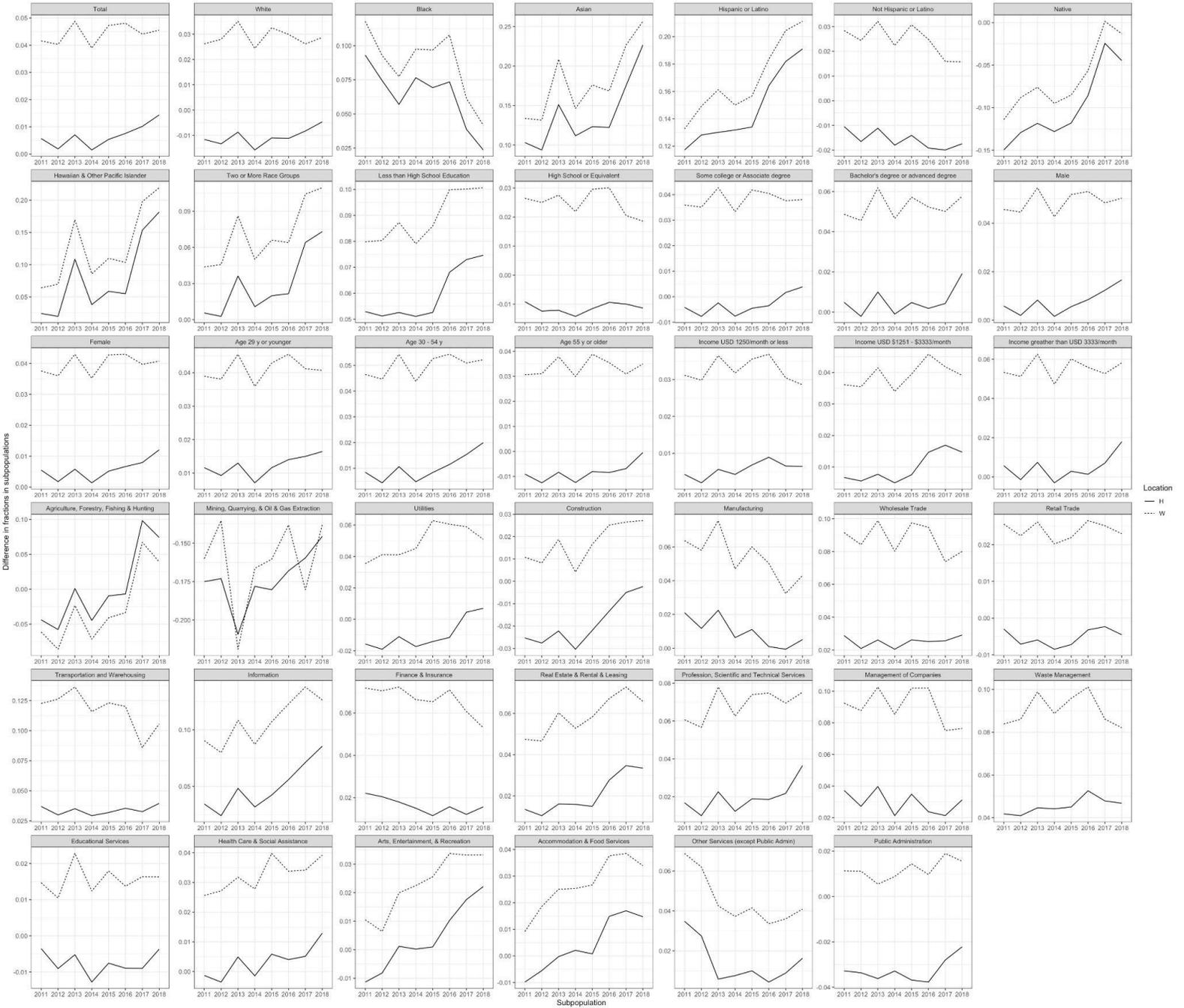

*Figure S8.3.2*: Difference in the fraction of the total population and subpopulations in tracts corresponding to the top and bottom decile based on H and W $PM_{2.5}$ concentrations over the years 2011-2018.

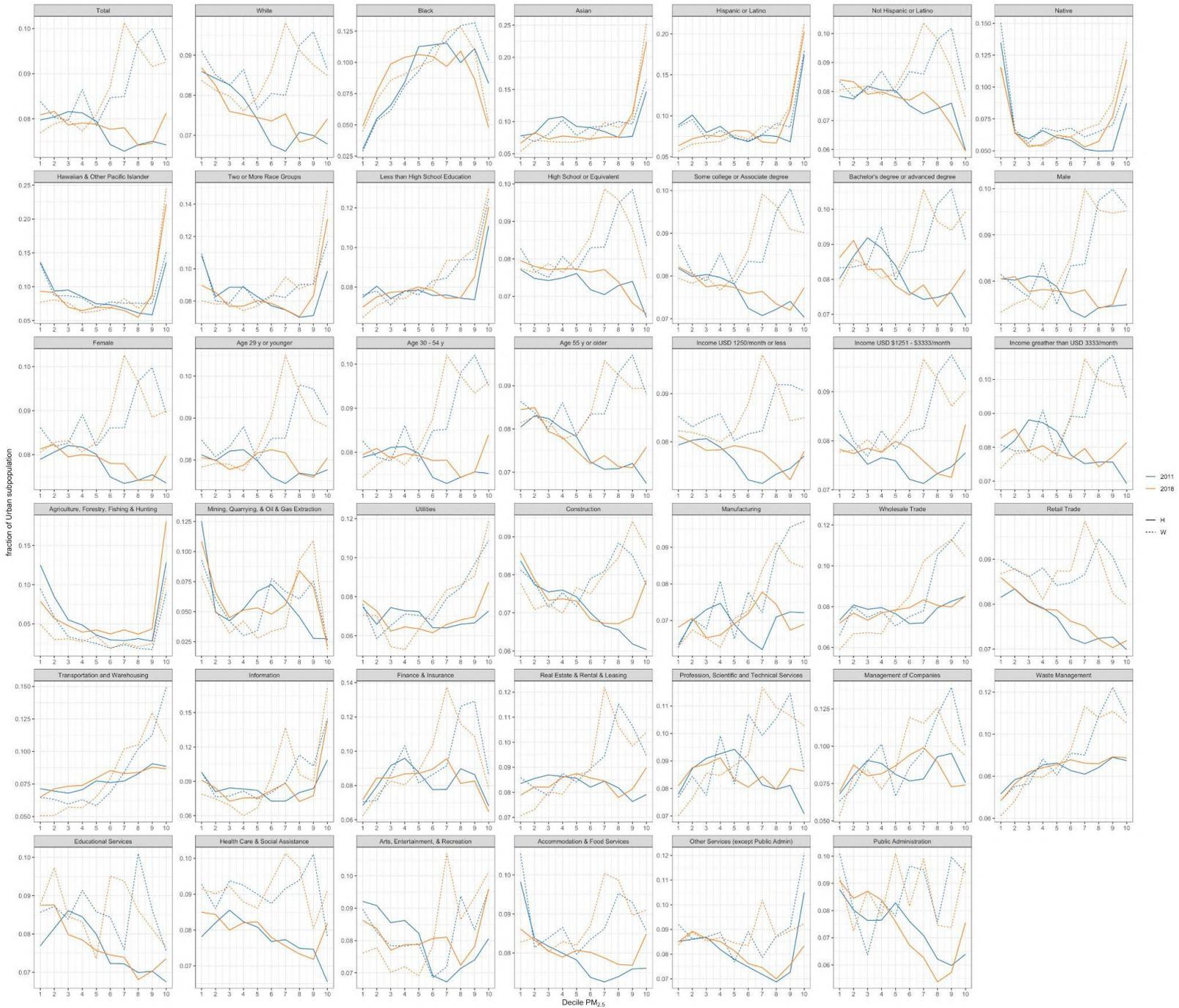

*Figure S8.3.3:* Fraction of the total population and subpopulations corresponding to different deciles of H and W PM$_{2.5}$ concentrations years 2011-2018 for urban areas

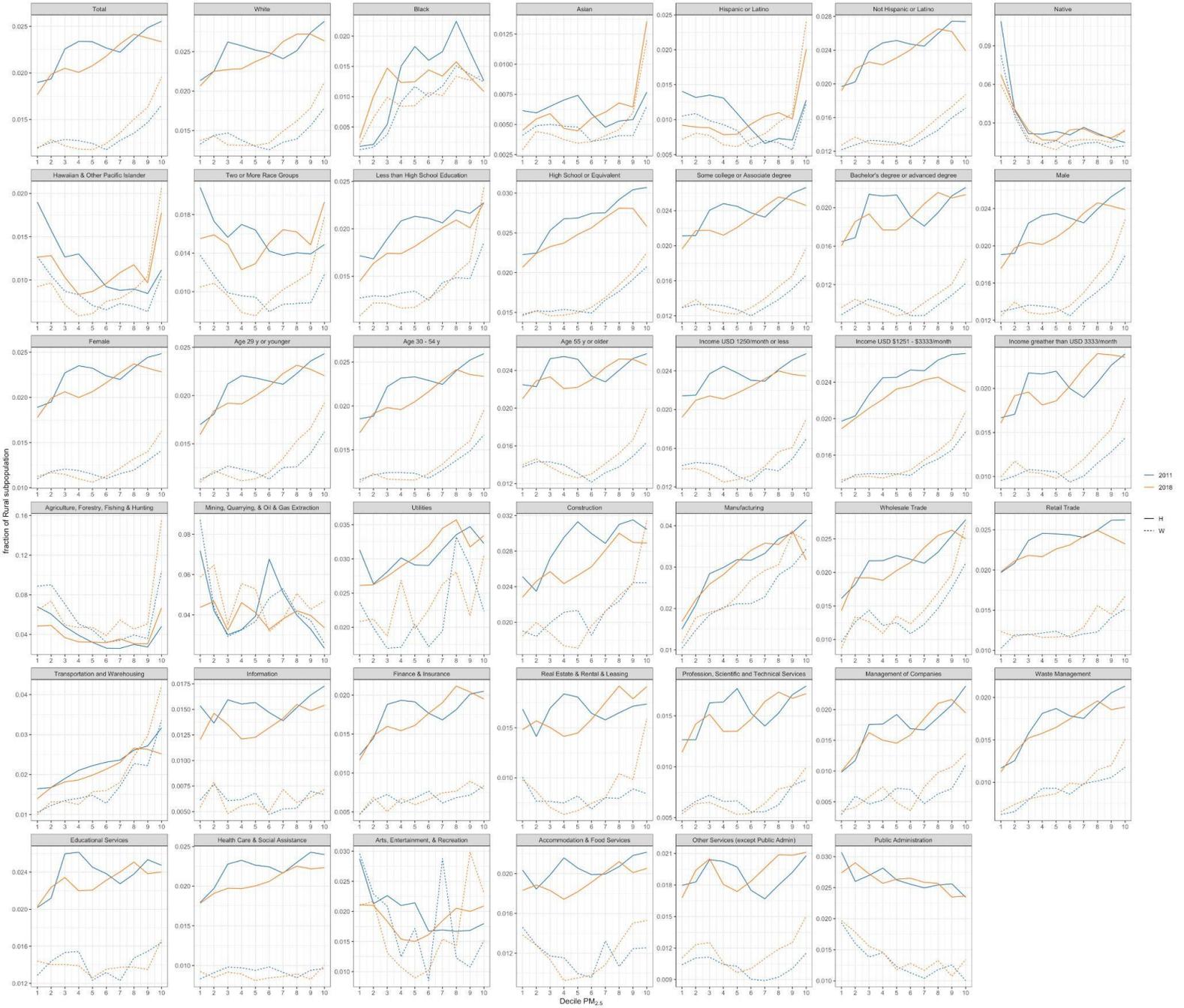

***Figure S8.3.4:*** *Fraction of the total population and subpopulations corresponding to different deciles of H and W $PM_{2.5}$ concentrations years 2011-2018 for rural areas*

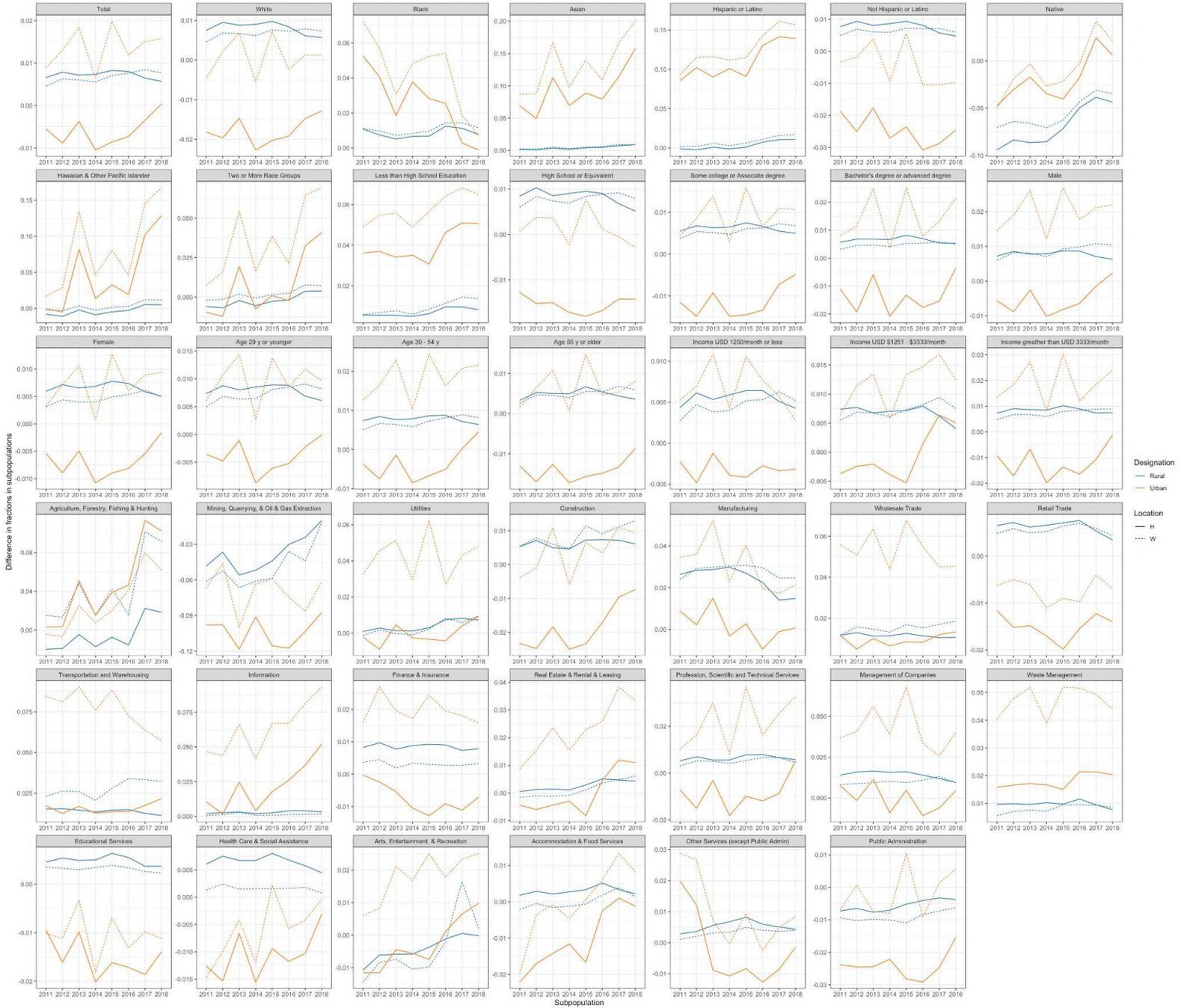

*Figure S8.3.5*: Difference in the fraction of the total population and subpopulations in tracts corresponding to the top and bottom decile based on H and W PM$_{2.5}$ concentrations over the years 2011-2018.

## S8.4: Atkinson Index

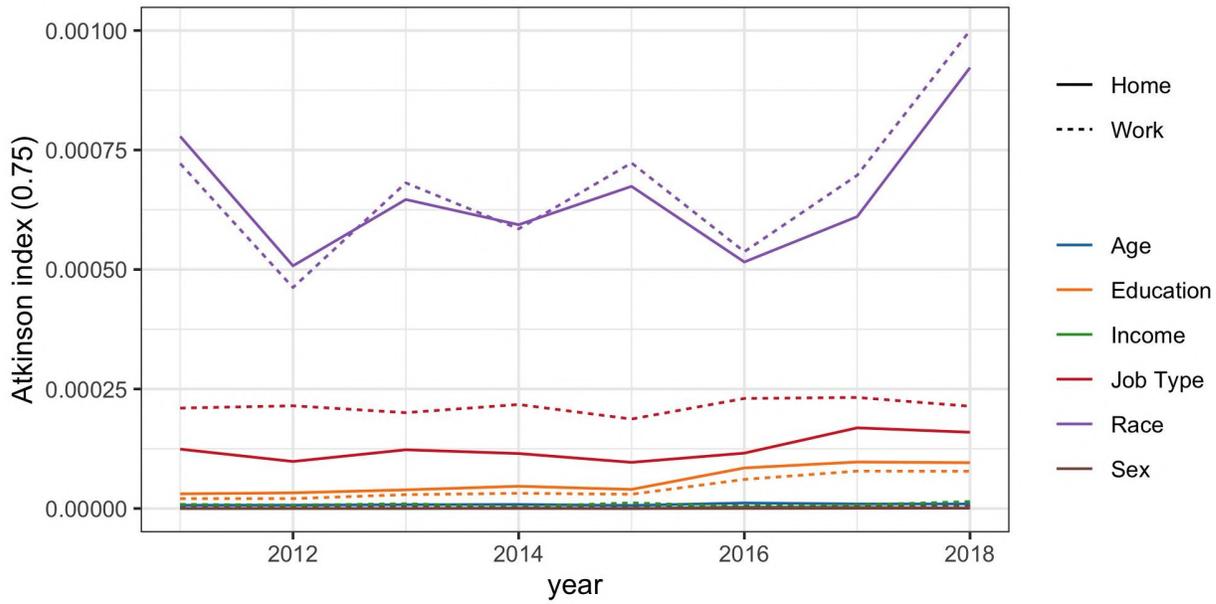

***Figure S8.4.1****: Atkinson Index for the years 2011 - 2018 (inequality aversion parameter = 0.75) displaying between-group inequality in H and W for the years 2011 - 2018*

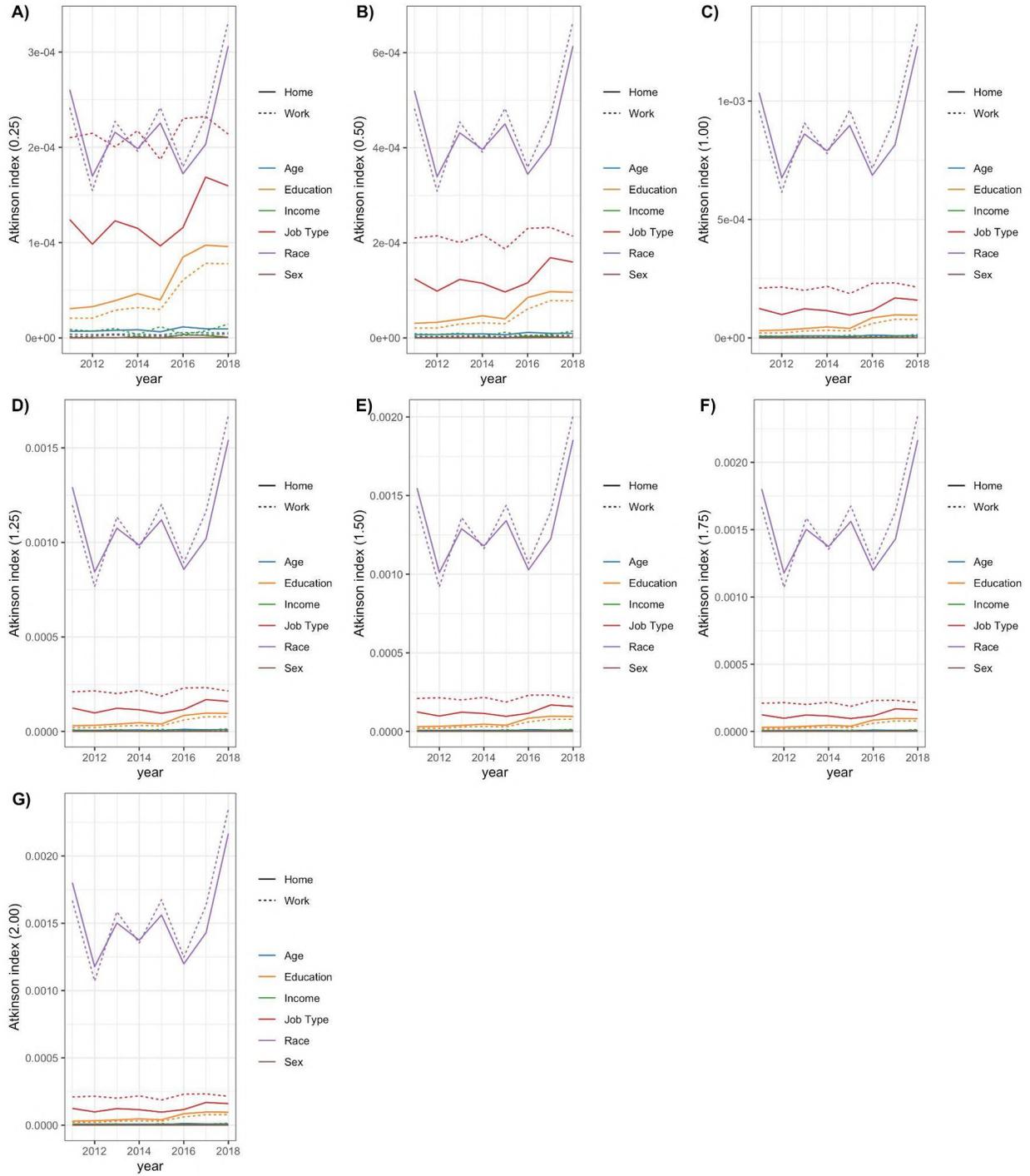

***Figure S8.4.2****: Atkinson Index for the years 2011 - 2018 (inequality aversion parameters = 0.25, 0.50, 1.0, 1.25, 1.50, 1.75, 2.0) displaying between-group inequality in H and W for the years 2011 - 2018*

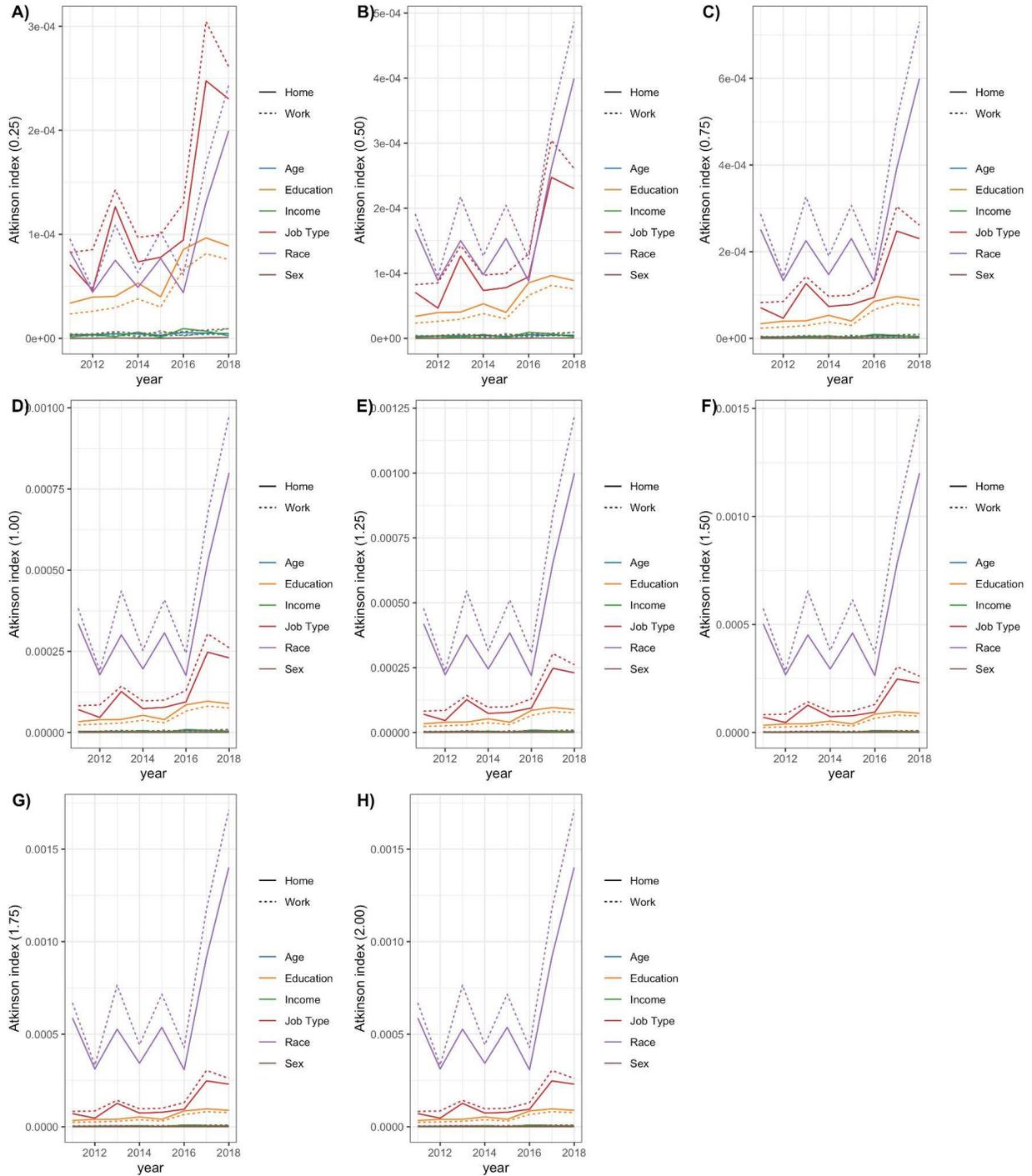

***Figure S8.4.3***: *Atkinson Index for the years 2011 - 2018 (inequality aversion parameters = 0.25, 0.50, 0.75, 1.0, 1.25, 1.50, 1.75, 2.0) displaying between-group inequality in H and W for the years 2011 - 2018 for urban populations, alone*

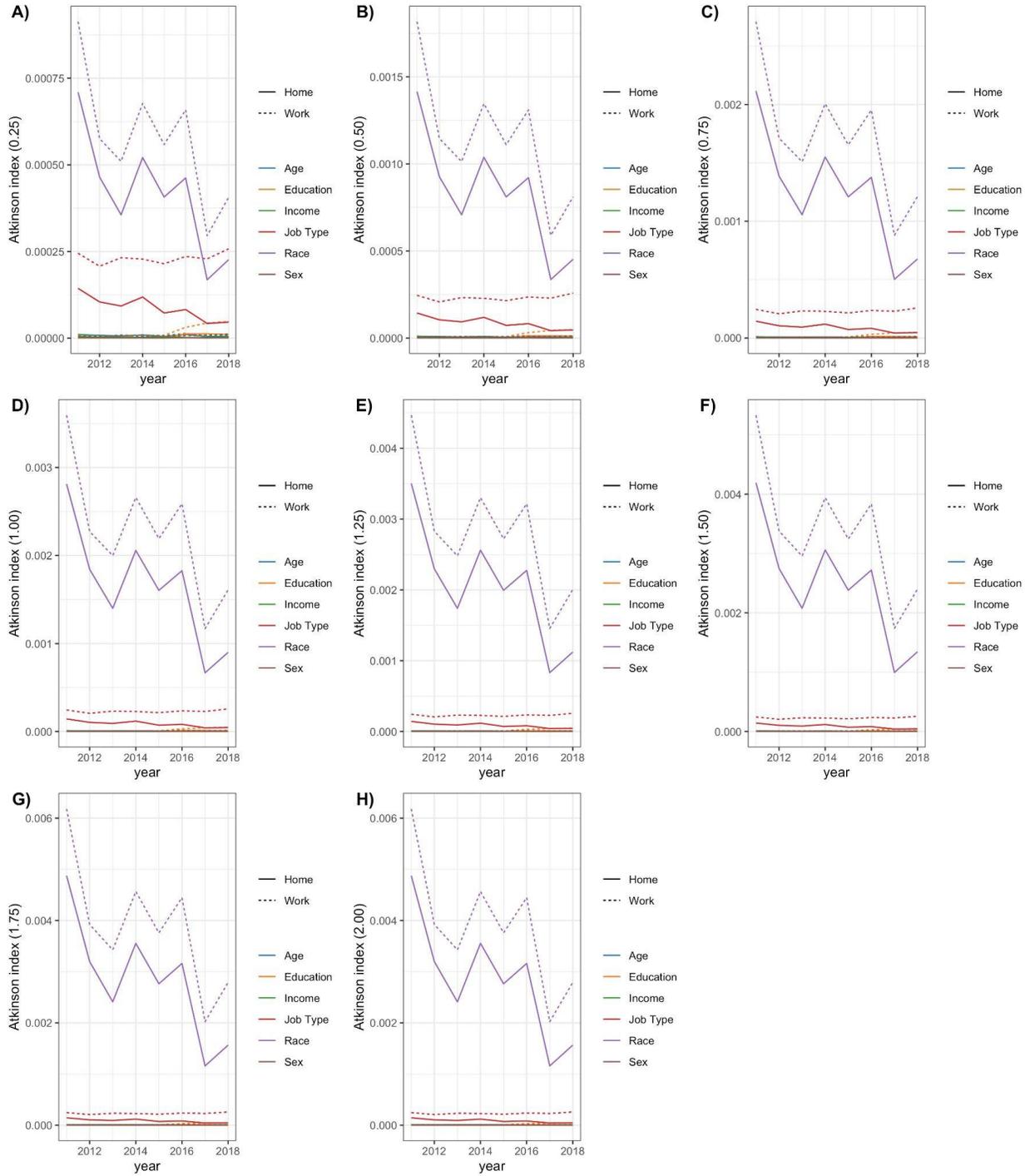

*Figure S8.4.4*: Atkinson Index for the years 2011 - 2018 (inequality aversion parameters = 0.25, 0.50, 0.75, 1.0, 1.25, 1.50, 1.75, 2.0) displaying between-group inequality in H and W for the years 2011 - 2018 for rural populations, alone.

## S8.5: Disparities relative to policy thresholds

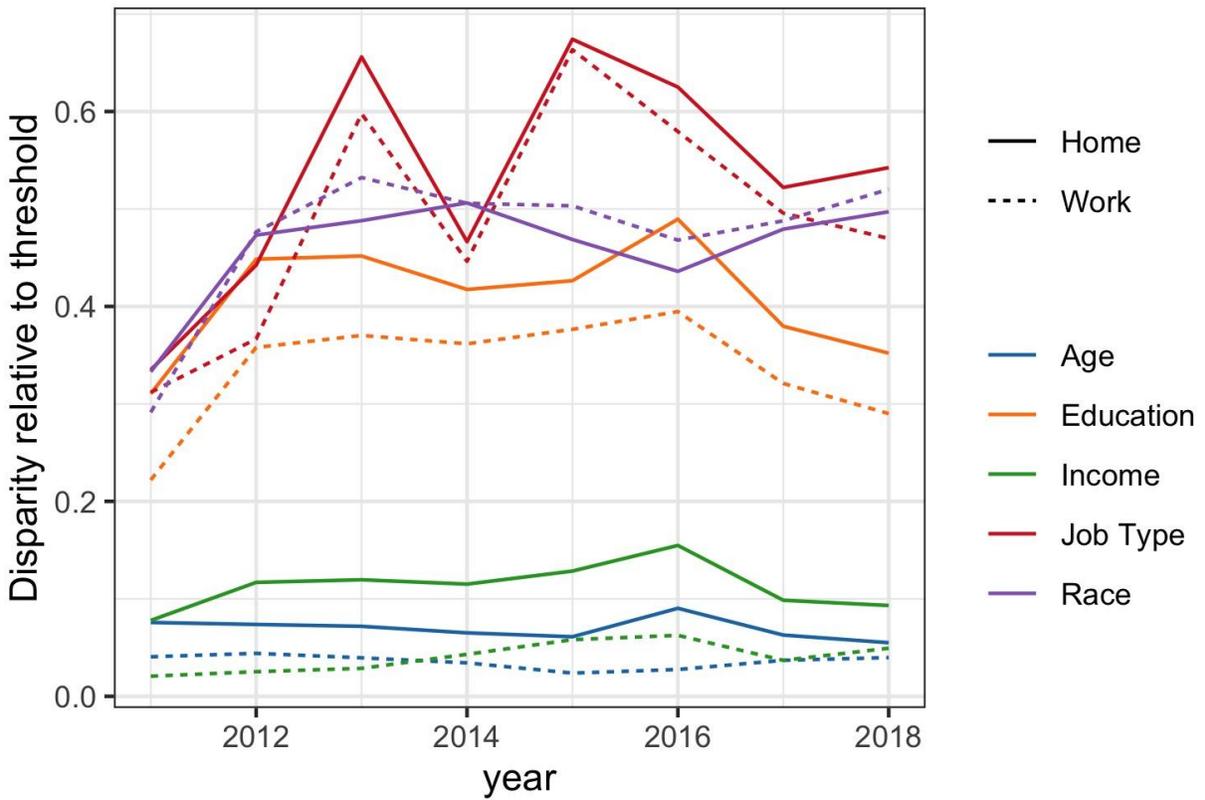

***Figure S8.5.1***: *Disparities relative to policy threshold of 12 µg/m³*

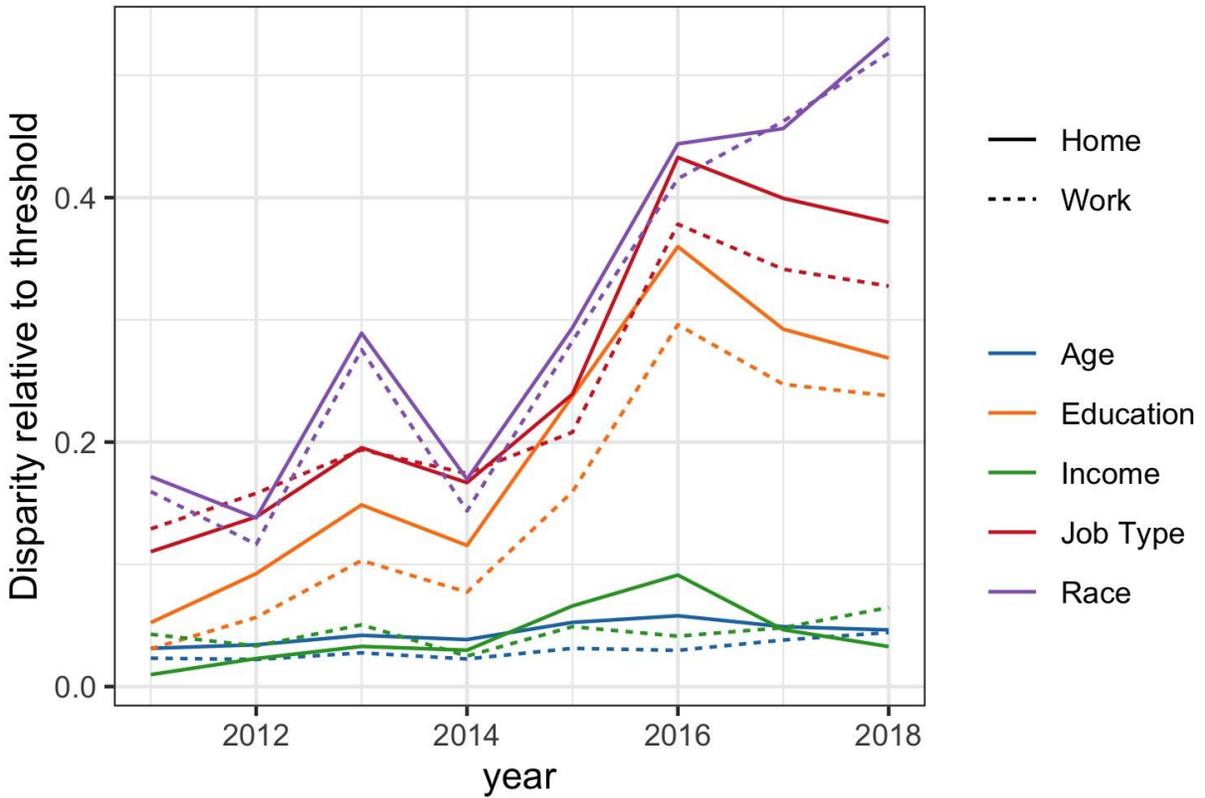

*Figure S8.5.2*: Disparities relative to policy threshold of 10 µg/m$^3$

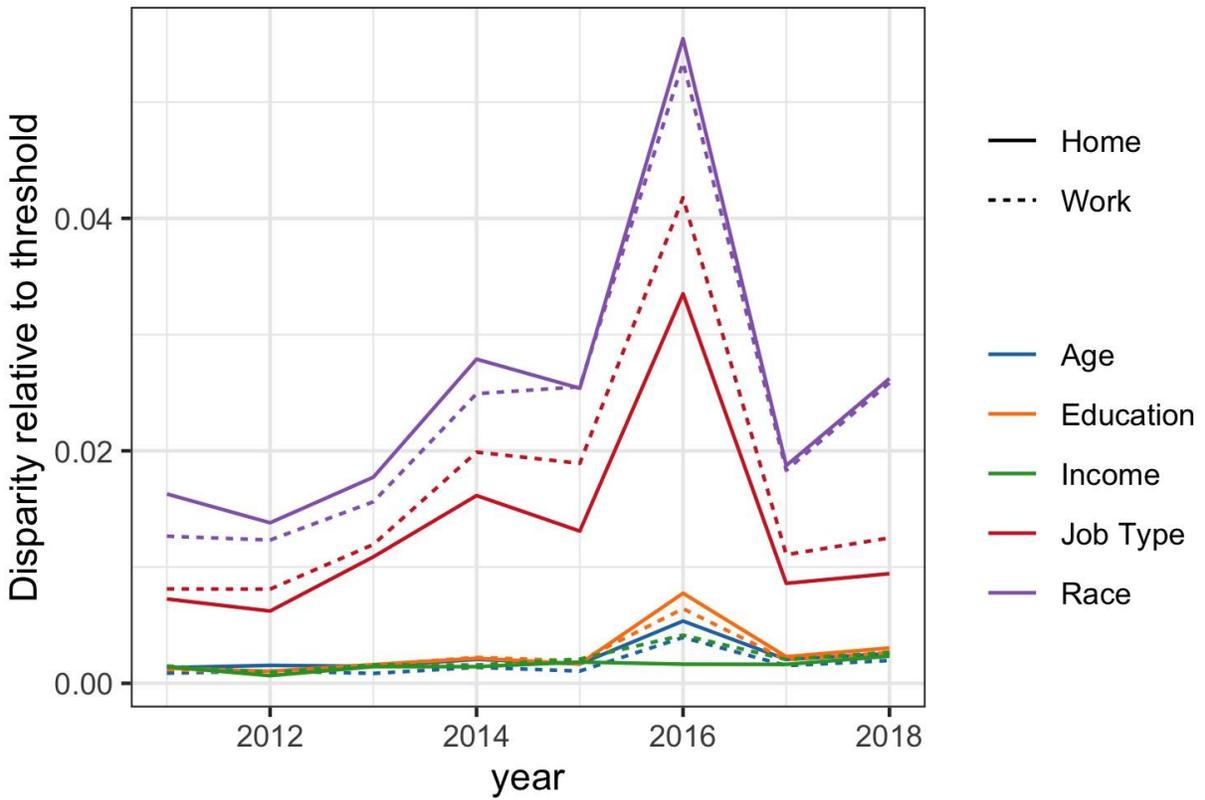

*Figure S8.5.3: Disparities relative to policy threshold of 5 µg/m³*

*Table S8.5.1: % Population exposed to H, W > 12 µg/m³*

|  | 2011 | | 2012 | | 2013 | | 2014 | | 2015 | | 2016 | | 2017 | | 2018 | |
| --- | --- | --- | --- | --- | --- | --- | --- | --- | --- | --- | --- | --- | --- | --- | --- | --- |
|  | Home | Work | Home | Work | Home | Work | Home | Work | Home | Work | Home | Work | Home | Work | Home | Work |
| All | 6.3% | 7.9% | 3.4% | 4.1% | 3.2% | 3.9% | 3.8% | 4.2% | 2.7% | 3.0% | 2.2% | 2.6% | 4.2% | 5.0% | 4.4% | 5.3% |
| Race | | | | | | | | | | | | | | | | |
| white | 5.8% | 7.4% | 3.2% | 3.9% | 3.0% | 3.6% | 3.5% | 4.0% | 2.5% | 2.8% | 2.0% | 2.4% | 3.9% | 4.7% | 4.1% | 5.0% |
| Black | 6.4% | 8.1% | 2.5% | 2.7% | 2.2% | 2.4% | 2.6% | 2.7% | 1.9% | 2.0% | 1.5% | 1.6% | 2.4% | 2.8% | 2.5% | 2.9% |
| Asian | 13.2% | 15.0% | 9.2% | 10.8% | 8.6% | 11.2% | 10.8% | 11.8% | 6.8% | 8.2% | 4.8% | 6.3% | 10.2% | 12.6% | 11.0% | 14.2% |
| Hispanic | 16.1% | 16.6% | 11.9% | 12.4% | 11.3% | 11.7% | 12.8% | 12.9% | 9.2% | 9.5% | 7.8% | 8.2% | 12.7% | 13.8% | 12.7% | 13.8% |
| Native | 8.0% | 9.2% | 5.8% | 6.7% | 5.4% | 6.0% | 6.1% | 6.6% | 4.4% | 4.8% | 3.8% | 4.3% | 6.7% | 7.7% | 7.0% | 7.9% |
| Hawaiian or Pacific Islander | 12.2% | 13.8% | 8.5% | 10.0% | 8.4% | 10.2% | 9.6% | 10.5% | 6.7% | 7.8% | 5.0% | 6.2% | 9.7% | 11.7% | 10.9% | 13.3% |
| Income | | | | | | | | | | | | | | | | |
| Income 1 (lowest) | 6.7% | 7.7% | 3.7% | 4.0% | 3.5% | 3.7% | 4.1% | 4.4% | 2.8% | 3.0% | 2.2% | 2.5% | 4.2% | 4.8% | 4.4% | 5.0% |
| Income 2 | 6.7% | 8.0% | 3.8% | 4.2% | 3.5% | 3.9% | 4.2% | 4.4% | 3.0% | 3.2% | 2.5% | 2.8% | 4.6% | 5.2% | 4.9% | 5.4% |
| Income 3 (highest) | 5.8% | 8.0% | 3.0% | 4.1% | 2.8% | 3.9% | 3.4% | 4.1% | 2.3% | 2.9% | 1.8% | 2.5% | 3.8% | 5.0% | 4.0% | 5.4% |
| Education | | | | | | | | | | | | | | | | |
| No highschool | 9.9% | 11.2% | 6.4% | 7.0% | 6.0% | 6.6% | 6.8% | 7.2% | 4.8% | 5.2% | 4.1% | 4.6% | 7.0% | 8.0% | 7.2% | 8.2% |
| Highschool | 5.4% | 7.1% | 2.8% | 3.4% | 2.6% | 3.1% | 3.1% | 3.4% | 2.2% | 2.5% | 1.8% | 2.1% | 3.4% | 4.1% | 3.6% | 4.4% |
| Some College | 6.0% | 7.8% | 3.2% | 4.0% | 3.0% | 3.7% | 3.6% | 4.1% | 2.5% | 2.9% | 2.0% | 2.5% | 3.9% | 4.8% | 4.1% | 5.1% |
| Advanced | 5.8% | 7.7% | 3.0% | 4.0% | 2.8% | 3.7% | 3.4% | 4.1% | 2.3% | 2.8% | 1.7% | 2.3% | 3.7% | 4.9% | 4.0% | 5.3% |
| Sex | | | | | | | | | | | | | | | | |
| Female | 6.3% | 7.6% | 3.4% | 3.9% | 3.1% | 3.6% | 3.7% | 4.0% | 2.6% | 2.8% | 2.1% | 2.4% | 4.1% | 4.8% | 4.3% | 5.1% |
| Male | 6.4% | 8.3% | 3.5% | 4.4% | 3.3% | 4.1% | 3.9% | 4.5% | 2.7% | 3.2% | 2.2% | 2.8% | 4.2% | 5.3% | 4.5% | 5.6% |
| Age | | | | | | | | | | | | | | | | |
| ≤ 29 years | 6.7% | 7.8% | 3.6% | 3.9% | 3.4% | 3.7% | 3.9% | 4.1% | 2.8% | 3.0% | 2.3% | 2.5% | 4.3% | 4.9% | 4.5% | 5.2% |
| 30 - 54 years | 6.4% | 8.2% | 3.5% | 4.3% | 3.3% | 4.0% | 3.9% | 4.4% | 2.7% | 3.1% | 2.2% | 2.7% | 4.2% | 5.2% | 4.5% | 5.5% |
| ≥ 55 yeas | 5.7% | 7.5% | 3.1% | 4.0% | 2.9% | 3.7% | 3.5% | 4.1% | 2.5% | 3.0% | 1.9% | 2.5% | 3.8% | 4.8% | 4.1% | 5.1% |
| Job-Type | | | | | | | | | | | | | | | | |

| Industry | | | | | | | | | | | | | | | | |
|---|---|---|---|---|---|---|---|---|---|---|---|---|---|---|---|---|
| Agriculture | 14.3% | 14.4% | 10.1% | 9.1% | 13.6% | 15.1% | 11.8% | 12.5% | 11.5% | 13.0% | 8.6% | 9.5% | 13.8% | 15.7% | 15.1% | 15.4% |
| Mining | 2.5% | 2.5% | 1.7% | 1.8% | 1.6% | 1.7% | 1.8% | 1.9% | 1.5% | 1.6% | 1.3% | 1.2% | 1.5% | 1.4% | 1.4% | 1.3% |
| Utilities | 6.4% | 9.4% | 4.1% | 6.4% | 3.7% | 6.0% | 4.1% | 5.5% | 3.0% | 4.5% | 2.4% | 4.2% | 4.2% | 6.8% | 4.6% | 7.2% |
| Construction | 5.3% | 6.7% | 3.0% | 3.7% | 2.9% | 3.4% | 3.5% | 3.9% | 2.5% | 2.8% | 2.1% | 2.3% | 3.8% | 4.3% | 4.2% | 5.1% |
| Manufacturing | 6.2% | 8.5% | 3.5% | 4.9% | 3.3% | 4.6% | 3.7% | 4.7% | 2.7% | 3.8% | 2.2% | 3.1% | 3.9% | 5.1% | 4.0% | 5.6% |
| Wholesale | 7.3% | 10.6% | 4.2% | 6.3% | 4.0% | 5.7% | 4.8% | 6.2% | 3.4% | 5.0% | 2.8% | 4.6% | 5.0% | 6.8% | 5.1% | 7.2% |
| Retail | 6.0% | 7.0% | 3.2% | 3.4% | 3.0% | 3.2% | 3.6% | 3.8% | 2.5% | 2.6% | 2.0% | 2.1% | 3.8% | 4.3% | 4.0% | 4.4% |
| Transportation & Warehousing | 7.5% | 13.5% | 3.9% | 6.1% | 3.7% | 5.2% | 4.3% | 5.4% | 3.2% | 4.5% | 2.8% | 4.5% | 4.8% | 6.9% | 5.0% | 7.4% |
| Information | 9.2% | 12.0% | 4.7% | 6.9% | 3.9% | 4.7% | 5.5% | 7.0% | 3.0% | 3.2% | 2.3% | 4.3% | 6.2% | 9.3% | 6.5% | 10.4% |
| Finance & Insurance | 5.5% | 7.0% | 2.6% | 3.4% | 2.2% | 3.1% | 2.8% | 3.1% | 1.8% | 2.0% | 1.4% | 1.6% | 3.0% | 4.0% | 3.1% | 4.1% |
| Real estate | 6.7% | 7.9% | 3.6% | 3.9% | 3.3% | 3.6% | 4.1% | 4.1% | 2.6% | 2.8% | 2.1% | 2.1% | 4.5% | 5.2% | 4.6% | 5.3% |
| Professional, Scientific, Technical Services | 5.8% | 7.9% | 2.9% | 3.9% | 2.6% | 3.6% | 3.3% | 4.1% | 2.0% | 2.8% | 1.5% | 2.1% | 3.5% | 5.2% | 3.8% | 5.3% |
| Management of companies | 6.2% | 8.5% | 2.9% | 3.9% | 2.7% | 4.2% | 3.2% | 3.8% | 2.2% | 2.7% | 1.7% | 2.3% | 3.4% | 5.0% | 3.7% | 5.2% |
| Waste Management | 7.3% | 9.1% | 4.0% | 4.8% | 3.6% | 4.3% | 4.4% | 4.8% | 3.0% | 3.6% | 2.6% | 3.0% | 4.6% | 5.6% | 4.8% | 6.0% |
| Education | 5.7% | 6.6% | 3.1% | 3.4% | 2.9% | 3.3% | 3.4% | 3.6% | 2.5% | 2.7% | 1.9% | 2.2% | 3.8% | 4.1% | 4.0% | 4.3% |
| Healthcare | 5.5% | 6.5% | 2.8% | 2.9% | 3.1% | 3.2% | 3.7% | 3.9% | 2.6% | 2.4% | 2.1% | 2.0% | 4.1% | 4.5% | 4.5% | 4.8% |
| Arts, Entertainment & Recreation | 6.9% | 8.3% | 4.0% | 4.4% | 3.6% | 4.2% | 4.6% | 4.7% | 3.0% | 3.6% | 2.4% | 3.2% | 4.9% | 5.6% | 5.1% | 5.7% |
| Accommodation | 6.5% | 7.1% | 3.5% | 3.3% | 3.2% | 3.3% | 4.0% | 3.8% | 2.7% | 2.6% | 2.2% | 2.1% | 4.4% | 4.8% | 4.6% | 4.8% |
| Other services | 9.3% | 10.6% | 5.8% | 6.3% | 3.1% | 3.4% | 3.9% | 3.9% | 2.6% | 2.7% | 2.1% | 2.1% | 4.1% | 4.6% | 4.4% | 4.9% |
| Public Administration | 5.6% | 8.0% | 3.3% | 5.0% | 3.1% | 4.2% | 3.5% | 4.5% | 2.5% | 3.1% | 2.0% | 3.1% | 3.7% | 4.6% | 3.9% | 5.1% |